\PassOptionsToPackage{table}{xcolor}
\documentclass[twocolumn,tighten]{aastex63}

\usepackage{amsmath}
\usepackage{xspace}
\usepackage{pifont}

\newcommand{\keyw}[1]{\textcolor{gray}{#1}}

\usepackage{longtable}
 
\begin{document}
\title{Porous Dust Particles in Protoplanetary Disks: Application to the HL Tau Disk}

\correspondingauthor{Shangjia Zhang}
\email{shangjia.zhang@unlv.edu}

\author[0000-0002-8537-9114]{Shangjia Zhang}
\affiliation{Department of Physics and Astronomy, University of Nevada, Las Vegas, 4505 S. Maryland Pkwy, Las Vegas, NV, 89154, USA}
\affiliation{Nevada Center for Astrophysics, University of Nevada, Las Vegas, Las Vegas, NV 89154, USA}

\author[0000-0003-3616-6822]{Zhaohuan Zhu}
\affiliation{Department of Physics and Astronomy, University of Nevada, Las Vegas, 4505 S. Maryland Pkwy, Las Vegas, NV, 89154, USA}
\affiliation{Nevada Center for Astrophysics, University of Nevada, Las Vegas, Las Vegas, NV 89154, USA}

\author[0000-0003-4902-222X]{Takahiro Ueda}
\affiliation{National Astronomical Observatory of Japan, Osawa 2-21-1, Mitaka, Tokyo 181-8588, Japan}
\affiliation{Max-Planck Institute for Astronomy (MPIA), K{\"o}nigstuhl 17, D-69117 Heidelberg, Germany}

\author[0000-0003-4562-4119]{Akimasa Kataoka}
\affiliation{National Astronomical Observatory of Japan, Osawa 2-21-1, Mitaka, Tokyo 181-8588, Japan}

\author[0000-0002-5991-8073]{Anibal Sierra}
\affiliation{Departamento de Astronom\'{i}a, Universidad de Chile, Camino El Observatorio 1515, Las Condes, Santiago, Chile}

\author[0000-0003-2862-5363]{Carlos Carrasco-Gonz\'{a}lez}
\affiliation{Instituto de Radioastronom\'{i}a y Astrof\'{i}sica (IRyA), Universidad Nacional Aut\'{o}noma de M\'{e}xico (UNAM), Mexico}

\author[0000-0003-1283-6262]{Enrique Mac\'{i}as}
\affiliation{ESO Garching, Karl-Schwarzschild-Str. 2, 85748, Garching bei Munchen, Germany}

\begin{abstract}
Dust particle sizes constrained from dust continuum and polarization observations by radio interferometry are inconsistent by at least an order of magnitude. Motivated by porous dust observed in small Solar System bodies (e.g., from the \textit{Rosetta} mission), we explore how the dust particle's porosity affects the estimated particle sizes from these two methods. Porous particles have lower refractive indices, which affect both opacity and polarization fraction. With weaker Mie interference patterns, the porous particles have lower opacity at mm wavelengths than the compact particles if the particle size exceeds several hundred microns.
Consequently, the inferred dust mass using porous particles can be up to a factor of six higher. The most significant difference between compact and porous particles is their scattering properties. The porous particles have a wider range of particle sizes with high linear polarization from dust self-scattering, allowing mm-cm-sized particles to explain polarization observations. With a Bayesian approach, we use porous particles to fit HL Tau disk's multi-wavelength continuum and mm-polarization observations from ALMA and VLA. The moderately porous particles with sizes from 1 mm-1 m can explain both continuum and polarization observations, especially in the region between 20-60 au. If the particles in HL Tau are porous, the porosity should be from 70\% to 97\% from current polarization observations. We also predict that future observations of the self-scattering linear polarization at longer wavelengths (e.g., ALMA B1 and ngVLA) have the potential to further constrain the particle's porosity and size.

\end{abstract}


\section{Introduction \label{sec:intro}}

Dust plays a fundamental role in the secular evolution of the protoplanetary disk by setting thermal structure and participating in dynamics as they aggregate and collapse to form planetesimals, eventually planets \citep{lesur22}. As dust particles are the primary source of opacity in protoplanetary disks, their presumed opacity is used to convert the radio continuum emissions to the disk dust mass. However, the opacity itself depends on the particle properties (such as composition, size, temperature, and distribution). Recently, ALMA continuum polarization observations provide an independent constraint on dust particle properties. Both dust continuum observation and polarization observation should work in synergy, increasing our knowledge of dust properties (e.g., a review from \citealt{miotello22}).

\subsection{Particle Size Problem}
If the polarization is due to dust self-scattering, ALMA (sub)mm polarization observations can tightly constrain the particle size. In order to produce polarization with dust self-scattering, the radiation field needs to be anisotropic so that the polarization due to light from different directions does not cancel out. This condition makes inclined and substructured disks ideal for observing self-scattering polarization \citep{kataoka15, kataoka16, yang16}. One of the best examples for protoplanetary disks are HL Tau \citep{stephens14, stephens17}; HD 142527 \citep{kataoka16b, ohashi18}; IM Lup \citep{hull18}; CW Tau, DG Tau \citep{bacciotti18}; AS 209 \citep{mori19}; and HD 163296 \citep{dent19, ohashi19}. Among these disks, the dust continuum polarization fractions are detected as $\sim$1\% at ALMA bands 6 and 7 ($\sim$1.3 mm, and 0.87 mm, respectively). Another condition requires strong dust scattering at mm wavelengths, and the scattered light needs to be polarized. For compact particles, the scattering albedo becomes substantial if the particle size $a\gtrsim \lambda/2\pi$, but the fraction of the polarized scattered light sharply drops to zero beyond that same particle size. These competing effects result in only a narrow range of particle sizes that can produce detectable polarization at a particular wavelength. In other words, the polarization fraction of the total intensity is very sensitive to the particle size. The inferred particle sizes are around 100 $\mu$m. particle sizes larger or smaller than this produce much lower linearly polarized emissions. However, this value is at least an order of magnitude less than the particle size constrained from multi-wavelength continuum observations at radio bands (e.g., a review from \citealt{andrews20}).

Traditionally, multi-wavelength dust continuum observations are used to constrain the particle size. If the disk is optically thin, the spectral index ($\alpha = \mathrm{dln}I(\nu)/\mathrm{dln\nu}$) between two wavelengths reveals the slope of dust absorption opacity ($\beta = \mathrm{dln}\mathrm{\kappa_{abs}}(\nu)/\mathrm{dln}\nu$). In the Rayleigh-Jeans limit, $\alpha=\beta$+2. With an assumption of dust composition, the particle size can be inferred from $\beta$. The typical $\beta$ from diffuse clouds is $\sim$1.7 from far-infrared and submillimeter observations \citep{finkbeiner99, li01}, whereas $\beta$ from most of the protoplanetary disks at radio bands are significantly $\lesssim$1, indicating mm to cm sized particles (e.g., \citealt{dalessio01, calvet02, draine06}). Detailed radio observations that resolve the disk show that $\alpha$ increases with radius \citep{perez15, tazzari16} and within dark gaps \citep{huang18}, suggesting particle sizes are smaller in the outer disk and the gaps if the emission is optically thin, consistent with the theoretical prediction of dust radial drift and dust trapping (e.g., \citealt{weidenschilling93}). Nevertheless, particles are still $\sim$1 mm even at these lower ends.

However, if the disk is optically thick, $\alpha$ will be $\sim$2 and cannot be used to constrain the particle size. Dust self-scattering can weaken the dust's thermal emission and affect the $\alpha$ value, even decreasing the $\alpha$ value below the blackbody value \citep{miyake93, liu19, zhu19b, sierra20}. The scattering albedo affects the thermal emission, which makes the particle size constraint more difficult with just dual-wavelength observations \citep{huang20}. In such a case, observations of three or more wavelengths are fit in tandem \citep{carrasco19, ueda20, macias21, sierra21, ueda21, ueda22, guidi22}. With scattering considered, the smallest particle sizes can be as low as several hundred microns in some regions, but mm-cm-sized particles are still valid solutions in most of the disks \citep{ueda20, ueda21, ueda22, sierra21, guidi22}.

Modeling the disk in detail with both radio continuum SEDs (Spectral Energy Distributions) and polarization observations can give us more stringent constraints on the particle sizes and other disk properties. \citet{lin20} model HD 163296 with the spectral index between ALMA bands 6 and 7 and polarization at band 7 to show that purely 90 $\mu$m sized particles can explain both types of observations. \citet{ueda21} study the particle size in the HL Tau's inner disk ($\lesssim$20 au) by fitting SEDs from 4 ALMA/VLA bands and polarization measurements from 3 ALMA bands. They consider dust vertical settling with mm-sized particles concentrated in the midplane and 100 $\mu$m-sized particles in the disk atmosphere. They find that mm-sized particles can still be present in the disk as long as they are highly settled and hidden in the optically thick midplane. Small 100-$\mu$m-sized particles in the atmosphere contribute to the detectable polarization emissions \citep{ueda21, sierra20, brunngraber21}.

\subsection{Consideration of Porosity}
In this paper, we extend the work by \citet{ueda21} and study both the inner and outer regions of the HL Tau disk, focusing on the influence of particle porosity on SED and polarization fittings. This is inspired by recent small Solar System objects missions (e.g., the \textit{Rosetta} mission). Cometary nuclei are considered to be the remnants of the planetesimals \citep{weissman20} and are highly porous \citep{groussin19}. The km-sized cometary nucleus 67P/Churyumov–Gerasimenko has a porosity of 70\%-85\% constrained by measuring the dust's permittivity \citep{kofman15, herique16} and the nucleus density \citep{jorda16, patzold16} from the \textit{Rosetta} mission. In addition, porous particles are detected \citep{bentley16, mannel16, mannel19} and account for more volume fractions than the compact particles among the collected samples \citep{merouane16}. While it is not clear whether the micro-porosity solely contributes to the total porosity, voids should be no larger than $\sim$ 9 meters from radar measurements \citep{ciarletti17}, and models \citep{blum17, burger23} with a combination of particle porosity (micro-porosity; \citealt{weidling09, zsom10}) and random packing of the particles \citep{skorov12, fulle17} can closely explain the observations from the \textit{Rosetta} mission.\footnote{We declare nomenclature on dust particles following \citet{guttler19}. We use \textit{particle} for any unspecified dust particle; \textit{grain} (or \textit{monomer}) as the smallest solid particle with homogeneous composition (compact with zero porosity). A \textit{dense aggregate} is an assemblage of rigidly joint grains, with porosity $<$10\%. A \textit{porous agglomerate} is composed of grains or dense aggregates with porosities between 10\%-99\%. A \textit{fluffy/fractal agglomerate} shows a fractal and dendritic nature with porosity $>$99\%. In this paper, we focus on \textit{dense aggregates} and \textit{porous agglomerates}. At a population level, we use \textit{dust} to refer to the solid component in protoplanetary disks that contrasts the gas component.}

Theory and experiments support the existence of porous particles in protoplanetary disks as micron-sized dust particles coagulate to form fluffy agglomerates \citep{ossenkopf93, weidenschilling93, wurm98, kempf99, krause04, okuzumi12, kataoka13, krijt15, estrada16, estrada22a, estrada22b}. Experiments reveal that the mm-cm-sized (pebble-sized) agglomerates that cannot surpass the bouncing barrier experience compression and eventually reach an equilibrium with porosity $\sim$ 64\% \citep{weidling09, zsom10}. Observationally, near-infrared scattered-light imaging starts to shed light on the small particle's porosity in protoplanetary disks. Recent analyses on the polarized scattered light phase functions observed by VLT/SPHERE reveal that micron-sized dust particles are porous in all 10 disks in their sample \citep{ginski23}. Particularly in IM Lup disk, particles are found to be fractal agglomerates with sizes $\sim$ 2 $\mu$m composed of $\sim$ 200 nm monomers \citep{tazaki23}.

For our problem, porous particles may help alleviate the size inconsistency in two ways. First, moderately porous particles (porous agglomerates) have a weaker size dependence of the self-scattering polarization fraction at mm wavelengths \citep{tazaki19, brunngraber21}, so the particle size does not need to be precisely 100 $\mu$m to match the high polarization fraction in observations. Second, the absorption and scattering opacities of the compact particles have strong interference patterns when 2$\pi a$ $\gtrsim\lambda$. The absorption opacity of a single-sized particle oscillates with sizes and wavelengths in high amplitude due to the interference. 
These interference features are typically at mm-cm wavelengths for mm-cm-sized particles, providing a wide range of spectral index values at mm wavelengths. Thus, the favorable particle sizes from SED fittings often fall into mm-cm size regime. Porous particles have weaker interference features in the absorption opacity \citep{kataoka14}, making it more difficult to constrain the particle size in the mm-cm size range by fitting the spectral index.
Allowing a non-zero porosity leaves more freedom in fitting SED and mm polarization observations in tandem. While we only focus on the impact of spherical porous particles in the current paper, we note that non-spherical compact particles \citep{kirchschlager20} and non-spherical porous particles \citep{kirchschlager19} are also able to change the inferred opacity and polarization.

As the optical properties of dust particles are the key to understanding the fitting results, in Section \ref{sec:dust}, we give a holistic view of dust properties on wavelengths, sizes, and levels of porosity. In Section \ref{sec:analytical}, we present the analytical fitting of SEDs considering different levels of porosity. In Section \ref{sec:MCRT}, we pick some best-fit models to run Monte Carlo Radiative Transfer (MCRT) to further compare with SEDs and polarization observations. 
In Section \ref{sec:dustmass}, we emphasize the significant impact of porous particles on dust mass estimation. In Section \ref{sec:f0.01}, we test whether an even lower value of filling factor can reproduce two types of observations. In Section \ref{sec:slope}, we discuss the impact of a different particle size distribution. In Section \ref{sec:constraints}, we use current polarization observations to constrain the possible range of the particle porosity in HL Tau. In Sections \ref{sec:predictions}, we lay out the approach to testing the prediction in the future. We also compare our results with recent near-infrared scattered light observations and the \textit{Rosetta} mission in Section \ref{sec:connection}. In Section \ref{sec:summary}, we use Table \ref{tab:summary} and Figure \ref{fig:money_plot} to summarize all possible solutions of particle sizes and porosities for HL Tau. Readers who prefer to skip the technical details can directly proceed to this section. We conclude our paper in Section \ref{sec:conclusion}.

\section{Particle Optical Properties\label{sec:dust}}
Particle optical properties are generated by mixing different materials using the Effective Medium Approximation (EMA; see Appendix \ref{sec:applicability} for its validity). This process is essentially averaging the complex refractive indices of each component with certain rules to obtain a composite refractive index, $m(\lambda)$=$n(\lambda)+ik(\lambda)$. We adopt the DSHARP composition, which contain 20\% water ice \citep{warren08}, 33\% astronomical silicates \citep{draine03}, 7\% troilite \citep{henning96}, and 40\% refractory organics\footnote{Including more absorptive carbonaceous materials can change the refractive index significantly and impact the polarization fraction \citep{yang20}. However, we only explore the composition with refractory organics in our paper.} \citep{henning96} in mass fractions \citep{birnstiel18}. To include porosity, we mix the DSHARP composition with vacuum using the Burggeman rule. The fraction of the vacuum is controlled by the porosity, $\mathcal{P}$, the volume fraction of the vacuum in the dust particle. Another widely used parameter is the filling factor, $f$, where $f$ = 1-$\mathcal{P}$. Lastly, we calculate the particle optical properties using the Mie theory with the refractive index as an input.

We describe the behavior of four crucial particle optical properties generated by the Mie theory to predict the dust continuum emission and mm polarization for various levels of porosity. Understanding their behaviors helps fit and interpret observations in Sections \ref{sec:analytical} and \ref{sec:MCRT}. The four properties are absorption opacity ($\kappa_{\mathrm{abs}}$), effective scattering opacity ($\kappa_{\mathrm{sca, eff}}$)\footnote{Particularly, the scattering opacity is the integrated value along different scattering angles. We use the effective scattering opacity $\kappa_{\mathrm{sca, eff}}$, rather than $\kappa_{\mathrm{sca}}$, in that when $a\gtrsim\lambda$/2$\pi$, the angular dependent scattering opacity has a strong peak within a small angle in the forward direction. Since the forward scattering does not change the direction of the light, it is a common practice to use a forward-scattering adjustment factor $g$ to truncate the values within a small angle close to the forward scattering. In this sense, $\kappa_{\mathrm{sca, eff}}$ is a better approximation than the un-adjusted value, $\kappa_{\mathrm{sca}}$. The same goes to the effective albedo $\omega_{\mathrm{eff}}$.}, effective albedo ($\omega_{\mathrm{eff}}$), and polarization fraction at 90$^\circ$ ($P$). The definition of these parameters can be found in Appendix \ref{sec:keyparameters}. $\kappa_{\mathrm{abs}}$, $\kappa_{\mathrm{sca, eff}}$, and $\omega_{\mathrm{eff}}$, along with density and temperature, determine the continuum emission. The product of $\omega_{\mathrm{eff}}$ and $P$ indicates the total linear polarization fraction \citep{kataoka15}.

On the particle size constraint from SEDs, the increase of the opacity due to Mie interference alters the opacity index ($\beta$), leading to different interpretations of the spectral index ($\alpha$) based on compact and porous particle assumptions. The changing of $\kappa_{\mathrm{sca, eff}}$ and $\omega_{\mathrm{eff}}$ further complicates the interpretation when the disk is optically thick.

In the following, we use an example with $af$ = 160 $\mu$m to introduce the behavior of $\kappa_{\mathrm{abs}}$ and $\kappa_{\mathrm{sca, eff}}$. We use $af$ to characterize the particle size since $a$ and $f$ are fully degenerate for $\kappa_{\mathrm{abs}}$, except at the Mie interference regime.

\subsection{Opacities at \texorpdfstring{$af$}{Lg} = 160 \texorpdfstring{$\mu$}{Lg}m \label{sec:example}}

\begin{figure*}[t!]
\includegraphics[width=\linewidth]{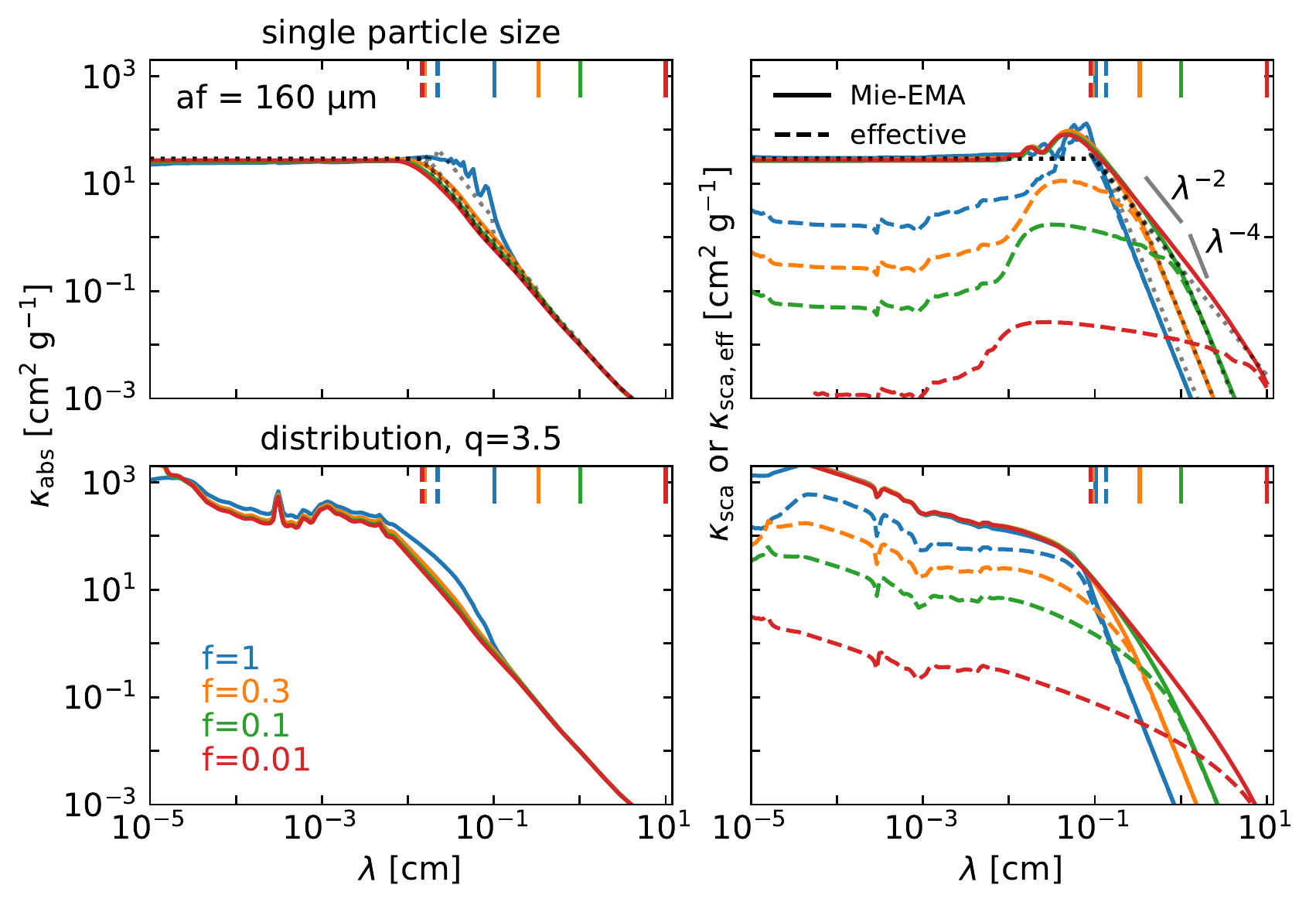}
\figcaption{\label{fig:demo} The wavelength dependence of the absorption opacity ($\mathrm{\kappa_{abs}}$), un-adjusted scattering opacity or effective scattering opacity ($\kappa_{\mathrm{sca}}$ or $\kappa_{\mathrm{sca,eff}}$) are shown from left to right. Top panels are single-particle-size values, whereas bottom panels are the values with particle size distribution $n(a)\propto a^{-3.5}$. Four filling factors are 1 ({\color{blue}blue}), 0.3 ({\color{orange}orange}), 0.1 ({\color{green}green}) and 0.01 ({\color{red}red}). Solid lines are the total scattering opacities calculated from the Mie theory ($\kappa_{\mathrm{sca}}$), whereas the dashed lines are scattering opacities with forward scattering truncated ($\kappa_{\mathrm{sca,eff}}$). The dotted lines are analytical approximations of $\mathrm{\kappa_{abs}}$ and $\mathrm{\kappa_{sca}}$ in \citet{bohren83, kataoka14}. The vertical solid lines represent $x$=1, delineating Rayleigh and Mie regimes. The vertical dashed lines are locations where $kx=3/8$ (for absorption) or $(n-1)x=1$ (for scattering), delineating Mie's optically thin and thick regimes for the dust particles. $x$ is the size parameter; $x=2\pi a/\lambda$ or $2\pi a_{\mathrm{max}}/\lambda$; see Appendices \ref{sec:keyparameters}, \ref{sec:analyticalapprox}, and \ref{sec:scat}. The {\color{green}green} vertical dashed line is almost at the same location as the {\color{red}red} one.}
\end{figure*}

For absorption opacities $\kappa_{\mathrm{abs}}$, the values for various filling factors are similar, except at $a\sim\lambda/2\pi$, where only the compact particle has higher opacity due to the Mie interference. This is demonstrated in the upper left panel of Figure \ref{fig:demo}, where it shows single-sized DSHARP opacities for $af$ = 160 $\mu$m along wavelengths for various filling factors. $f$ = 1 is the compact particles (represented by {\color{blue}blue} lines), whereas $f$=0.3, 0.1, and 0.01  are porous agglomerates represented by {\color{orange}orange}, {\color{green}green}, and {\color{red}red} lines, respectively.

The upper right panel of Figure \ref{fig:demo} shows the un-adjusted scattering opacities, $\kappa_{\mathrm{sca}}$ (solid lines) and effective scattering opacities adjusted for the forward scattering, $\kappa_{\mathrm{sca, eff}}$ (dashed lines). We see that the effective scattering opacities have much lower values at shorter wavelengths compared to $\kappa_{\mathrm{sca}}$. For $\kappa_{\mathrm{sca}}$, $a$ and $f$ are also degenerate for different filling factors at short wavelengths ($\lambda \lesssim 2\pi a$). However, the opacities are separated between various levels of porosity at longer wavelengths ($\lambda \gtrsim 2\pi a$). The effective scattering opacities, $\kappa_{\mathrm{sca, eff}}$, are well separated at all wavelengths since the dependence of forward scattering does not have degeneracy between $a$ and $f$.

In reality, particles with a size distribution, rather than a single size, are a better approximation in protoplanetary disks. Thus, we generate opacity with a power-law size distribution, $n(a)\propto a^{-q}$, often used in the literature. We take the size distribution index $q$=3.5 as a typical ISM value \citep{mathis77}. The minimum particle size, $a_{\mathrm{min}}$, is 0.1 $\mu$m. At mm-cm wavelengths, $a_{\mathrm{min}}$ is not important in averaging the opacity if $a_{\mathrm{min}}$ $\ll$ $a_{\mathrm{max}}$ and $q <$ 4 (the limit where the mass is dominated by the dust particles in the top-heavy mass distribution). The resulting absorption and scattering opacities are shown at the bottom panels of Figure \ref{fig:demo}. Since the small particles have higher opacities at shorter wavelengths, the resulting opacities are higher than the values shown on the top panel at short wavelengths. Other than the short wavelength regime, the overall trends of the opacity are similar to the single-sized opacity.

We note that the particle size distribution is not well-constrained from observations. While simple steady state solutions between coagulation and fragmentation lead to a power-law distribution with $q$=3.5 \citep{dohnanyi69, tanaka96}, simulations with dust growth-fragmentation in disks give a complicated distribution, which cannot be described as a power-law \citep{birnstiel11}. We use $q$=3.5 as a fiducial model in this paper, and discuss the impact of $q$ in Sections \ref{sec:slope} and \ref{sec:constraints}.

We also want to mention that absorption opacity ($\kappa_{\mathrm{abs}}$) alone strongly impacts the \textit{dust mass} estimation from continuum observations. When $\lambda \sim 2\pi a$, the $\kappa_{\mathrm{abs}}$ for compact particles is around six times higher than that for porous particles (for $f$=0.1). This opacity increase is due to Mie interference unique to the compact particles with a higher refractive index. Thus, with mm-cm-sized particles, adopting compact particles instead of porous particles can increase the dust continuum opacity at radio bands by a factor of six or more and decrease the mass estimate by the same factor. We will use Section \ref{sec:dustmass} to emphasize its significance after discussing how it can alleviate the particle size problem.

In Appendices \ref{sec:analyticalapprox} to \ref{sec:appendixA}, we follow \citet{kataoka14} to develop a simple model to understand the particle opacity at various regimes. The simple model provides a clear physical picture obscured by the direct Mie calculations. These insights are general and not related to a specific dust composition mixture. The model is shown by the dotted lines in the upper panels of Figure \ref{fig:demo}. The vertical dashed lines also show three regimes for the model, delineated by the size parameter ($x$=$2\pi a/\lambda$), and the complex refractive index (see Appendices \ref{sec:keyparameters}, \ref{sec:analyticalapprox}, and \ref{sec:scat}). On the left of the vertical-dashed line, the particle is in \textit{Mie's optically thick regime}. Between the vertical-dashed line and the vertical solid line, the particle is in \textit{Mie's optically thin regime}\footnote{We note that the optical depth here is referred to the optical depth \textit{inside} a particle.}. On the right of the vertical-solid line, the particle is in \textit{Rayleigh regime}. The analytical approximation fits the DSHARP opacities ($\kappa_{\mathrm{abs}}$, and $\kappa_{\mathrm{sca}}$) perfectly at short and long wavelengths except for $\lambda\sim 2\pi a$ in Mie's optically thin regime.  A more detailed explanation on the analytical approximations can be found in Appendices \ref{sec:analyticalapprox} and \ref{sec:scat}.

\subsection{More Particle Sizes and Polarization Fraction \label{sec:parameterspace}}

\begin{figure*}[t!]
\includegraphics[width=\linewidth]{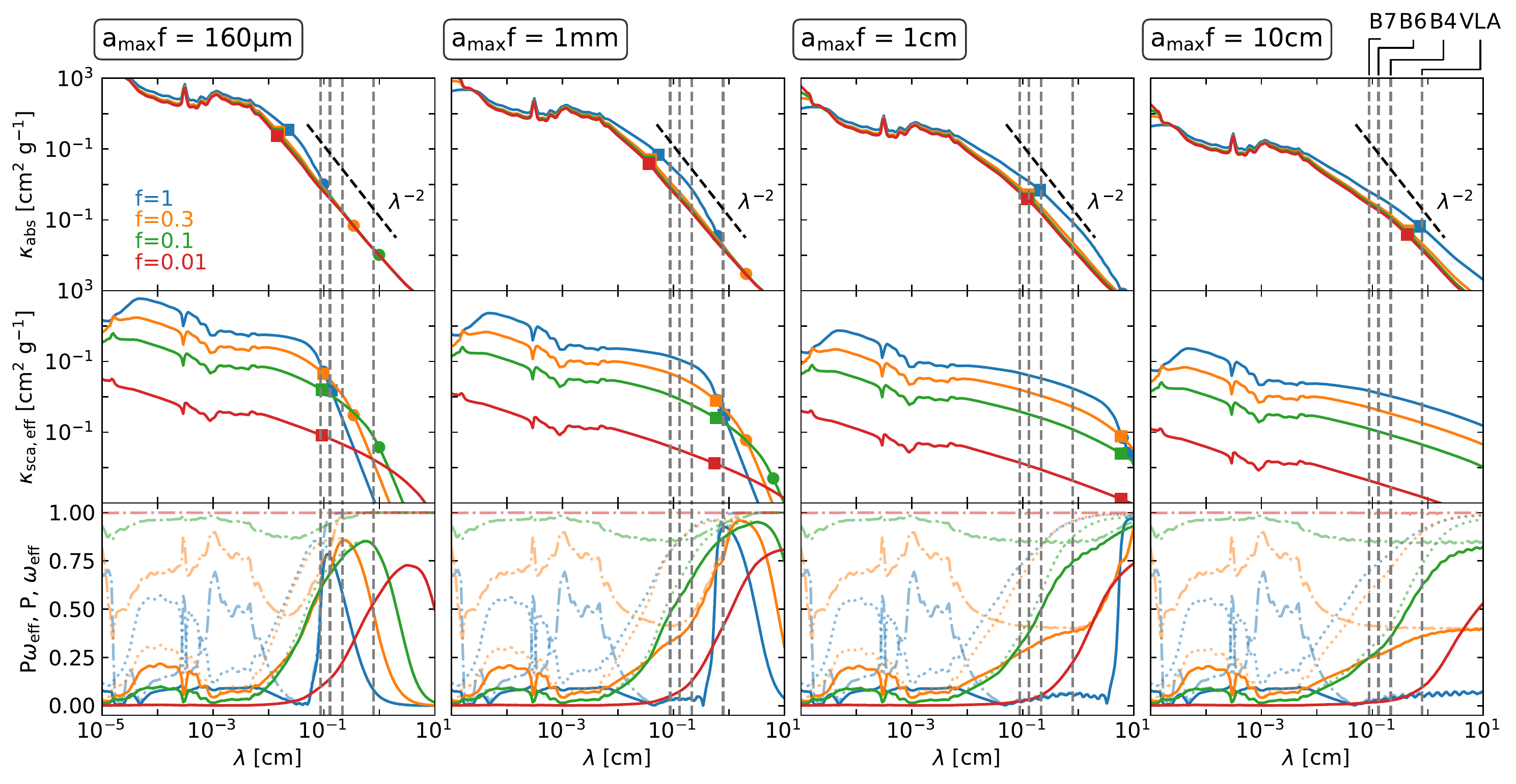}
\figcaption{Wavelength dependence of the absorption opacity (top panels), effective scattering opacity (middle panels) and $P\mathrm{\omega_{eff}}$ (solid lines, bottom panels), $P$ (dotted-dashed lines, bottom panels) and $\mathrm{\omega_{eff}}$ (dotted lines, bottom panels). From left to right, the maximum particle sizes $a_{\mathrm{max}}f$ are 160 $\mu$m, 1 mm, 1 cm and 10 cm. Four filling factors are 1 ({\color{blue}blue}), 0.3 ({\color{orange}orange}), 0.1 ({\color{green}green}) and 0.01 ({\color{red}red}). The circles (\ding{108}) represent $x$=1, delineating Rayleigh and Mie regimes. The squares (\ding{110}) are locations where $kx=3/8$ (for absorption) or $(n-1)x=1$ (for scattering), delineating Mie's optically thin and thick regimes for the dust particles. $x$ is the size parameter; $x=2\pi a_{\mathrm{max}}/\lambda$ (see Appendices \ref{sec:keyparameters}, \ref{sec:analyticalapprox}, and \ref{sec:scat}).
\label{fig:opacwav}}
\end{figure*}

We map out the optical properties for a wide range of $a_{\mathrm{max}}f$ with a particle size distribution $q=3.5$. In the top rows of Figure \ref{fig:opacwav}, we show the wavelength dependence of the particle absorption opacities, $\mathrm{\kappa_{abs}}$ for various $a_{\mathrm{max}}f$. From left to right, the maximum particle size $a_{\mathrm{max}}f$ increases from 160 $\mu$m to 10 cm. 
Additionally, four vertical lines indicate the observational wavelengths for ALMA bands 7 ($\sim$0.87 mm), 6 ($\sim$1.3 mm), 4 ($\sim$2.1 mm) and VLA Ka+Q band ($\sim$7.9 mm). Except for the small particles, $a_{\mathrm{max}}f$=160 $\mu$m (same as Figure \ref{fig:demo}), the compact particles have opacities several times higher than their porous counterparts at ALMA and VLA bands due to the interference pattern \citep{kataoka14, birnstiel18}. For these compact particles, the opacity index between different bands depends largely on the interference pattern and can vary far from $\lambda^{-2}$ represented by the dashed line. For example, if $a_{\mathrm{max}}f$ = 1 mm, the opacity indices between ALMA bands are less than two, whereas that between ALMA band 4 ($\sim$2.1 mm) and VLA band Ka+Q ($\sim$7.9 mm) can be much steeper than two. The feature disappears when the particle has a moderate porosity even if the filling factor $f$ is as high as 0.3. For the porous particles, the opacity approaches to $\lambda^{-2}$ at ALMA and VLA bands and its slope only starts to become shallower when $a_{\mathrm{max}}f$ $\gtrsim$ 1 cm. Compact and porous particles have such a significant difference at these wavelengths since they are within Mie's optically thin regime, where the boundaries between Rayleigh, Mie's optically thin, Mie's optically thick regimes are indicated by circles (\ding{108}) and squares (\ding{110}), respectively.

The second row of Figure \ref{fig:opacwav} shows the effective scattering opacity $\mathrm{\kappa_{sca,eff}}$. The effective scattering opacity decreases with increasing $a_{\mathrm{max}}f$. Similar to Figure \ref{fig:demo} ($a_{\mathrm{max}}$ = 160 $\mu$m), the effective scattering opacity at shorter wavelengths decreases as the particles become more porous. The more porous particles have higher effective scattering opacity at longer wavelengths ($\lambda > 2\pi a_{\mathrm{max}}$, i.e., the Rayleigh regime).

On the particle size constraint from polarization emission, both $\omega_{\mathrm{eff}}$ and $P$ become more favorable in producing a higher polarization fraction with porous particles, as shown in the third row of Figure \ref{fig:opacwav}. $P\omega_{\mathrm{eff}}$, $P$, and $\omega_{\mathrm{eff}}$ are represented by solid, dotted-dashed, and dotted lines, respectively. For compact particles, $\omega_{\mathrm{eff}}$ is only high when $a\sim \lambda/2\pi$ at a given wavelength. The range of particle sizes with high polarization fraction is wider for porous particles\footnote{The range of particle sizes with high polarization fraction is also wider for non-spherical particles \citep{kirchschlager20}, but with a different shape.}. $P$ also becomes constantly high along all the wavelengths \citep{tazaki19}, instead of a sharp drop-off when $a \gtrsim \lambda/2\pi$. Combining these two factors, we have a wider range of particle sizes to explain high polarization fractions observed in ALMA bands 6 and 7 (\citealt{tazaki19} and see Appendix \ref{sec:poldeg} for details). In other words, if particles are porous, an observed high polarization fraction at a certain wavelength can no longer translate to a tightly-constrained particle size, as it does for compact particles.

Take the $a_{\mathrm{max}}f$ = 160 $\mu$m case as an example, the compact particles only have high polarization fraction when $\lambda \sim$ 1 mm. As the particles become more porous, the window for high polarization fraction becomes much wider. At $f$=0.1 ({\color{green}green curve}), the $P\omega_{\mathrm{eff}}$ is above 0.5 for $\lambda$ from sub-mm to several cm. However, when the particles are too porous ($f$=0.01), the polarization fraction becomes lower within ALMA bands ({\color{red}red curve}) and can no longer explain observed high polarization fractions at bands 7 and 6, which is confirmed by MCRT calculations (see Figure \ref{fig:inferred_supplement} and \citealt{tazaki19, brunngraber21}). For compact particles, as the maximum particle size increases, the narrow window for high polarization fraction shifts to longer wavelengths, so the high polarization fraction is expected to be observed only at longer wavelengths. In contrast, the window shift is less crucial for porous particle with $f$=0.1. For instance, the polarization fraction can still be high at ALMA band 7 even if $a_{\mathrm{max}}f$ = 1 mm.

\subsection{Spectral Indices and Polarization Fractions Predicted by $\mathrm{\kappa_{abs}}$, $\mathrm{\kappa_{sca, eff}}$, $\omega_{\mathrm{eff}}$, and $P$}

Aside from predicting polarization fraction using $P\omega_{\mathrm{eff}}$, we can also predict the spectral index ($\alpha$; I$_\nu\propto \nu^{\alpha}$) using $\mathrm{\kappa_{abs}}$, $\mathrm{\kappa_{sca, eff}}$, and $\omega_{\mathrm{eff}}$ with the help of the radiative solution of the plane parallel approximation considering scattering \citep{miyake93, zhu19b, sierra19, carrasco19}. The analytical expression\footnote{We use the equation from \citet{sierra19}, where the radiative transfer equation is integrated. The equation from \citet{zhu19b} is similar but slightly different in that they use the Eddington approximation to find a solution to the emergent intensity.} is  
\begin{equation}
    I_\nu = B_\nu(T)\Big\{1-\mathrm{exp}\Big(-\frac{\tau_\nu}{\mu}\Big) + \omega_\nu F(\tau_\nu,\omega_\nu)\Big\},
    \label{eq:intensity}
\end{equation}
where the third term in the curly braces is the correction due to scattering. 
\begin{multline}
F(\tau_\nu, \omega_\nu) = \frac{f_1(\tau_\nu, \omega_\nu)+f_2(\tau_\nu, \omega_\nu)}{\mathrm{exp}(-\sqrt{3}\epsilon_\nu\tau_\nu)(\epsilon_\nu -1) - (\epsilon_\nu+1)},\\
f_1(\tau_\nu, \omega_\nu) = \frac{1-\mathrm{exp}\{-(\sqrt{3}\epsilon_\nu+1/\mu)\tau_\nu\}}{\sqrt{3}\epsilon_\nu\mu+1},\\
f_2(\tau_\nu, \omega_\nu) = \frac{\mathrm{exp}(-\tau_\nu/\mu)-\mathrm{exp}(-\sqrt{3}\epsilon_\nu\tau_\nu)}{\sqrt{3}\epsilon_\nu\mu-1}, \label{eq:F}
\end{multline}
where $\epsilon_\nu$ = $\sqrt{1-\omega_\nu}$. When the disk is very optically thick, $\omega_\nu F(\tau_\nu, \omega_\nu)$ is negative and contributes to the reduction of intensity below the blackbody radiation. To truncate forward scattering, we should use $\omega_\mathrm{eff}$ for $\omega_\nu$. Additionally, we should use $\kappa_\mathrm{sca,eff}$ to calculate the total optical depth, where
\begin{equation}
    \tau_\nu\ \mathrm{or}\ \tau_{\mathrm{tot}} = (\kappa_\mathrm{abs} + \kappa_\mathrm{sca,eff}) \Sigma_{d}, \label{eq:kappatot}
\end{equation}
where $\Sigma_{d}$ is the dust surface density.

\begin{figure*}[t!]
\includegraphics[width=\linewidth]{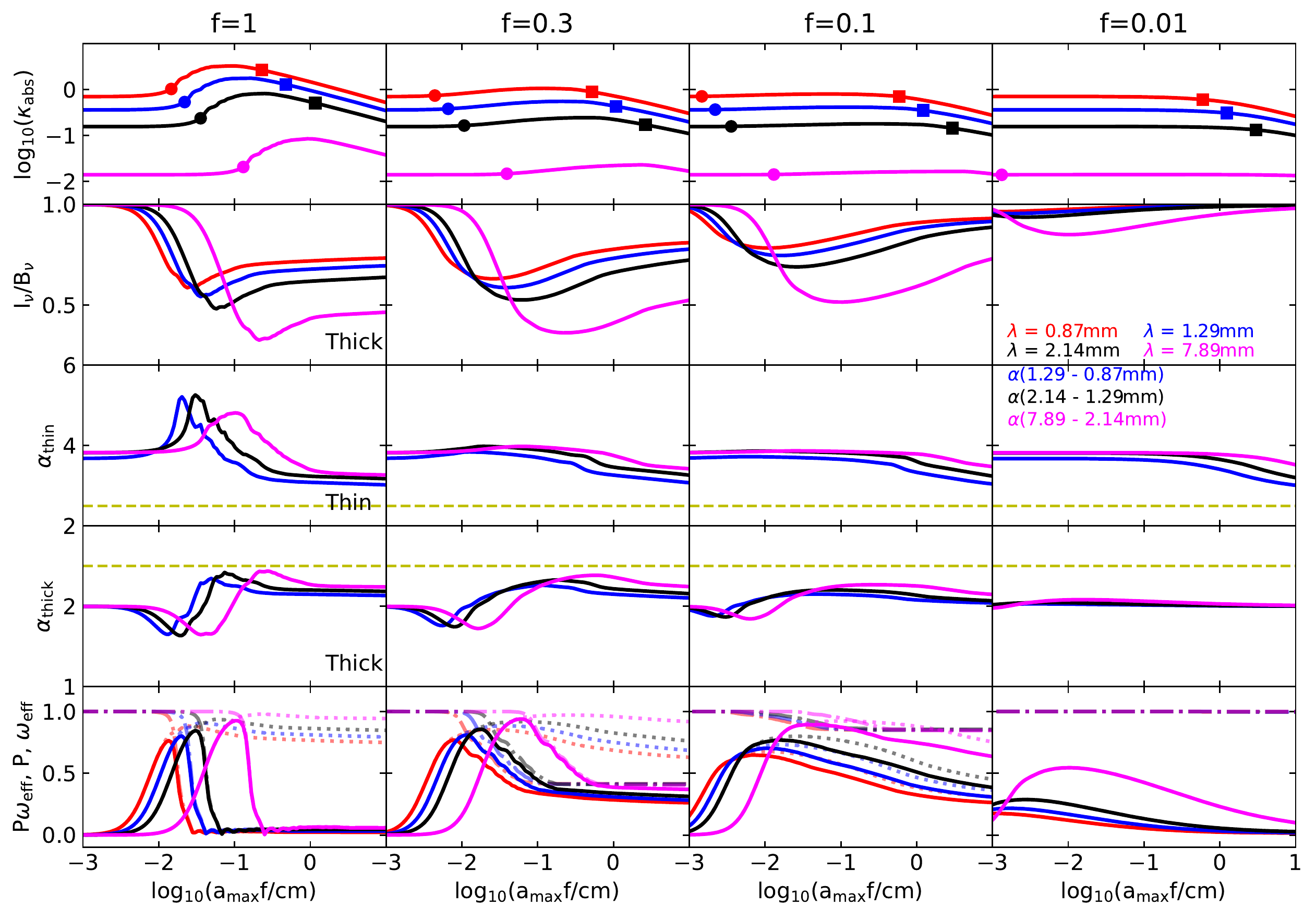}
\figcaption{From top to bottom: the size ($a_{\mathrm{max}}f$) dependence of the $\mathrm{\kappa_{abs}}$ (absorption opacity), $I_\nu$/$B_\nu$ (intensity reduction due to scattering compared to the blackbody radiation), $\alpha_{\mathrm{thin}}$, $\alpha_{\mathrm{thick}}$ (spectral indices at optically thin and thick regimes), and $P\mathrm{\omega_{eff}}$ (The product of the polarization fraction at 90$^{\circ}$, $P$ and the effective albedo, $\omega_{\mathrm{eff}}$). In the bottom panels, solid lines are $P\omega_{\mathrm{eff}}$, dotted lines are $\omega_{\mathrm{eff}}$, and dashed-dotted lines are $P$. From left to right, the filling factors are $f$=1, 0.3, 0.1, and 0.01. {\color{red}Red}, {\color{blue}blue}, {\color{black}black}, and {\color{magenta}magenta} curves represent these quantities at 0.87 mm, 1.29 mm, 2.14 mm, and 7.89 mm. The circles (\ding{108}) represent $x$=1, delineating Rayleigh and Mie regimes. The squares (\ding{110}) are locations where $kx=3/8$ (for absorption) delineating Mie's optically thin and thick regimes for the dust particles. Horizontal yellow dashed lines are $\alpha$=2.5 as a reference point between $\alpha_{\mathrm{thin}}$ and $\alpha_{\mathrm{thick}}$.
\label{fig:opacsize}}
\end{figure*}

The general trend for the spectral indices is that they become less sensitive to particle sizes as the particles become more porous. In Figure \ref{fig:opacsize}, we demonstrate the predictions of spectral indices using Equation \ref{eq:intensity} in both optically thin ($\tau$=0) and thick ($\tau$$\to$$\infty$) limits. All panels display the size dependence of particle optical properties with different filling factors (from left to right, the particles become more porous from $f$=1 to 0.01) and at different wavelengths ({\color{red}red}: 0.87 mm; {\color{blue}blue}: 1.29 mm; {\color{black}black}: 2.14 mm; {\color{magenta}magenta}: 7.89 mm), corresponding to ALMA bands 7, 6, 4, and VLA Ka+Q bands.

The spectral indices in the optically thin limit where the scattering is negligible are shown in the third row of Figure \ref{fig:opacsize}. The blackbody radiation is assumed to be at the Rayleigh-Jeans tail\footnote{Note that in protoplanetary disks' midplane the temperature can be at several tens of Kelvin. At ALMA bands, $h\nu$/$k$T $\lesssim$ 1, but not much less than unity, so the blackbody spectrum is not completely in the Rayleigh-Jeans tail, i.e., the wavelength dependence is shallower than B$_\nu$ $\propto$ $\nu^2$. Nevertheless, the factor is highly temperature dependent and does not deviate from the two significantly.}, so that $\alpha$=$\beta$+2; $\mathrm{\kappa_{abs}}(\nu)$ $\propto$ $\nu^{\beta}$, where $\beta$ is the opacity index. In this limit, the spectral index is only determined by the absorption opacity ($\kappa_{\mathrm{abs}}$) shown in the first row. For various porosities, the $\alpha$ changes from the ISM value ($\sim$ 3.7) at the small-particle end to lower values ($\sim$ 3.0) at the big-particle end. When the particle is compact, the spectral indices have peaks at $a_{\mathrm{max}} \sim \lambda/2\pi$ due to the interference feature in the absorption opacity. For ALMA bands, $\alpha$ can be as high as 5.0. Even with a moderate porosity, e.g., $f$=0.3, these peaks with very high values disappear. Instead, the spectral indices are similar (with a slight increase) to the ISM value until $a_{\mathrm{max}}f \sim $ several mm. After that, the spectral indices start to drop. This trend also happens in particles with smaller filling factors, but the drop of spectral indices is moving to larger particle sizes. The spectral index at the large particle extreme can be approximated using $\beta \approx (q-3)\beta_{\mathrm{ISM}}$ \citep{draine06}, where $q$=3.5 and $\beta_{\mathrm{ISM}}$=1.7. An important implication for these trends is that with porosity, the particle size becomes difficult to infer from the spectral index since they are less sensitive to $a_{\mathrm{max}}f$. For example, the spectral indices observed similar to the ISM values can even be due to $a_{\mathrm{max}}f \sim$ mm particles.

The spectral indices at the optically thick regime mainly depend on the effective albedo, $\mathrm{\omega_{eff}}$, and can be taken as the ratio of the logarithmic intensity reduction between two wavelengths plus two ($\alpha_{\mathrm{thick}}$ = dln(I$_\nu$/B$_\nu$)/dln$\nu$ + 2). If there were no scattering, the spectral index should be exactly two in the Rayleigh-Jeans approximation. 

We introduce the intensity reduction due to the dust scattering, $I_\nu/B_\nu$, shown in the second row of Figure \ref{fig:opacsize}. The reduction factor is what in the curly braces in Equation \ref{eq:intensity} and in the optically thick limit with the assumption $\mu$=1 (the disk is face-on)\footnote{Note that in some parameter space the intensity can be enhanced by scattering. If the disk is optically thin and its dust has a high albedo, the intensity with scattering can be higher than that without scattering at a face-on view \citep{sierra20}. This is because the scattering can remove the photons in the optically thick edge-on direction and put them into the optically thin face-on direction.}. The reduction is stronger if the disk is inclined \citep{zhu19b, sierra20}. For the compact particles, the intensity reduction starts at short wavelengths when $a_{\mathrm{max}}f <$ 100 $\mu$m. When $a_{\mathrm{max}}f >$ 1 mm, the intensity is reduced more at longer wavelengths. As the particles become more porous, the intensity is reduced less for a given $a_{\mathrm{max}}f$, and the particle size with the lowest reduction factor becomes smaller. However, the reduction is still substantial if $f\gtrsim$ 0.1. The scattering is not crucial for $f$ =0.01, expect for $\lambda$ = 7.89 mm and $a_{\mathrm{max}}f <$ 1 cm (with $I_\nu/B_\nu$ $\sim$ 90\%).

Now we are ready to present the spectral indices at the optically thick regime in the fourth row of Figure \ref{fig:opacsize}. For compact particles, the spectral indices are affected the most by the scattering. The spectral indices have a dip with the value $\sim$ 1.7 when $a_{\mathrm{max}}f$ is several 100 $\mu$m. As the particle size increases, the spectral indices increase and peak at 2.5 when $a_{\mathrm{max}}f$ is several mm. The trends at different wavelengths are similar. The peaks and dips are just shifted to larger sizes for longer wavelengths. As the particles become more porous, the dips shift to smaller particles, whereas the peaks shift to larger particles. In the meantime, the amplitudes of the varying spectral indices become smaller. For $f$= 0.01, the spectral indices for all particle sizes approach two for all wavelengths.

Overall, as the particle becomes more porous, the spectral index approaches ISM values at the optically thin limit and two at the optically thick limit in a vast parameter space.

To complete the prediction on polarization fraction, we show the proxy of polarization fraction, $P\mathrm{\omega_{eff}}$, in the bottom row of Figure \ref{fig:opacsize}. They are similar to the bottom row of Figure \ref{fig:opacwav}, but with $a_{\mathrm{max}}f$ as the x-axis. The trend is similar to Figure \ref{fig:opacwav}--the more porous particles have a wider range of $a_{\mathrm{max}}f$ to produce high polarization fraction at a give wavelength. For the compact particles, at each wavelength, the polarization fraction has a sharp peak around a specific particle size. The peak is around $a_{\mathrm{max}}=\lambda/ 2\pi$. For 0.87 mm and 1.29 mm, $a_{\mathrm{max}}f$ is around 100 $\mu$m, which explains why mm polarization observations tend to constrain the particle size to be 100 $\mu$m (e.g., \citealt{kataoka15, yang16}). With porosity, however, the window for an $a_{\mathrm{max}}f$ that can reproduce a certain polarization fraction becomes wider. 

While the wide window of $a_{\mathrm{max}}f$ to produce high polarization fraction makes it
difficult to constrain the particle size at a particular wavelength, we can potentially probe whether the particles are
porous through the wavelength dependence of the polarization fraction. For porous particles, as long as the particle size is larger than several 100 $\mu$m, the polarization fraction becomes larger for longer wavelengths. This contrasts with the 100 $\mu$m solution, in which the polarization fraction decreases with wavelengths and becomes negligible at VLA/ngVLA (Very Large Array/next-generation Very Large Array) bands ($\lambda$ $\gtrsim$ several mm). With highly porous particles ($f$=0.01), the polarization fraction is too low to explain the observed polarization fraction (see Figure \ref{fig:inferred_supplement} and \citealt{tazaki19}). This means that if particles are porous, we can constrain the porosity from current observations (see Section \ref{sec:constraints}). Changing the slope of the particle size distribution can quantitatively change these relations, but the overall trend is the same as long as $q<4$.

\section{Analytical Fitting on Continuum \label{sec:analytical}}

With a comprehensive understanding on the porous particles' optical properties in Section \ref{sec:dust}, we now fit the HL Tau continuum observations using a wide range of possible dust surface density, temperature, and particle sizes for both compact and porous particles. These best-fit models will be used as inputs for the MCRT simulations in Section \ref{sec:MCRT}.

\subsection{Observations}
We use the radial profiles reduced in \citet{carrasco19} for ALMA bands 7, 6, 4, and VLA Ka+Q bands (at $\sim$ 0.87, 1.3, 2.1, and 7.9 mm, respectively). All images have been convolved to the same angular resolution of $\sim$ 7.35 au (50 mas). Details of the observations and image reduction can be found therein. 

\subsection{Method}
We use a Bayesian approach, i.e., P(model$|$data) = P(data$|$model)P(model)/P(data), to find the best parameters for the model. The best model is the combination of parameters that maximizes the posterior, P(model$|$data). To write it out explicitly, 
\begin{equation}
\begin{split}
    \mathrm{P}(T, \Sigma_d, a_{\mathrm{max}}f| I_{\mathrm{B7}}, I_{\mathrm{B6}}, I_{\mathrm{B4}}, I_{\mathrm{VLA}}) \\
    \propto \mathrm{P}(I_{\mathrm{B7}}, I_{\mathrm{B6}}, I_{\mathrm{B4}}, I_{\mathrm{VLA}}|T, \Sigma_d, a_{\mathrm{max}}f)\\
    \mathrm{\times P}(T, \Sigma_d, a_{\mathrm{max}}f). \label{eq:bayesian}
\end{split}
\end{equation}
The first term on the right-hand side,
\begin{equation}
     \mathrm{P}(I_{\mathrm{B7}}, I_{\mathrm{B6}}, I_{\mathrm{B4}}, I_{\mathrm{VLA}}|T, \Sigma_d, a_{\mathrm{max}}f) \propto \mathrm{exp}(-\frac{\chi^2}{2}),\label{eq:likelihood}
\end{equation}
is the likelihood of the observations given a combination of parameters, where,
\begin{equation}
     \chi^2 = \sum_{i} w_i \Bigg( \frac{I_{\mathrm{obs,}i}-I_{\mathrm{m,}i}}{\sigma_i} \Bigg)^2. \label{eq:chi2}
\end{equation}
The $I_{\mathrm{m,i}}$ is the modeled intensity at $i$ band using Equation \ref{eq:intensity}, with inclination being 46.72$^{\circ}$. The $a_{\mathrm{max}}f$ enters the equation by determining the opacities and effective albedo ($\mathrm{\kappa_{abs}}$, $\mathrm{\kappa_{sca,eff}}$, and $\mathrm{\omega_{eff}}$). Note that we only take $q$=3.5 as the size distribution in the current section, but we will also present a case with $q$=2.5 in Section \ref{sec:slope}. With the easy-to-use \texttt{dsharp\_opac} package \citep{dsharp_opac}, the opacity with any composition and filling factors can be easily generated. Then they can be taken as the input for the analytical fitting. $I_{\mathrm{obs,}i}$ and $\sigma_i$ are the observed intensity and uncertainty at band $i$. The uncertainty is composed of the azimuthal variation $\Delta I_{\mathrm{obs,}i}$ and calibration error $\delta I_{\mathrm{obs,}i}$. The total uncertainty reads
\begin{equation}
    \sigma_i^2 = (\Delta I_{\mathrm{obs,}i})^2 + (\delta I_{\mathrm{obs,}i})^2. \label{eq:error}
\end{equation}
The calibration errors are set to be 10\%, 10\%, 5\%, 5\% for ALMA bands 7, 6, 4, and the VLA band. $w_i$ is the weight for each wavelength. Since the VLA band is a combination of Ka and Q bands, we give the weight as 2, whereas all ALMA bands have weights as 1.

The second term on the right-hand side of Equation \ref{eq:bayesian} is prior. In previous literature \citep{macias21, sierra21, ueda22}, the prior is assumed to be uniform (in logarithmic space) for $a_{\mathrm{max}}f$ and dust surface density within a certain range. We also set uniform priors for $a_{\mathrm{max}}f$ between 0.1 $\mu$m to 1 m, and dust surface density from $10^{-1.5}$ to $10^{1.5}$ $\mathrm{g\ cm^{-2}}$. In addition to the whole particle population model, we also separate the particle population into big particles or small particles to find the best model within these two populations, similar to \citet{macias21, sierra21, ueda22}. This is motivated by the dependence of the spectral index on the particle size. As seen from the third row in Figure \ref{fig:f1}, two solutions can be found for a given spectral index\footnote{This is because the optically-thin spectral indices at ALMA bands have maxima when the particle size is sub-millimeter for compact particles.}. The temperature prior ranges from 10 K to 300 K, but is not uniform, since for several regions in the disk, a uniform prior of temperature makes the temperature unconstrained. This is because temperature and surface density are correlated. The temperature can be fitted infinitely high while the surface density only decreases slightly. Hence, we assign a Gaussian prior with a large scattering to produce more smooth and reasonable temperature profiles. The prior temperature is centered on the following profile,
\begin{equation}
    T_{p}(r) = \mathrm{200\ K \times }(r/\mathrm{au})^{-0.4}, \label{eq:Tcenter}
\end{equation}
with scattering also being 1 $\times$ $T_p(r)$,
\begin{equation}
    \mathrm{P}(T, \Sigma_d, a_{\mathrm{max}}f) =\mathrm{exp}\Bigg(-\frac{1}{2}\frac{(T - T_p(r))^2}{T_p(r)^2}\Bigg). \label{eq:prior}
\end{equation}
In this way, the temperature is constrained in a reasonable range while the prior is still loose enough so that the temperature does not collapse onto the Gaussian prior center.
A temperature prior is also used in \citet{macias21} in which the prior is based on the expected temperature profile for a passively irradiated disk. 
We fit the intensity profiles from 1 au to 100 au, spacing every 1 au, following \citet{macias21, sierra21, ueda22}. We note that the neighboring radii are correlated due to the nature of radio images and the much larger beam size (7.35 au). The temperature grid is spaced every 5 K; the ratio between the neighbouring particle sizes is 1.08; and the ratio between the neighbouring dust surface density is 1.04. We also emphasize that the main focus of this current paper is to explain the observations of the outer HL Tau disk ($\gtrsim$ 20 au) since the inner optically thick part can be substantially affected by dust vertical settling \citep{ueda21}. Nevertheless, we will discuss the fitting result at face value throughout the disk, including the inner part.

\subsection{Results}
\subsubsection{f=1 (Compact Particles)}
\begin{figure*}[t!]
\includegraphics[width=\linewidth]{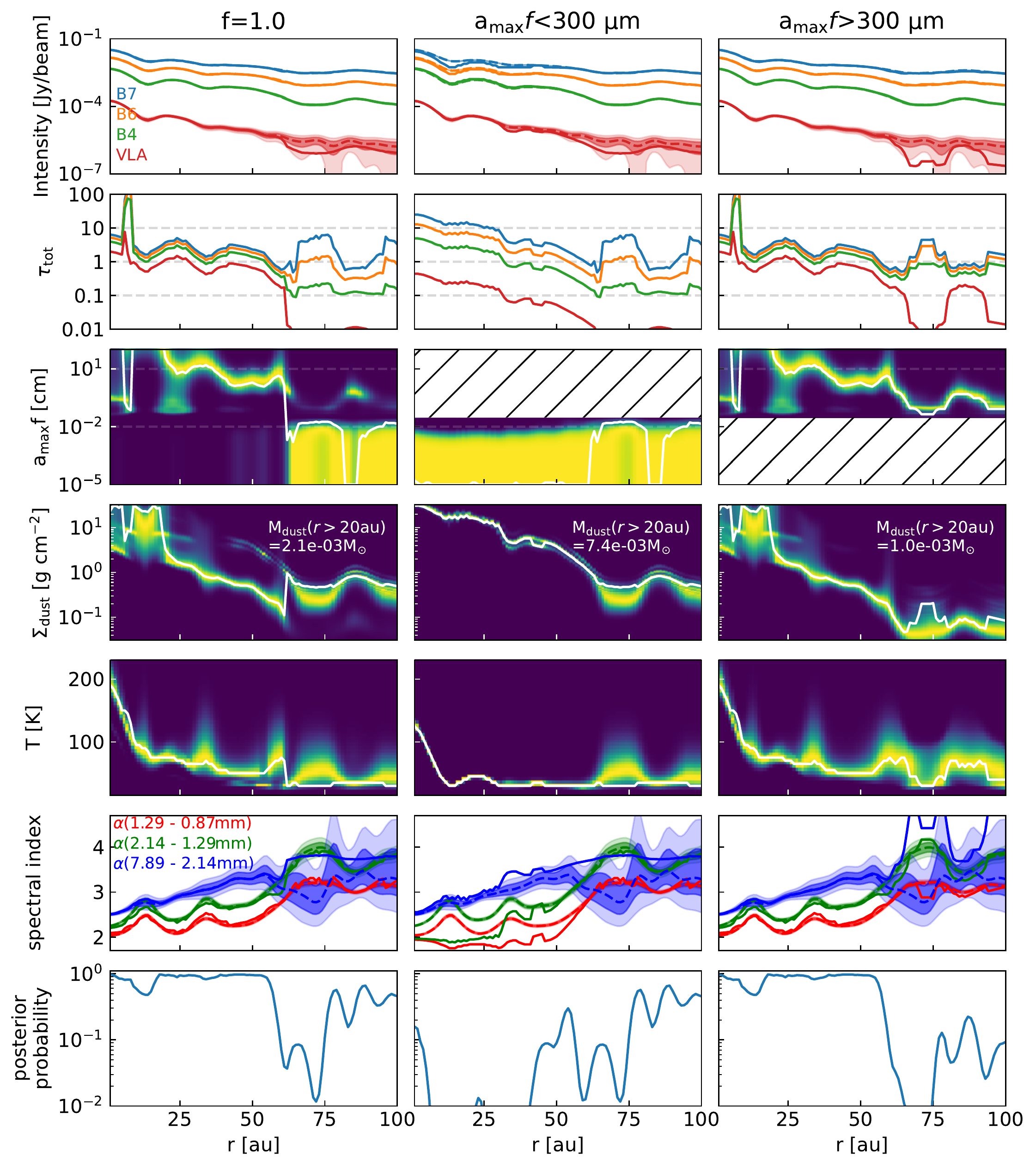}
\figcaption{From left to right: best fit analytical models considering particles with all sizes, only small particles ($a_{\mathrm{max}}f$ $<$ 300$\mu$m) and only big particles ($a_{\mathrm{max}}f$ $>$ 300$\mu$m). From top to bottom: the intensity from the best model; total optical depths at different wavelengths; marginalized posterior probability of the maximum particle size $a_{\mathrm{max}}f$, dust surface density, and temperature; the spectral indices; and the posterior probability of the best fit. For the intensity and spectral index plots, observations are in dashed lines with the shaded area as 1 $\sigma$ and 2 $\sigma$. For the total optical depth plots, horizontal dashed lines represent $\tau_{\mathrm{tot}}$ = 0.1, 1, and 10. For the a$_{\mathrm{max}}f$, $\Sigma_{d}$, and $T$, the best models are represented by white curves (the best models under three-dimensional parameter space). The dust mass beyond 20 au is listed on top of the surface density plots.
\label{fig:f1}}
\end{figure*}

Figure \ref{fig:f1} presents the continuum fitting results for compact particles. The left, middle and right panels are the results for the whole particle population, the small-particle-only ($a_{\mathrm{max}}f < $ 300 $\mu$m) and the big-particle-only ($a_{\mathrm{max}}f > $ 300 $\mu$m) priors. In the top panels, solid lines show the best-fit intensities, and the dashed lines with shaded areas are the observations with $1\sigma$ and $2\sigma$ uncertainties. The second row shows the best-fit total optical depth $\mathrm{\tau_{tot}}$, which is ($\mathrm{\kappa_{abs}+\kappa_{sca,eff}}$)$\Sigma_d$ (Equation \ref{eq:kappatot}). The third to fifth rows show the marginalized posterior for $a_{\mathrm{max}}f$, $\Sigma_d$ and $T$, respectively. The posterior is normalized by the maximum value within the map. The yellow regions represent a higher probability. The hatched regions are not used in the fitting. The white lines are the best fit models among the 3-D parameter space. The dust masses in each model are also listed in the panels. The sixth row shows the spectral indices for models (solid lines) and observations (dashed lines with shaded areas). The last row shows the value of the posterior probability using the parameters of the best-fit model. 

For the compact particles, the big-particle population fits better in the inner disk ($<$60 au), whereas the small-particle population fits better in the outer disk ($>$60 au). The quality of the fitting can be indicated by the intensity profiles, spectral indices, and posterior probabilities. It is noticeable that no model can perfectly fit the VLA data beyond 60 au due to the low signal-to-noise ratio at that region. Considering only small particles, we cannot tightly constrain the particle sizes. They can be between 0.1 $\mu$m and 100 $\mu$m. For small particles, the fitting within 60 au is not good for three reasons. (a) The best $a_{\mathrm{max}}f$ is at the parameter boundary (0.1 $\mu$m); (b) spectral indices at two ALMA bands deviate from the observations up to $\Delta \alpha$=1 (also indicated by intensities); (c) the posterior probability is very low. So the small-particle model is not preferred within 60 au. The size can be better constrained with the big-particle population. Within 20 au, the $a_{\mathrm{max}}f$ exceeds 1 m; for 20 au $\lesssim$ $r$ $\lesssim$ 40 au, the $a_{\mathrm{max}}f$ is around 10 cm; for 40 au $\lesssim$ $r$ $\lesssim$ 60 au, the $a_{\mathrm{max}}f$ prefers 1 cm; beyond that the size becomes 1 mm or smaller. The exceedingly large particle sizes within 40 au can be due to a unique opacity feature for the particle size distribution $q$=3.5. If $q$=2.5, the particle size can be around several mm within 60 au (see Section \ref{sec:slope}, and Figures \ref{fig:q=2.5} and \ref{fig:discussion}). The better constraint on $a_{\mathrm{max}}f$, closer matches on intensities and spectral indices, and a higher posterior probability indicate that big particles are better candidates within 60 au.

Big particles have higher opacities, so the inferred dust mass is much lower than using the small particles (7.4$\times$10$^{-3}$ $M_\odot$ vs. 1.0$\times$10$^{-3}$ $M_\odot$ beyond 20 au). For the small-particle model, the optical depths for the ALMA bands are $\gtrsim$ 0.1 across the disk, whereas the VLA band optical depth is two orders of magnitude lower. For the big-particle model, optical depths at four bands are all close to unity, except for regions beyond 60 au. For the small-particle model, if we assume the gas to dust mass ratio is 100:1, the gas disk mass is even comparable to the stellar mass, meaning that the disk is very gravitationally unstable. The temperature always has the opposite trend as the surface density, so the small-particle model prefers lower temperature. The incentive for separating small and big particles can been seen on the probability plots for $a_{\mathrm{max}}f$ (the third row). Considering both populations, it is clear that the probability has a valley around $a_{\mathrm{max}}f$ =300 $\mu$m along all radii. Above and below this value, the probabilities are non-zero.

\subsubsection{f=0.1 (Porous Particles)}

\begin{figure*}[t!]
\includegraphics[width=\linewidth]{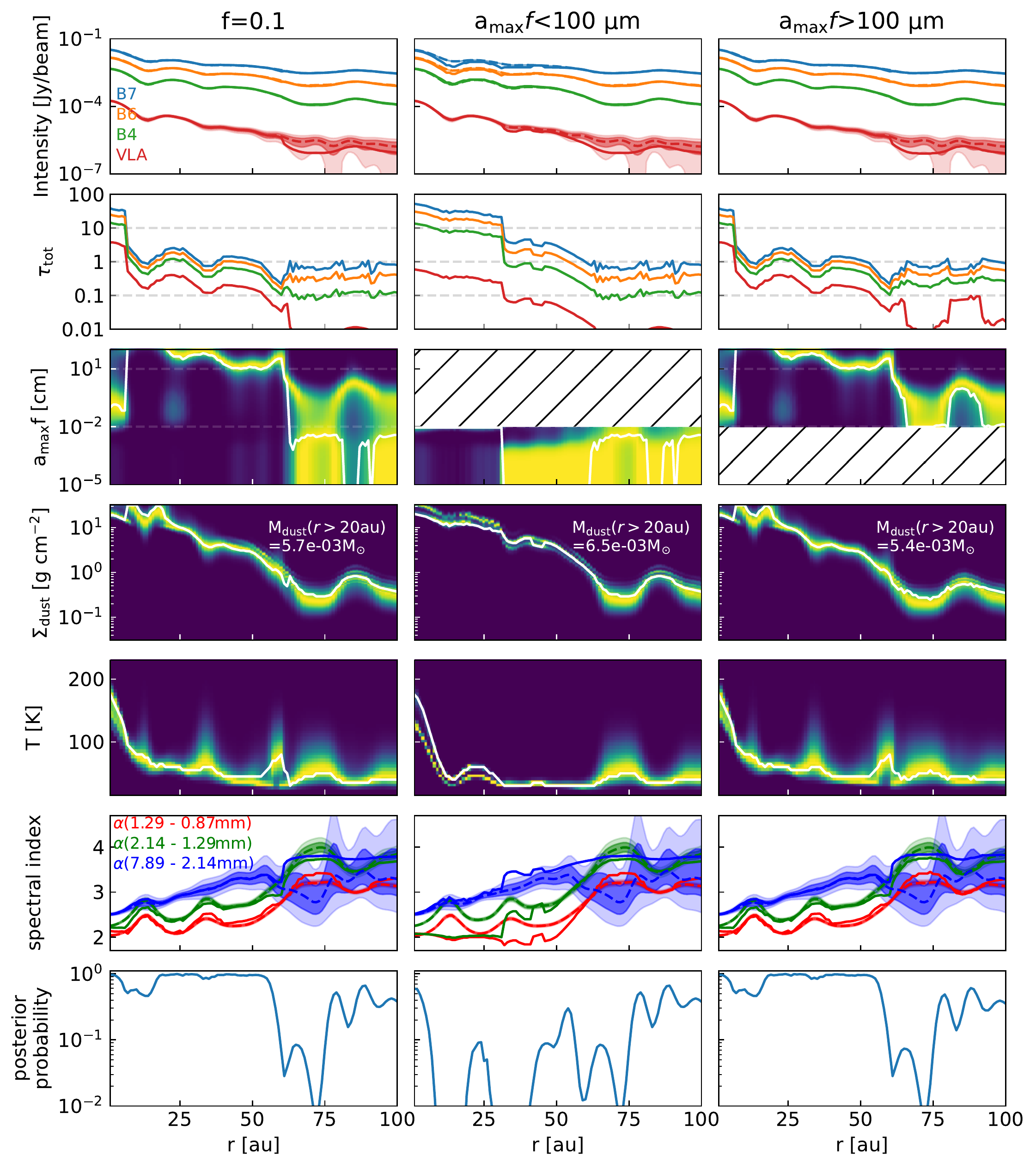}
\figcaption{Same as Figure \ref{fig:f1} but for porous particles with $f$=0.1. The particle size that separates big and small particles is $a_{\mathrm{max}}f$=100$\mu$m. 
\label{fig:f0p01}}
\end{figure*}

For porous particles with $f$=0.1, the inner disk still prefers large particles; the outer disk still prefers small particles. The models are shown in Figure \ref{fig:f0p01} with the same layout as Figure \ref{fig:f1}. For the small particles, the possible range of $a_{\mathrm{max}}f$ is still quite large, from 0.1 $\mu$m to 100 $\mu$m. The posterior probability is still low. For the big particles, the maximum particle sizes are higher than the compact counterparts. Within 20 au, the $a_{\mathrm{max}}f$ exceeds 1 m; for 20 au $\lesssim$ $r$ $\lesssim$ 40 au, the $a_{\mathrm{max}}f$ is around 1 m; for 40 au $\lesssim$ $r$ $\lesssim$ 60 au, the $a_{\mathrm{max}}f$ prefers 10 cm; beyond that the size becomes 1 mm or smaller. Again, the $q$=2.5 size distribution can bring the $a_{\mathrm{max}}f$ down to mm-cm (see Section \ref{sec:slope}, and Figures \ref{fig:q=2.5} and \ref{fig:discussion}).  Besides the size difference, small-particle and big-particle models predict similar surface density and temperature since the opacity for the big particles no longer has an interference pattern. The inferred masses for these two species are very high and similar to the small-particle model for compact particles. Keeping the particle size the same, using opacity of big porous particles can lead to a factor of 6.5 dust mass higher than their compact counterparts. \citet{guidi22} have a similar finding in HD 163296.  Even though the particle size is still difficult to constrain by these observations, different particle sizes lead to similar constraints on the dust surface density and temperature. 

Overall, the porous particles can also explain the continuum observations.

In Appendix \ref{sec:constraints_SED}, we extend SED fitting by examining additional cases with various combinations of filling factors and particle size distribution slopes. We also discuss more derived quantities, such as gas-to-dust mass ratio and Stokes number from SED fitting.

\section{MCRT fitting and polarization \label{sec:MCRT}}

We run Monte Carlo Radiative Transfer (MCRT) simulations with two goals. One is to test the 1D analytical fittings on continuum emissions (Section \ref{sec:analytical}). In the 1D analytical approximation (Equation \ref{eq:intensity}), $\kappa_{\mathrm{sca,eff}}$ is the same for any scattering angle (the total scattering opacity is scaled to an isotropic-equivalent value). In MCRT simulations, we use the angle-dependent scattering matrix to capture the anisotropic multiple scattering.  Since many models predict marginally optically thick optical depths in HL Tau within $r$ $\lesssim$ 60 au, the full treatment of radiative transfer is necessary. The other is to generate linear self-scattering polarization maps to compare with current polarization data and make predictions at longer wavelengths. The metric for polarization fraction, $P\omega_{\mathrm{eff}}$ (introduced in Section \ref{sec:dust}) only approximates the case with isotropic single scattering, so the observed polarization fraction from multiple scattering needs to be calculated from a realistic 3D disk setup. Furthermore, even with the exact same particle size and composition, different optical depths and inclinations can lead to different polarization fractions.

\subsection{MCRT Setup \label{sec:MCRTsetup}}
Radiative transfer simulations are performed with the Monte Carlo radiative transfer code RADMC-3D \citep{dullemond12}. The detail of the method can be found in \citet{ueda21, zhang21}. Here we only provide a summary. 

\textit{Vertical Settling.} We use many particle species to simulate vertical settling. The particle size distribution is logarithmically divided into 15 particle size bins per decade. The vertical dust density follows Gaussian profiles, 
\begin{equation}
    \rho_d (r, a) = \frac{\Sigma_d(r, a)}{\sqrt{2\pi} h_d (r, a)} \mathrm{exp}\Bigg(-\frac{z^2}{2h_d(r, a)^2} \Bigg), \label{eq:rhod}
\end{equation}
where $z$ is the vertical height, and $h_d$ is the dust scale height. It is assumed to be in a mixing-settling equilibrium \citep{dubrulle95, youdin2007},
\begin{equation}
    h_d (r, a) = h_g (r) \Bigg(1+\frac{\mathrm{St}}{\alpha_t}\frac{1+2\mathrm{St}}{1+\mathrm{St}} \Bigg)^{-1/2}, \label{eq:hd}
\end{equation}
where $h_g$ is the gas scale height given as $h_g(r)$ = $c_s(r)/\Omega_K(r)$. $c_s$ is the sound speed. Since we assume the temperature is constant on the vertical direction (i.e., a vertically isothermal disk), $c_s$ is only dependent on $r$. The Stokes number, St=St($r$, $a$), is 
\begin{equation}
    \mathrm{St}  = \frac{\pi}{2} \frac{\rho a}{\Sigma_{g}(r)} = \frac{\pi}{2} \frac{\rho_{\mathrm{int}}af}{\Sigma_{g}(r)}. \label{eq:st}
\end{equation}
This means that different particle species have different scale heights, which depend on the particle size $af$, particle internal density $\mathrm{\rho_{int}}$, local gas surface density $\Sigma_g$ and turbulence\footnote{we use subscript `t' for the turbulence $\alpha$ to distinguish it from the spectral index $\alpha$.} $\alpha_{t}$.
We adopt $\alpha_t$ = $10^{-4}$, and $\Sigma_g$ = 1000 ($r$/au)$^{-0.5}$ g cm$^{-2}$. This gas surface density profile is the same as the one in \citet{ueda21}. We want to emphasize that the gas surface density here only affects the degree of settling through St. Since St$/\alpha_t$ determines the settling, for a fixed degree of settling (i.e., fixed St$/\alpha_t$), a different $\Sigma_{g}$ essentially means a different $\alpha_t$. Hence, this adopted $\Sigma_{g}$ does not prevent us from assuming different $\Sigma_g$ in Section \ref{sec:constraints}. While the vertical settling is crucial for the inner disk $\lesssim$ 20 au \citep{sierra20, ueda21}, it is not as important in the outer disk, where it is less optically thick. Thus, we only use one fiducial $\alpha_t$, which is small enough to enable dust settling. Although dust settling has little effect on the SED and polarization of the outer disk, we include it to keep our models self-consistent. We can also capture the transition between optically thin and optically thick regions. For the DSHARP composition, the internal density, $\rho_{\mathrm{int}}$ = 1.675 g cm$^{-3}$. By definition of the Stokes number (Equation \ref{eq:st}), we can see that $af$ not only determines the opacity, but also St, therefore the vertical settling.

\textit{Radially Varying of $a_{\mathrm{max}}f$.} To simulate the dust with a size distribution and varying $a_{\mathrm{max}}f$ at different radii, we give each dust species a certain weight at each radius, ensuring the particle size distribution still follows a power-law, $n(a) \propto a^{-q}$, with maximum particle size as $a_{\mathrm{max}}(r)$. The dust surface density for the $i$-th species reads, 
\begin{equation}
    \begin{split}
    \Sigma_{{d,} i}(r, a_i) &= \Sigma_{{d}}(r) \frac{a_{i+1}^{-q+4}-a_{i}^{-q+4}}{a_{\mathrm{max}}^{-q+4}(r)-a_{\mathrm{min}}^{-q+4}},\\
    &\mathrm{for}\ r\ \mathrm{where}\ a_{\mathrm{max}}(r) > a_{i+1},\\
    \Sigma_{{d,} i}(r, a_i) &= 0, \\
    &\mathrm{for}\ r\ \mathrm{where}\ a_{\mathrm{max}}(r) <= a_{i+1},
    \end{split}
\end{equation}
where the fraction after $\Sigma_{{d}}(r)$ is the mass fraction of the $i$-th species, and the sum of all species equals the total dust surface density.

We use the best-fit\footnote{``Best-fit'' means the most-probable solution under some constraints. For example, the small-particle models can have lower posterior probabilities within 60 au for the continuum observations than the big-particle models.} radial profiles of $a_{\mathrm{max}}f$, $\Sigma_{\mathrm{d}}$, and $T$ from Section \ref{sec:analytical} (Figures \ref{fig:f1} and \ref{fig:f0p01}) as the inputs of MCRT calculations. There are six models in total. Namely, they are whole-population, small-particle-only and big-particle-only models with $f=$ 1 and 0.1. These models have a fiducial size distribution $q$=3.5. Additionally, we also run several cases with $f=0.01$ and $q=2.5$, which will be presented in Sections \ref{sec:f0.01} and \ref{sec:slope}.

To produce images, we incline and position the disk the same way as HL Tau. Since we find the best-fit parameters using the data with resolution as 7.35 au \citep{carrasco19}, we compare the synthetic images directly with continuum images, without convolving them the second times. A more proper way is to deconvolve the observations, find the intrinsic parameters, use these intrinsic values as inputs to produce MCRT images, then convolve them with the observational resolution. While we have taken the beam effect into account, a different order from the proper way might lead to some error, but they should be less significant than completely ignoring the beam effect when deep gaps are present as discussed in \citet{lin20}. To compare with polarization images, the MCRT images are smoothed with 30 au 2D Gaussian kernel, similar to the resolutions in observations \citep{kataoka2017, stephens17}. We still use DSHARP opacity, but use \texttt{Optool} \citep{optool} to generate opacity including scattering matrices thanks to its faster speed.

\subsection{MCRT Results \label{sec:MCRTresults}}
\subsubsection{Continuum Radial Profiles}
\begin{figure*}[t!]
\includegraphics[width=\linewidth]{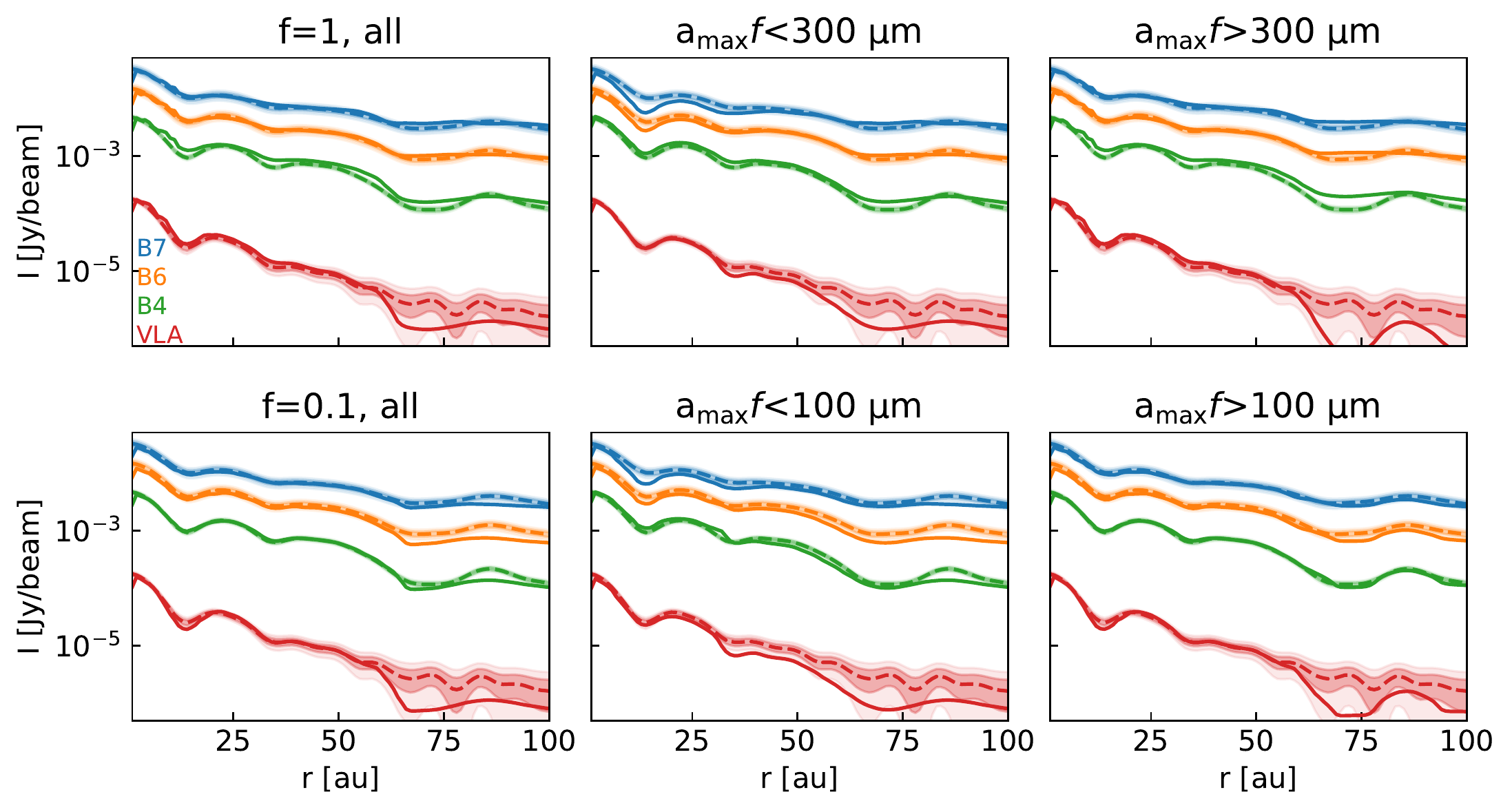}
\figcaption{Intensities of RADMC-3D (MCRT) models with different levels of dust porosity and sizes. Top: compact particles considering the whole particle population, small-particle population ($a_{\mathrm{max}}f<$ 300 $\mu$m) and big-particle population ($a_{\mathrm{max}}f>$ 300 $\mu$m). Bottom: porous particles ($f$=0.1) considering the whole particle population, small-particle population ($a_{\mathrm{max}}f<$ 100$\mu$m) and big-particle population ($a_{\mathrm{max}}f>$ 100$\mu$m).
\label{fig:radmcintens}}
\end{figure*}

Figure \ref{fig:radmcintens} shows radial profiles of MCRT models with six best-fit models in Figures \ref{fig:f1} and \ref{fig:f0p01} in ALMA bands 7, 6, 4 and VLA band Q+Ka (from top to bottom in each panel). They are generated by azimuthally averaging images produced by MCRT. 

The first row shows the cases for compact particles ($f$=1). These radial profiles agree with the analytical results in Figure \ref{fig:f1}.  From left to right, the dust includes the whole population, small-particle population and big-particle population, with respect to solutions in Figure \ref{fig:f1}. The continuum emissions deviate from the observations within 60 au for the small-particle model. The small-particle model fits the outer disk ($\gtrsim$60 au) better, whereas the big-particle model fits the inner disk better. The whole population model combines the strengths of these two models. 

As for the porous particles ($f$=0.1), the MCRT (Figure \ref{fig:radmcintens} lower panels) also agrees with the analytical results in Figure \ref{fig:f0p01} largely, but the big-particle model fits the whole disk better. The small particle model under-predicts the emissions from several gaps and rings. Comparing the compact and porous models (top-left and bottom-left), the porous model fits the SEDs better. No models can predict VLA observations beyond 60 au accurately due to low signal-to-noise ratio for those observations in that region.

\begin{figure*}[t!]
\includegraphics[width=\linewidth]{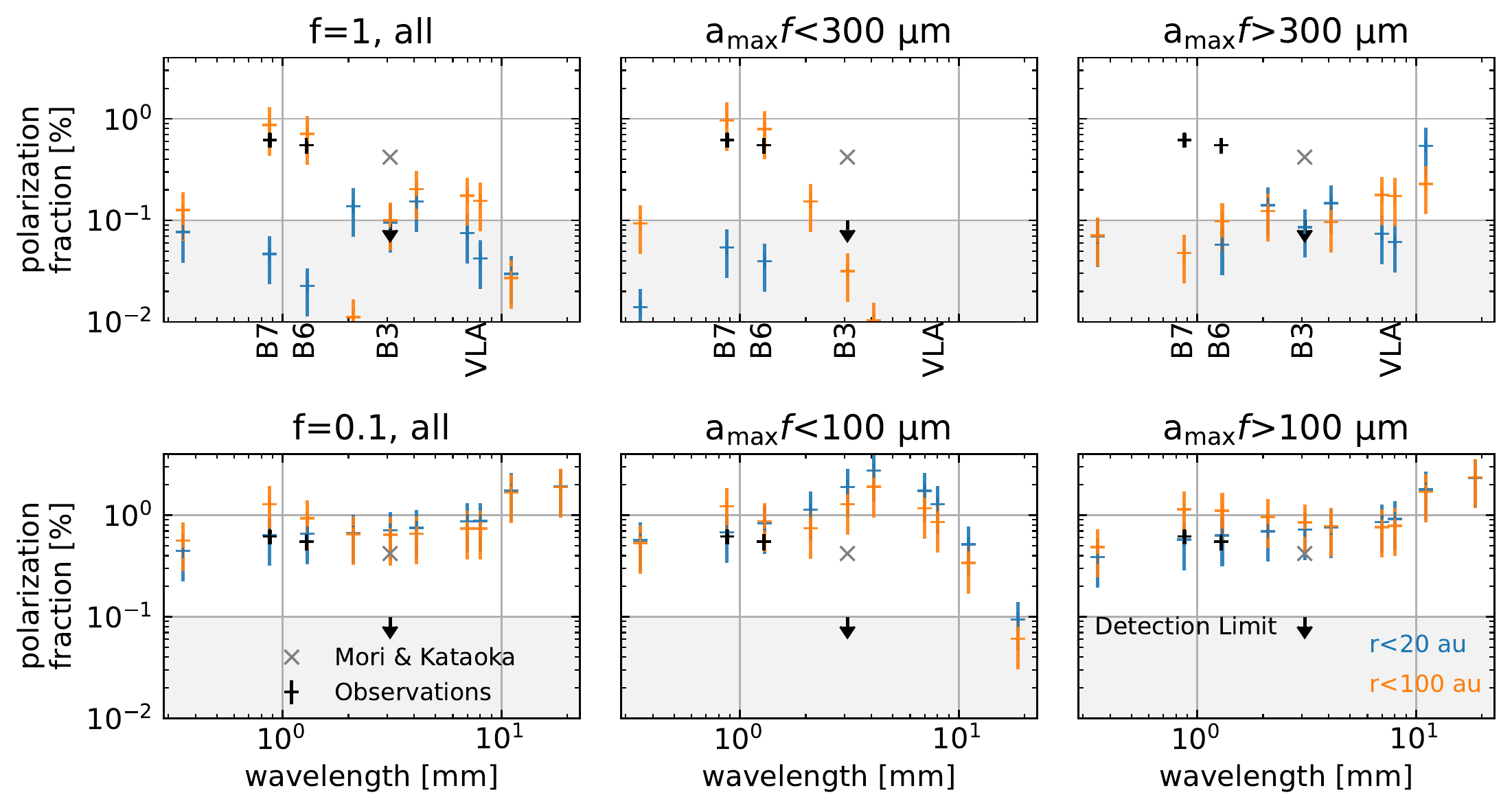}
\figcaption{The wavelength dependence of the linear polarization fraction at the center 20 au ({\color{blue}blue `+'}) and within 100 au ({\color{orange}orange `+'}) from RADMC-3D (MCRT) models and ALMA observations. The uncertainties of these models are taken to be 50\%, as different MCRT simulations show this level of variation \citep{kataoka15}. The existing observations are marked in black. The {\color{gray}gray `$\times$'} indicates the self-scattering component of the linear polarization from \citet{mori21}. The layout is the same as Figure \ref{fig:radmcintens}.
\label{fig:polint}}
\end{figure*}

\subsubsection{Integrated Polarization Fractions}
Figure \ref{fig:polint} presents the linear polarization fractions integrated in the central 20 au ({\color{blue}blue `+'}) and within 100 au ({\color{orange}orange `+'}). The uncertainties of these models are taken to be 50\%, as different MCRT codes show this level of variation for the polarization fraction \citep{kataoka15}. The observation values in the central 20 and 100 au with error bars \citep{kataoka2017, stephens17} are plotted in black. The observed polarization fractions within 100 au are very close\footnote{0.61\% and 0.58\% at B7; 0.53\% and 0.50\% at B6; and below the detection limit (0.1\%) at B3 for integrated polarization fractions within 20 au and 100 au, respectively.} to those in the central 20 au. We just use one label to represent observed values from both regions. The polarization fraction is computed from the spatially integrated (within the region of interest) $I$, $Q$, and $U$ emissions to average out the azimuthal polarization, focus on the component parallel to the minor axis, and boost the signal-to-noise ratio. A quick summary of observations is the following. The polarization fractions are $\sim$ 0.5\% at bands 7 and 6 ($\sim$0.87 and 1.3 mm). The linear polarization fraction at band 3 ($\sim$3.1 mm) is non-detection, with the upper limit as 0.1\%. However, the self-scattering polarization fraction can still be as high as 0.4\% indicated by `$\times$' since the polarization due to radiative alignment can cancel some of the self-scattering component \citep{mori21}. Owing to this complication at band 3, we mainly focus on comparing with ALMA bands 7 and 6 observations.

\textit{Compact Particles.} The upper panels of Figure \ref{fig:polint} show models with compact particles ($f$=1). All three models ({\color{blue}blue} markers) fall short of polarization fractions to explain the bands 7 and 6 data in the inner disk by one order of magnitude. However, with a slight change of parameters, both small-particle and big-particle models can explain the polarization of the inner disk ($\lesssim$ 20 au).

For the small compact particle model ($\lesssim$20 au), the low polarization fraction is because the best-fit $a_{\mathrm{max}}f$ is around 0.1 $\mu$m in the inner disk, very far from $a_{\mathrm{max}}f$ $\sim$ 100 $\mu$m that can explain the high polarization fractions. Since 100-$\mu$m-sized particles are the preferred particle size in previous literature to explain band 7 polarization data \citep{kataoka15, kataoka16, yang16}, we want to confirm they can produce enough polarization fractions in our setup. In the meantime, we want to test the quality of the SED fitting using 100-$\mu$m-sized particles compared to the small-particle model with $a_{\mathrm{max}}f$ around 0.1 $\mu$m. In Appendix \ref{sec:af100um} and Figure \ref{fig:100um}, we confirm that 100-$\mu$m-sized compact particles can explain the polarization fractions as previous studies, and the quality of the SED fitting is similar to that of the 0.1-$\mu$m-sized particles.

For the optically-thick inner region ($\lesssim$20 au), \citet{ueda21} has demonstrated that vertical settling can hide big particles in the midplane. The small particles on the surface can provide enough polarization. When the settling is very strong $\alpha_t \lesssim 10^{-5}$ (stronger than 10$^{-4}$ adopted here), the $a_{\mathrm{max}}f$ = 1 mm case can still provide polarization fractions high enough at bands 7 and 6 at the inner disk \citep{ueda21}.

For the outer disk's integrated polarization fractions ($<$ 100 au), the small compact particles ({\color{orange}orange} markers) can match the observed polarization fractions at bands 7 and 6, since the best solution of $a_{\mathrm{max}}f$ is around 100 $\mu$m in this region (see Figure \ref{fig:f1} middle column). The compact-big-particle models cannot produce observed polarization fractions. Since the outer disk is most likely to be optically thin, the big particles cannot be hidden in the midplane. Thus, the compact-big-particle solution is ruled out.

\textit{Porous Particles.} The lower panels of Figure \ref{fig:polint} show cases with porous particles ($f=0.1$). The polarization fractions of both inner and outer disk are much higher for big porous particles compared to big compact particles. Even for $a_{\mathrm{max}}f$ $\sim$ 1 m, the polarization fractions can match observations at bands 7 and 6. Note that these moderately porous particles (porous agglomerates) have similar properties to the large irregular particles, which can produce high polarization fractions even though their size is very large \citep{lin23}. All three models can match the polarization fractions across the disk. For all three models, the outer disks have similar but higher polarization fractions than the inner disks. We find that this is because the outer disk ($\gtrsim$20 au) has optical depths closer to unity, whereas the inner disk ($\lesssim$20 au) has $\tau_{\mathrm{tot}} \gg$1 (see Figure \ref{fig:f0p01} right column second row). The polarization is the strongest when the optical depth is around unity \citep{yang17}. Several more detailed tests show that the polarization fraction is high enough to match observations as long as $a_{\mathrm{max}}f$ $\gtrsim$ 100 $\mu$m. Overall, the prediction for porous big particles is that at longer wavelengths, the linear polarization fractions are higher than or comparable to bands 7 and 6 ($\sim$ 0.5\%).

\subsubsection{Polarization Fractions along Major and Minor Axes at ALMA B7}
\begin{figure*}[t!]
\includegraphics[width=\linewidth]{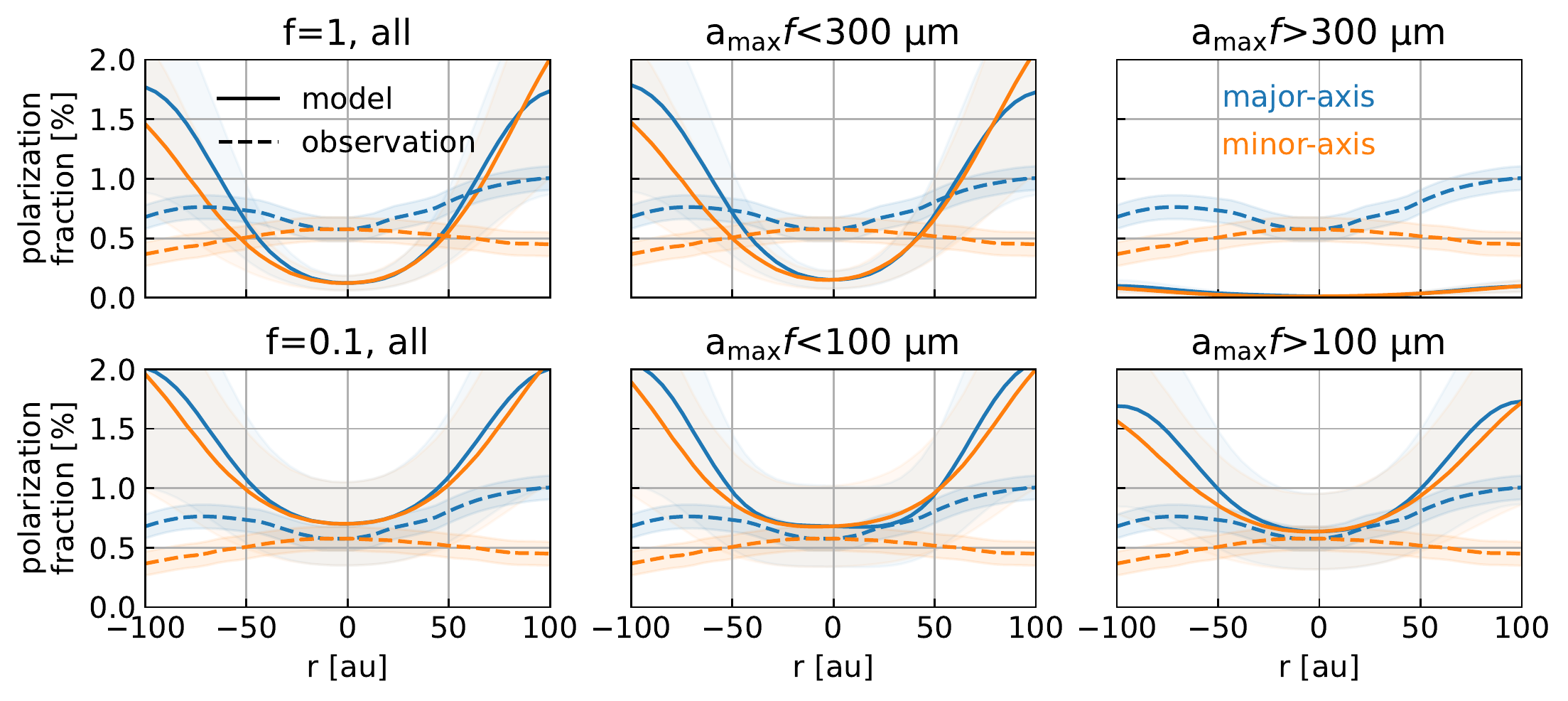}
\figcaption{The polarization fraction along the major ({\color{blue}blue}) and minor ({\color{orange}orange}) axes at ALMA band 7 from RADMC-3D (MCRT) models and the ALMA observation. The solid curves are models and dashed curves are the observation. The layout is the same as Figures \ref{fig:radmcintens} and \ref{fig:polint}.
\label{fig:polb7}}
\end{figure*}

Figure \ref{fig:polb7} presents the linear polarization fraction at ALMA band 7 along major and minor axes as measured in \citet{hull18}. The layout for six models are the same as Figures \ref{fig:radmcintens} and \ref{fig:polint}. Now the polarization fraction is calculated at the sub-beam level as Figure 3 in \citet{hull18}. The dashed lines are observations ({\color{blue}blue}: major axis; {\color{orange}orange}: minor axis) and solid lines are models. The uncertainties of the observations are taken to be 0.1\% as the receiver calibration errors. The model uncertainties are 50\% of the central values due to the differences in various MCRT codes \citep{kataoka15}.

For compact particles, polarization fractions for big particles are too low to explain the observation. Since the outer disk ($r$ $\gtrsim$ 30 au) of HL Tau is optically thin, settling cannot explain the observed polarization fractions that are still high at bands 7, so the compact big particles solution can be ruled out in the outer disk. The small-particle model produces very low polarization in the inner disk and a strong linear polarization fraction at the outer disk. This is because the particle size for the small-particle model in the inner disk is $\sim$ 0.1 $\mu$m but becomes $\sim$ 100 $\mu$m in the outer disk (the same reason for explaining the difference of integrated polarization fractions between the inner and outer disk for the compact-small-particle model in Figure \ref{fig:polint}; see Appendix \ref{sec:af100um}).

If the particle is porous, the parameter space of $a_{\mathrm{max}}f$ to explain the high linear polarization fraction becomes larger. Both big-particle and small-particle models can explain the high polarization fraction across the disk, as does their combination. Even though the models over-predict the polarization fraction at the outer disk, we treat them as good matches considering there are larger uncertainties at the outer disk, e.g., contamination from the outflow \citep{stephens17}. This means that with porosity, a wide parameter space of $a_{\mathrm{max}}f$ ($\gtrsim 100 \mu$m) can explain the polarization observations, which is already reflected in the analytical results of $P\mathrm{\omega_{eff}}$ in Figures \ref{fig:opacwav} and \ref{fig:opacsize}.

\section{Discussion \label{sec:discussion}}

\subsection{Dust Mass\label{sec:dustmass}}
Dust mass is one of the most crucial properties of the protoplanetary disks. While our paper focuses on reproducing continuum (SED) and polarization observations jointly with both particle's absorption and scattering features, we want to emphasize that the absorption opacity of the porous particles alone can have crucial impact on the dust mass estimation solely from single-band continuum observations.

Recent SED fitting on hundreds of (spatially-unresolved) protoplanetary disks from optical to mm-wavelengths \citep{ribas20, rilinger22, xin23, kaeufer23} show that the dust masses are on average several times higher than those inferred from single-band (sub)mm fluxes (e.g., \citealt{andrews13, ansdell16, pascucci16, cieza19, grant21}). This inconsistency is mainly attributed to the optically-thin assumption used in these mm single-band studies. With a more proper radiative transfer treatment of the optically thick disks (e.g., using Equation \ref{eq:intensity} or full MCRT), the inferred dust mass can be several times higher \citep{liu19, zhu19b, liu22}. 

However, perhaps not all protoplanetary disks are massive enough to be optically thick. If dust particles are porous in protoplanetary disks, the dust mass inferred from mm dust emissions can still be underestimated by using the compact particle's opacity at mm wavelengths even in the optically thin region. As indicated by Figures \ref{fig:f1} and \ref{fig:f0p01}, the dust mass of HL Tau beyond 20 au inferred by porous-particle whole population model is three times higher than compact-particle whole population model. If we compare the big-particle populations, the difference can be a factor of six. This is simply due to the Mie interference pattern of the absorption opacity when $\lambda \sim 2\pi a$. Since it is a common practice to presume the dust emissions observed at mm wavelengths are also produced by particles around mm, the opacities are always taken at or near the peak of the Mie interference pattern. This increase of the opacity can be around a factor of six, as shown by Figures \ref{fig:demo}, \ref{fig:opacwav}, and \ref{fig:opacsize}, which naturally explains why the dust mass is underestimated by that amount. The SED fitting across the whole spectrum does not suffer from this issue, since the opacities at other wavelengths (where $\lambda$ $\ll$ $a_{\mathrm{max}}$) do not show Mie interference pattern, so the compact and porous absorption opacities are almost identical (Figure \ref{fig:demo} left panels). Even if these wide-wavelength-range SED fitting adopts the compact particles' opacity, only one or several fluxes at mm wavelengths are affected by the Mie interference pattern. This does not lead to a significant bias on the dust mass with the contribution of other tens to hundreds of data points spanning several orders of magnitudes of wavelengths.

\subsection{More Porous Particles with $f$=0.01 \label{sec:f0.01}}

We want to emphasize that we only study porous particles that belong to \textit{dense aggregates} or \textit{porous agglomerates} with fractal number $d_f$ $\sim$ 3 and $f \gtrsim$ 0.01 \citep{guttler19} in this paper. More porous particles ($f<0.01$) are \textit{fluffy agglomerates} that have fractal number $d_f$ $\sim$ 2 and different optical properties (e.g., \citealt{tazaki16, tazaki18}). They are inefficient at producing mm dust self-scattering polarization \citep{tazaki19, tazaki19a}.

\begin{figure}[t!]
\includegraphics[width=\linewidth]{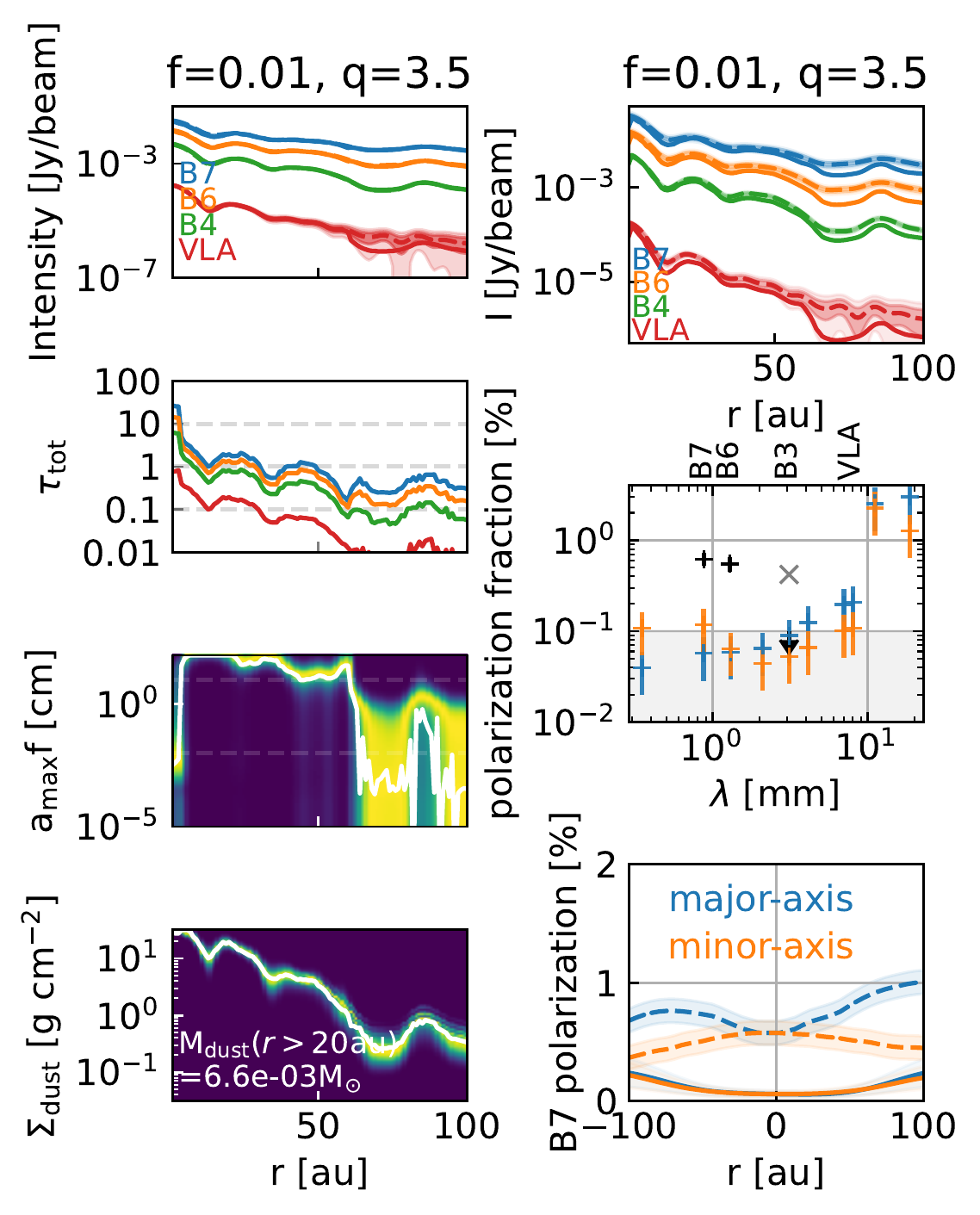}
\figcaption{The analytical fitting of continuum SEDs (left) and the MCRT (right) results for $f$=0.01 case. From top to bottom, the analytical fittings are the intensities, total optical depths, $a_{\mathrm{max}}f$, and the dust surface densities. The MCRT results are the intensities, integrated polarization fractions at different wavelengths, and the major and minor axes cuts of the polarization fractions at ALMA band 7.
\label{fig:inferred_supplement}}
\end{figure}

In this subsection, we test whether particles with $f$=0.01 (the lower bound of porous agglomerates' filling factor) can explain SED and polarization observations together. First, we try analytical fitting of SEDs (see Figure \ref{fig:inferred_supplement} left column). The fitting is as good as the case with $f=0.1$ (Figure \ref{fig:f0p01} first column). For simplicity, we omit the temperature, spectral index, and posterior probability rows. The $a_{\mathrm{max}}f$ is $\sim$ 1 m for $r \lesssim$ 20 au; 10 cm-1 m between 20-60 au; and $\sim$ 100 $\mu$m beyond 60 au. The dust mass beyond 20 au is 6.6$\times$10$^{-3}$ M$_\odot$, which is also similar to the case with $f=0.1$.

Then we run a RADMC-3D (MCRT) simulation for the best-fit model from the analytical fitting and present the result in Figure \ref{fig:inferred_supplement} right column. The first row shows that the radial profiles of the intensities under-predict the observations by a small fraction. The second and third rows show that the model predicts non-detectable polarization fractions at ALMA bands 7 and 6. For this reason, it is not a valid solution in HL Tau disk. The very low values of the polarization fractions from the MCRT result can be predicted from $P\omega_{\mathrm{eff}}$ in Figures \ref{fig:opacwav},  \ref{fig:opacsize}, and \ref{fig:discussion2} (also see \citealt{tazaki19, brunngraber21}).

\subsection{Impact on Different Particle Size Distribution Slope\label{sec:slope}}
The particle size distribution slope $q$ is not well-constrained from observations. It may vary from the ISM value (or from the steady state solution between coagulation and fragmentation) of 3.5 or even does not follow a power-law due to the dust growth \citep{birnstiel12}. On the observational side, \citet{macias21} allow $q$ as another fitting parameter and varying with radius to fit TW Hydrae disk. For that disk, the tightly constrained $q$ value is larger than 3.5 beyond 20 au and can reach 4 at 50 au. 

Here we try to study the impact of $q$ by adopting a shallower slope $q$=2.5, which is more consistent with the dust growth model and SED constraints \citep{dalessio01}. On the practical side, \citet{sierra21, macias21} have shown that with $q$=3.5, the optically thin spectral indices are always greater than 3, so they cannot find a reasonable solution of $a_{\mathrm{max}}$ or otherwise the size is tremendously large. Adopting $q$=2.5 instead, the solution can be found around several mm (see Figure 4 in \citealt{sierra21}).

\begin{figure*}[t!]
\includegraphics[width=\linewidth]{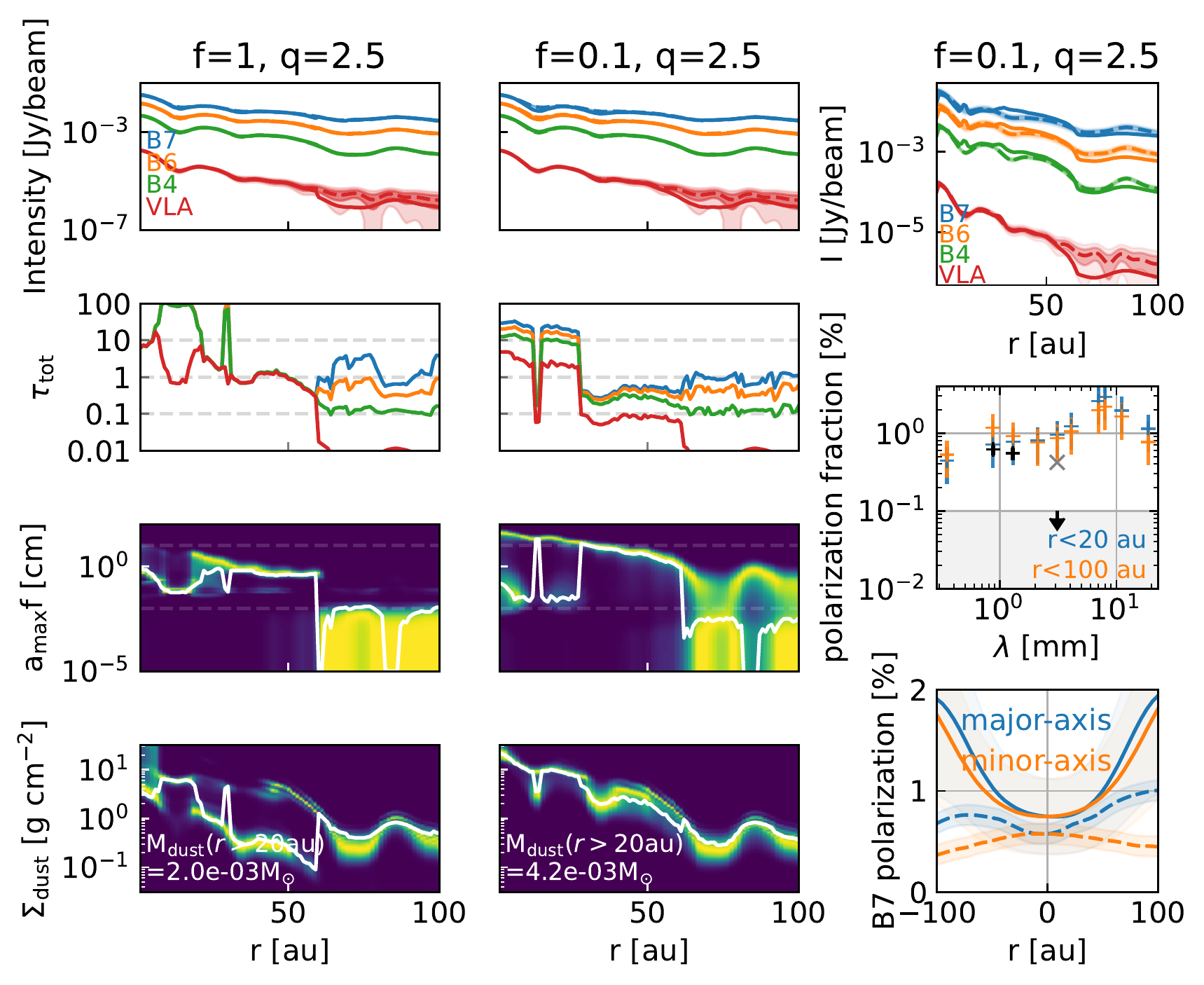}
\figcaption{The SED fitting of $q$=2.5, $f$=1 and  $q$=2.5, $f$=0.1 cases, and the MCRT fitting of the $q$=2.5, $f$=0.1 case. The layout is similar to Figure \ref{fig:inferred_supplement}.
\label{fig:q=2.5}}
\end{figure*}

The left two columns of Figure \ref{fig:q=2.5}  show the analytical fitting of continuum SEDs for $f$=1, $q$=2.5 and $f$=0.1, $q$=2.5. It is clear that both cases have much smaller particle sizes compared to their $q$=3.5 counterparts (Figures \ref{fig:f1} and \ref{fig:f0p01}). This difference is due to the different opacity indices of two particle size distribution slopes as pointed out by \citet{sierra21, macias21}. For the compact case ($f$=1), $a_{\mathrm{max}}f$ is 0.1-1 cm within 60 au, in contrast with 1 cm-1 m solution for $q$=3.5. Small particles ($\sim$100 $\mu$m) are preferred beyond 60 au.  The total dust mass beyond 20 au stays almost the same (2.0 $\times$ 10$^{-3}$ M$_{\odot}$ compared to 2.1 $\times$ 10$^{-3}$ M$_{\odot}$ in Figure \ref{fig:f1}) when $q$ changes from 3.5 to 2.5.

Similarly, for the porous case ($f$=0.1, $q=$2.5),  $a_{\mathrm{max}}f$ is 100 $\mu$m-10 cm within 60 au. The inner 30 au seems to prefer both 100$\mu$m and 10 cm sized particles. Small particles ($\sim$100 $\mu$m) are also preferred beyond 60 au. Overall, the particle size decreases with increasing radius. The total dust mass beyond 20 au decreases slightly from 5.7 $\times$ 10$^{-3}$ M$_{\odot}$ (in  Figure \ref{fig:f0p01}) to 4.2$\times$ 10$^{-3}$ M$_{\odot}$.

The third column of Figure \ref{fig:q=2.5} shows the MCRT fitting results. The MCRT model over-predicts the intensities within 60 au and under-predicts them beyond 60 au. The polarization fractions are very similar to the case with $q$=3.5, $f$=0.1 (Figures \ref{fig:polint} and \ref{fig:polb7}). This model can match the observed polarization fractions at ALMA bands 7 and 6. We do not show the MCRT results for $q$=2.5, $f$=1 case, since they cannot produce enough polarization fractions and the fractions are very similar to the $q$=3.5, $f$=1 case (Figures \ref{fig:polint}, and \ref{fig:polb7}).

Overall, the $q$=2.5, $f$=0.1 model can reproduce both SED and polarization observations. Admittedly, the SEDs from the MCRT do not fit the observations as good as the $q$=3.5, $f$=0.1 model (Figure \ref{fig:radmcintens}) without further tuning. It also predicts smaller particle sizes and slightly less dust masses.

\subsection{Constraints on Filling Factor and Size Distribution Slope from Current Polarization Observations\label{sec:constraints}}

With accurate MCRT models, we already know that $f$=0.1 cases can explain both SEDs and polarization fractions (Figures \ref{fig:radmcintens}, \ref{fig:polint}, and \ref{fig:polb7}). However, owing to the computational cost of MCRT calculations, we only generate discrete values of the filling factors ($f$ = 1, 0.1, and 0.01). Using the proxy for polarization fraction, $P\omega_{\mathrm{eff}}$, we can further constrain the possible values of $f$ in a continuous range from ALMA bands 7 and 6 polarization observations.

We caution that $P\omega_{\mathrm{eff}}$ can be only used as a reference, since the observed polarization fractions are always smaller than $P\omega_{\mathrm{eff}}$ due to complicated anisotropic multiple scattering. The exact ratio between the MCRT result and $P\omega_{\mathrm{eff}}$ depends on various parameters (e.g., $a_{\mathrm{max}}f$, $q$, dust surface density, disk inclination, etc.). We adopt the conversion factor $C$=2\% (observed polarization fraction is $CP\omega_{\mathrm{eff}}$) from \citet{kataoka16}, which are calibrated for an HL Tau MCRT model with $a_{\mathrm{max}}f$ $\sim$ 100 $\mu$m, $q$=3.5, and $f$=1. This exact suite of parameters is different from our setups. Ideally, we should calibrate as many models as possible. However, this cannot be realized in our situation. As shown in the analytical fitting (Figure \ref{fig:discussion}), $a_{\mathrm{max}}f$ is always changing with radius. Since the observational beams are large for current polarization observations ($\sim$ 30 au), the contribution of polarization from different particle sizes in different regions are highly mixed. Only an MCRT model with an observational setup can provide a definite value of the polarization fraction.

\begin{figure}[t!]
\includegraphics[width=\linewidth]{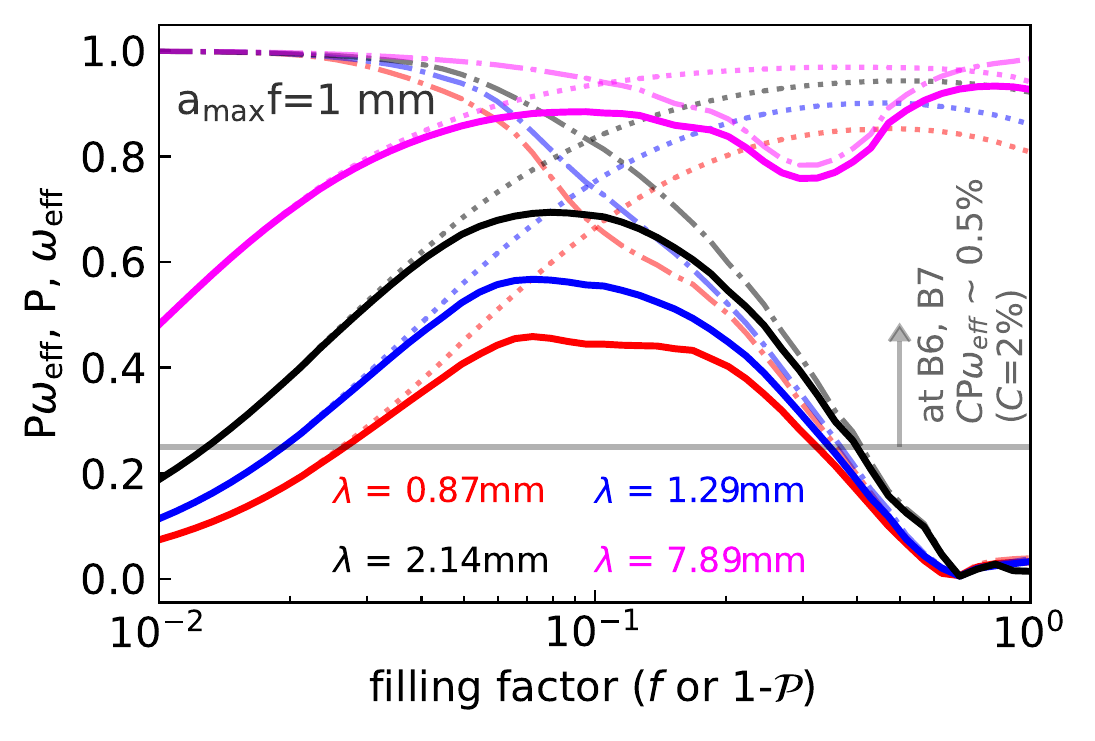}
\figcaption{$P\omega_{\mathrm{eff}}$ of the particles against different filling factors for $af$ = 1 mm, in a similar manner as Figure \ref{fig:opacsize}. {\color{red}Red}, {\color{blue}blue}, {\color{black}black}, and {\color{magenta}magenta} curves represent these quantities at 0.87 mm, 1.29 mm, 2.14 mm, and 7.89 mm. Solid lines are $P\omega_{\mathrm{eff}}$, dotted lines are $\omega_{\mathrm{eff}}$, and dashed-dotted lines are $P$. The shallow horizontal gray line indicates the observational linear polarization values at ALMA B6, B7 transferred to $P\omega_{\mathrm{eff}}$ with a normalization constant $C=$ 2\% (e.g., \citealt{kataoka16}). Other combinations of $af$ and $q$ can be found in Figure \ref{fig:af_supplement}.
\label{fig:discussion2}}
\end{figure}

Figure \ref{fig:discussion2} shows $P\omega_{\mathrm{eff}}$ for $a_{\mathrm{max}}f$ = 1 mm, $q$=3.5 against the filling factors at various wavelengths ({\color{red}red}: 0.87 mm, {\color{blue}blue}: 1.29 mm,, {\color{black}black}: 2.14 mm, and {\color{magenta}magenta}: 7.89 mm). Except for $\lambda$= 7.89 mm, the polarization fraction has a peak around $f$=0.1 and drops to a low value when $f$ is small or approaches unity. For ALMA B6 and B7 observations, the observational linear polarization fraction is around 0.5\%. Converting to $P\omega_{\mathrm{eff}}$ with conversion factor $C$=2\%, we find that $P\omega_{\mathrm{eff}}$ should be above 25\% (the gray horizontal line). This means that the filling factor $f$ should be from 0.03 to 0.3, or the porosity $\mathcal{P}$ should be 70-97\%. 

\begin{figure*}[t!]
\includegraphics[width=\linewidth]{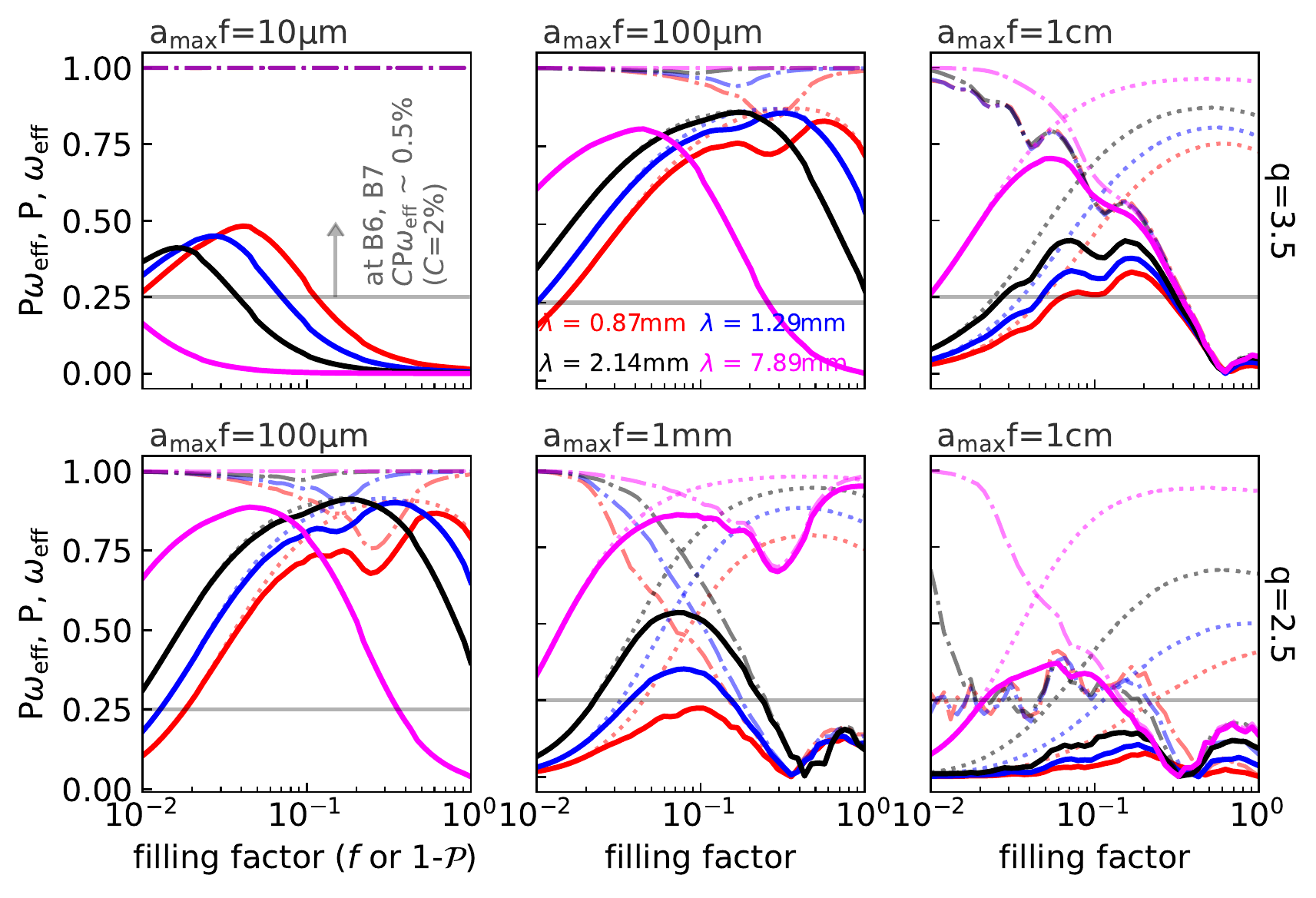}
\figcaption{$P\omega_{\mathrm{eff}}$ of the particles against different filling factors for different $af$, and size distributions $q$ that complements Figure \ref{fig:discussion2}. The top panels are with $q$=3.5 and $af$=10 $\mu$m, 100 $\mu$m, and 1 cm. The bottom panels are with $q$=2.5 and $af$=100 $\mu$m, 1 mm, and 1 cm.
\label{fig:af_supplement}}
\end{figure*}

The slope of the particle size distribution also affects the polarization fraction. Typically, a smaller $q$ produces weaker polarization fractions, since a smaller $q$ means less contribution from the small particles that contribute to the high polarization. Fixing $a_{\mathrm{max}}f$ and $f$, we expect a larger $q$ can fit the observed high polarization fractions better. Figure \ref{fig:af_supplement} shows $P\omega_{\mathrm{eff}}$ for two different particle size distribution slopes $q$ (3.5 and 2.5) and various particle sizes $a_{\mathrm{max}}f$. For $q$=3.5 (top panels), when $a_{\mathrm{max}}f$ = 1 cm, filling factor $\sim$ 0.1 can still explain the high polarization fraction observed at band 6 and 7. When $a_{\mathrm{max}}f$ = 10 $\mu$m, only $f$ $\lesssim$ 0.1 can explain the observed polarization fraction. For $a_{\mathrm{max}}f$ = 100 $\mu$m, both $q=$3.5 (top-middle panel) and 2.5 (bottom-left panel) can explain the observed polarization fraction as long as $f\gtrsim$ 0.02. For $q$=2.5 (bottom panels), $a_{\mathrm{max}}f$=1 mm can marginally match the observed polarization fraction only when $f\sim$ 0.1. For $a_{\mathrm{max}}f$=1 cm, the polarization fractions are too low to explain the observations.

We note again that MCRT simulations are needed to compare with observations closely. As Figure \ref{fig:q=2.5} shown, the $q$=2.5, $f$=1 MCRT model has $a_{\mathrm{max}}f \gtrsim$ 1 cm between 30-60 au, so the polarization fractions in that region are expected to be low according to Figure \ref{fig:af_supplement} bottom right panel. However, the polarization fractions (within 20 au or 100 au) of the MCRT model can still match the observational polarization fractions at ALMA bands 7 and 6. This is because the polarization fractions are mixed with regions that can produce high polarization fractions for $r\lesssim$ 30 au and $r\gtrsim$ 60 au, where $a_{\mathrm{max}}f \sim$ 100 $\mu$m.

\subsection{Prediction for Future Observations \label{sec:predictions}}

Higher resolution polarization observations at bands 7 and 6 can better constrain the particle properties. The current resolution for polarization observations is $\sim$ 30 au. This blends the regions with different particle sizes together (perhaps also $q$ and $f$). For example, as Section \ref{sec:constraints} mentioned, the $q=2.5$ models should produce less polarization fractions than the $q=3.5$ when $a_\mathrm{max}f$ = 1 cm. However, this distinction can be blended by the particles with 100 $\mu$m sizes that can produce higher polarization fractions in neighbouring regions. Higher resolution of the polarization observations at bands 7 and 6 can possibly separate the contributions from different parts of the disk.

Even with current angular resolution, we can still learn more on the porosity and particle size if we observe at longer wavelengths. Since only big porous particles can explain both SED and polarization observations between 20 to 60 au, the self-scattering polarization fractions for future longer wavelength observations (ALMA band 1 and ngVLA polarization observations) should be comparable to or higher than 0.5\%. If not, one would need to come up with other mechanisms to explain the inconsistency between SED fittings and mm polarization observations. 

Ideally, it is easy to test the prediction of the self-scattering polarization at longer wavelengths. The self-scattering polarization is either higher than 0.5\% (pointing to large, porous particles) or non-detectable (pointing to small particles). However, the situation is complicated by the ever-stronger azimuthal polarization pattern due to the dust thermal polarization at longer wavelengths \citep{stephens17, kataoka2017}. The self-scattering dominates at ALMA bands 7 and 6, whereas the dust thermal polarization dominates at band 3. The thermal polarization is thought to come from the particle alignment, but the underlying mechanism is unclear. Fortunately, the elusive mechanism does not prevent the morphological studies of the observations. The polarization can be decomposed by self-scattering and particle alignment components under some assumptions. After subtracting the dust alignment component, even at band 3, a moderate amount of polarization ($\sim$0.4\%) should come from the self-scattering to explain the observation \citep{yang19, mori21, lin22a}. In principle, we can follow these studies to probe the self-scattering component at longer wavelengths. If the self-scattering component is non-detected, the particle size should be small. Otherwise, it should be large and porous. However, one more complication comes from the assumption of small particle sizes in the Rayleigh regime ($a_{\mathrm{max}}\lesssim 100 \mu$m) used in these previous studies on the particle alignment. While the patterns are easier to understand in the Rayleigh regime, the size for the big particle is in the Mie regime, where the polarization properties for particle alignment mechanisms are very different \citep{guillet20}. Under the large and porous particle scenario, one must carefully study the particle alignment polarization pattern before separating out the self-scattering component at longer wavelengths.

On the observational side, one needs to carefully characterize the free-free emission near the star and subtract its flux contribution before calculating the polarization in the inner disk \citep{carrasco16, carrasco19}. For the outer disk where the emission is weaker, the challenge comes from the requirement of high signal-to-noise ratio (500 for 1\% of polarization) and observations with high enough resolution to separate the free-free emission near the star.

\subsection{Connection to the Near-Infrared Protoplanetary Disk Observations and the Comet 67P Measurements from the Rosetta Mission\label{sec:connection}}

Near-infrared scattered light observations are sensitive to micron-sized dust grains/aggregates/agglomerates in the upper atmosphere (several gas scale heights) of the protoplanetary disks. Recently, \citet{ginski23} reveal the porosity of the micron-sized particles among 10 disks observed by VLT/SPHERE by deriving the full polarizing scattering phase function of such particles at multiple wavelengths.
Micron-sized particles are found to be porous in all the disks in the sample. From the shape of the phase function, they separate the disks into two categories. Category I is linked to fractal agglomerates, whereas Category II is consistent with moderately porous agglomerates with porosity $\sim$ 50\%.

Since the mm-cm-sized particles observed in radio observations should have been formed from these micron-sized particles in the past, the mm-cm-sized particles have to grow with some levels of porosity in the beginning. If they formed a long time ago (fast particle growth), they may have formed from fractal agglomerates (Category I). If they formed recently (slow particle growth), they may have been formed from already processed particles with moderately porous agglomerates (Category II). In either cases, we should expect that the mm-cm-sized particles to form with some levels of porosity. This is in agreement with our finding that mm-cm-sized particles are porous in HL Tau (a relatively young protoplanetary disk).
From mm self-scattering linear polarization, the level of porosity is roughly constrained to be 70\%-97\% using $P\omega_{\mathrm{eff}}$ (Figure \ref{fig:discussion2}). At face value, this level of porosity is lower than the Category I fractal agglomerates and higher than the Category II porous agglomerates. However, Category II micron-sized particles alone can be the constituents of the mm-cm-sized porous particles, since the inter micron-sized particles vacuum region can also contribute to the bulk porosity.

Comets are originated from a population of kilometer-sized icy planetesimals that formed in the solar nebula beyond the snow line. A review by \citet{blum22} describes a plausible scenario to form comets as the following steps.

In protoplanetary disks (or the solar nebula, specifically), sub-$\mu$m sized particles mainly went through hit-and-stick processes to form up to mm-sized (or even larger) fractal agglomerates with $>$95\% porosity \citep{okuzumi12, estrada16, estrada22a, estrada22b}. As the fractal agglomerates grew larger and more massive, the increasing collision energy led to compaction of the fractal agglomerates with a higher fractal number and less porosity. Eventually, the growth met the bouncing barrier, where the dust could not grow further and mm-dm-sized pebbles with porosity $\sim$ 60\% formed \citep{guttler10, zsom10, lorek18}. Then, these pebbles could concentrate in eddies, vortices, pressure bumps \citep{johansen14} and then by streaming instability \citep{youdin2005}. Once the criterion for gravitational instability was met, the pebble cloud collapsed and planetesmials formed \citep{johansen14}. After the gas depletion, these planetesmials went through radiogenic heating \citep{mousis17, lichtenberg21, golabek21} and collisional evolution \citep{jutzi15, jutzi17, schwartz18, golabek21} and became comets. For low mass planetesimals (with radii $\lesssim$ 50 km), the integrity of the pebbles can be preserved due to less collisions in the free-fall phase, surface impact during formation, and weaker hydrostatic pressures inside the planetesimals \citep{bukhari17, wahlberg17, blum18}.

The comet 67P/Churyumov–Gerasimenko is a km-sized comet and may well preserve the pristine pebbles through the dust growth. Various instruments on the \textit{Rosetta} mission provide accurate measurements on the dust porosity (see a review by \citealt{guttler19}). Measurements on the bulk density \citep{jorda16, patzold16} and the particle's permittivity \citep{kofman15, herique16, burger23} indicate the nucleus has porosity from 70\%-85\%. This is consistent with our constraint on the porosity in HL Tau from mm dust self-scattering polarization observations. However, macro-porosity (space between particles, pebbles, or boulders) might also contribute to the bulk porosity, even though the void should be no larger than 9 m \citep{ciarletti17}. \citet{blum17, burger23} explain the bulk porosity as the combination of 60\% micro-porosity  as the result of compaction at the bouncing barrier \citep{weidling09, zsom10} and 80\% macro-porosity due to the random packing of these dust particles \citep{skorov12, fulle17}. If the porosity $\sim$ 60\% due to compaction is applicable to the whole disk, the porosity of 70\%-97\% in HL Tau might indicate the dust is still growing and has not finished the compaction. On the other hand, the exact value of the porosity due to the compaction does depend on the location, dust-to-ice ratio, and the disk model. The quoted value is from the experiments that mimic the condition at 1 au with minimum mass solar nebula \citep{zsom10}. For larger radii to the star or increasing dust-to-ice ratio, the porosity due to the compaction can be larger \citep{lorek18}. Overall, we find a good agreement with the measurements on the porosity of comet 67P from the \textit{Rosetta} mission.

\section{Summary of the Constraints and Predictions \label{sec:summary}}

In Table \ref{tab:summary}, we collect all the results in this paper and in \citet{ueda21} to provide a tabular summary of all the possible solutions of HL Tau's particle properties from both SED and polarization observations. We also provide the polarization fraction predictions at longer wavelengths (e.g., ALMA B1 and ngVLA) for these possible solutions. We divide the HL Tau disk into three regions: within 20 au, where the disk is optically thick; 20-60 au, the major focus of our paper, where the disk has $\tau_{\mathrm{tot}}\lesssim$ 1; and 60-100 au, where the signal-to-noise ratio for the VLA data is low. In Table \ref{tab:summary}, each line presents a combination of porosity and particle size, following their viabilities of fitting current SED and polarization observations (marked by crosses or ticks), and predictions for polarization fractions at longer wavelengths. Only when two {\color{green}green} ticks are on the same line, it is a viable solution. 

\subsection{1-20 au}
Within 20 au, we find that big particles fit the SEDs better than small particles, regardless of the porosity (Figures \ref{fig:f1}, \ref{fig:f0p01}, and \ref{fig:radmcintens}). For compact particles, small particles are needed to explain the high polarization fractions (Figures \ref{fig:polint}, and \ref{fig:polb7}). For porous particles, both small and big particles with $f=0.1$ can explain the high polarization fractions (Figures \ref{fig:polint}, and \ref{fig:polb7}). For $f=0.01$, neither small nor big particles can reproduce high polarization fractions (Figure \ref{fig:inferred_supplement}). In summary, the only viable solution is big porous particles with $f$=0.1. We predict that polarization fractions are high ($\gtrsim$ 0.5\%) at longer wavelengths (e.g., ALMA B1 and ngVLA). However, \citet{ueda21} show that vertical settling can make small compact particles a viable solution as well. Since the optical properties between small compact and small porous particles are similar, we believe small porous particle is also a viable solution (for this reason, we also list the prediction for this model, even though we mark a cross under the SED column). For these two small particle solutions, we expect non-detectable polarization fractions ($<$0.1\%) at longer wavelengths. Thus, future polarization observations at longer wavelengths can distinguish the big and small particle solutions.

\subsection{20-60 au}
Between 20-60 au, the situation is very similar to the case within 20 au. The only difference is that the optical depth becomes smaller in this region, so the particle vertical settling cannot hide big particles in the midplane. Thus, the only viable solution is $f=0.1$ porous big particles, which is underlined in the table. With the help of $P\omega_{\mathrm{eff}}$ in Figure \ref{fig:discussion2}, we can optimistically loosen the possible $f$ to be from 0.03-0.3.

\begin{figure*}[t!]
\includegraphics[width=\linewidth]{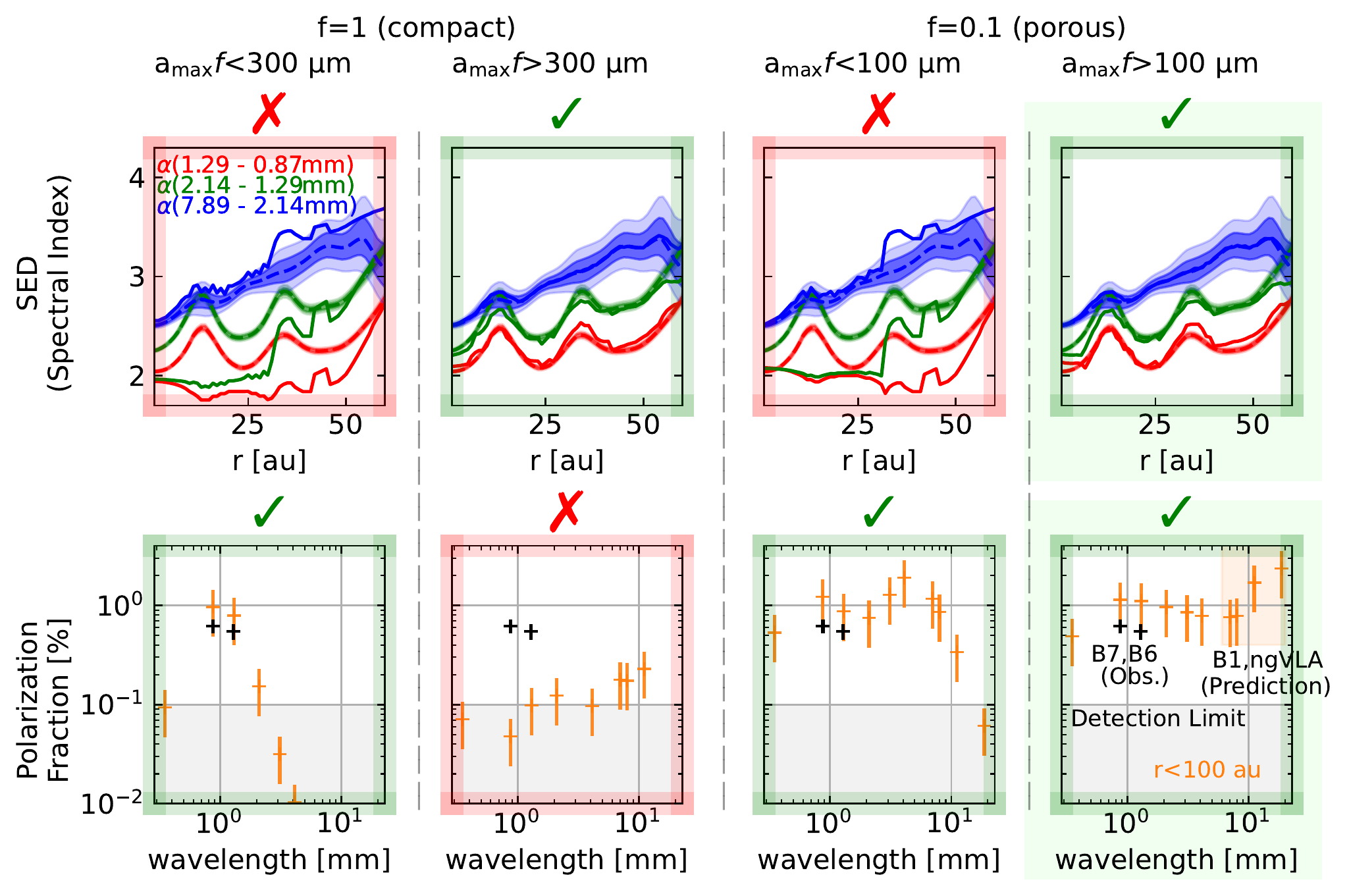}
\figcaption{A summary of the constraints on the porosity and particle size from SED and polarization observations. Top row shows the spectral indices (from Figures \ref{fig:f1} and \ref{fig:f0p01}) and the bottom row shows the integrated polarization fractions within 100 au at different wavelengths (from Figure \ref{fig:polint}). `{\color{green}\ding{51}}' means allowable solution, and `{\color{red}\ding{55}}' means a poor fitting.
\label{fig:money_plot}}
\end{figure*}

Since the region between 20-60 au is of our primary focus, we use Figure \ref{fig:money_plot} to visualize the process that eliminates other possible situations. The left two columns show compact cases with small and big particle models for $f$=1. The right two panels show porous cases for $f$=0.1. In the upper row, we take the spectral indices in Figures \ref{fig:f1} and \ref{fig:f0p01} to demonstrate the quality of SED fitting. In the lower row, we use the integrated polarization fractions within 100 au to demonstrate the quality of polarization fraction fitting, and also the predictions at longer wavelengths. If the model matches the observations, we mark the plot in {\color{green} green} and labeled it as a {\color{green} green} tick. If the model does not match the observations, we mark the plot in {\color{red} red} and labeled it as a {\color{red} red} cross. It is clear that only the $a_{\mathrm{max}}f$ $>$ 100 $\mu$m, $f$=0.1 model can explain both SED and polarization observations. At ALMA B1 and ngVLA wavelengths, the self-scattering linear polarization fraction should be high ($\gtrsim$ 0.5\%).

\subsection{60-100 au}
The constraint is not strong between 60-100 au due to the poor signal-to-noise ratio of VLA data. Still, we describe the results at face value. In general, this region prefers small particles. For $f$=1, small particles can explain both SED and polarization observations (Figures \ref{fig:f1}, \ref{fig:radmcintens}, \ref{fig:polint}, and \ref{fig:polb7}). This model predicts non-detectable polarization fraction at longer wavelengths. Big particles fail to explain both types of observations. For $f$=0.1, both small and big particles can explain both types of observations (Figures \ref{fig:f0p01}, \ref{fig:radmcintens}, \ref{fig:polint}, and \ref{fig:polb7}). Whether polarization fractions are high or not at longer wavelengths can distinguish the particle size. For $f$=0.01, both big and small particles can explain the SEDs, similar to the $f$=0.1 cases. However, both of the particle sizes fail to explain the polarization fractions at ALMA bands 7 and 6 (Figure \ref{fig:inferred_supplement}).

\begin{deluxetable*}{llllll}
\tabletypesize{\scriptsize}
\tablecaption{Summary of the parameter space exploration \label{tab:summary}}
\tablehead{
\colhead{region} &
\colhead{porosity} & \colhead{particle size} & \colhead{SED} & 
\colhead{polarization} &
\colhead{polarization prediction}
\\
\colhead{} &
\colhead{$\mathcal{P}$=1-$f$} & \colhead{a$_{\mathrm{max}}f$ (mm)} & \colhead{(Fig. \ref{fig:f1}, \ref{fig:f0p01}, \ref{fig:radmcintens}, \ref{fig:inferred_supplement})} & 
\colhead{(Fig. \ref{fig:polint}, \ref{fig:polb7}, \ref{fig:100um})} &
\colhead{at longer $\lambda$}
}
\colnumbers
\startdata
\textless{}20 au & 0\%  & $\le$0.3       & {\color{red}\ding{55}} ({\color{green}\ding{51}}\tablenotemark{*})               & {\color{green}\ding{51}}  ({\color{green}\ding{51}}p\tablenotemark{*})  & $<$0.1\%         \\
 &          & \textgreater{}0.3           & {\color{green}\ding{51}} ({\color{green}\ding{51}}p\tablenotemark{*})    & {\color{red}\ding{55}}  ({\color{green}\ding{51}}\tablenotemark{*})      &    $\gtrsim$0.5\% \\
           & 90\%   & $\le$0.1       & {\color{red}\ding{55}}              & {\color{green}\ding{51}}  & $<$0.1\%           \\
 &          & \textgreater{}0.1       & {\color{green}\ding{51}}     & {\color{green}\ding{51}}  & $\gtrsim$0.5\%            \\
           & 99\%   & $\le$0.1       & {\color{red}\ding{55}}              & {\color{red}\ding{55}}  &                           \\
 &          & \textgreater{}0.1       & {\color{green}\ding{51}}     & {\color{red}\ding{55}}  &                            \\
\hline\hline
20-60 au     & 0\%  & $\le$0.3        & {\color{red}\ding{55}}   & {\color{green}\ding{51}}           &                          \\
 &          & \textgreater{}0.3        & {\color{green}\ding{51}}               &  {\color{red}\ding{55}}   &                    \\
 & 90\%   & $\le$0.1       & {\color{red}\ding{55}}              & {\color{green}\ding{51}}           &                  \\  \textbf{(best solution)}       &          &  \textbf{\textgreater{}0.1 }     & {\color{green}\ding{51}}              & {\color{green}\ding{51}}           &  \textbf{$\gtrsim$0.5\%} \\
\hline
 & 99\%   & $\le$0.1       & {\color{red}\ding{55}}              & {\color{red}\ding{55}}           &                  \\         &          &  \textgreater{}0.1      & {\color{green}\ding{51}}              & {\color{red}\ding{55}}           &                  \\
\hline\hline
60-100 au (low SNR)  & 0\%  & $\le$0.3        & {\color{green}\ding{51}}     & {\color{green}\ding{51}}             &   $<$0.1\%       \\
               &          & \textgreater{}0.3        & {\color{red}\ding{55}}           &    {\color{red}\ding{55}}   &                \\
                 & 90\%   & $\le$0.1       & {\color{green}\ding{51}}       & {\color{green}\ding{51}}           & $<$0.1\% \\
                 &          & \textgreater{}0.1       & {\color{green}\ding{51}}     & {\color{green}\ding{51}}  &    $\gtrsim$0.5\%        \\
                 & 99\%   & $\le$0.1       & {\color{green}\ding{51}}       & {\color{red}\ding{55}}           &                 \\
                 &          & \textgreater{}0.1       & {\color{green}\ding{51}}     & {\color{red}\ding{55}}  &                           \\                 
\hline\hline
\enddata
\tablecomments{Green ticks ({\color{green}\ding{51}}) are allowable solutions, whereas red crosses ({\color{red}\ding{55}}) are excluded due to the lack of polarization, or they are less favorable solutions in continuum fitting. The prediction is on the polarization fraction of the self-scattering component of the future polarization observations at longer wavelengths (e.g., ALMA band 1 and ngVLA). The best solution of HL Tau outer disk with sufficient observed SNR (20-60 au) is underlined and in \textbf{bold}.}
\tablenotetext{*}{The inner 20 au can be affected by the vertical settling and optical depth effects. We combine results from \citet{ueda21}, where ``p'' (preferred) means that the fitting requires less fine-tuning comparing between big and small-particle models \citet{ueda21}.}
\end{deluxetable*}

\section{Conclusion \label{sec:conclusion}}

We use porous particles to explain the inconsistency of the particle size constraints between the radio SEDs and mm polarization observations. We first explore the optical properties of particles with different porosities at different wavelengths. Then we test out our finding using the HL Tau disk. We use a Bayesian approach to fit the multi-wavelength continuum observations (SEDs) under different filling factors, taking the particle size, dust surface density, and temperature to be free parameters. With the obtained best solutions, we run Monte Carlo Radiative Transfer (MCRT) to make a closer comparison with SED (at ALMA bands 7, 6, 4, and VLA Q+Ka band) and polarization observations (at ALMA bands 7 and 6).
Our main findings are as follows:

\begin{enumerate}
\item  The porous particles have closer-to-unity refractive indices than compact particles. This makes the Mie interference pattern at mm wavelengths for mm-cm-sized particles disappear (Figures \ref{fig:demo}, \ref{fig:opacwav}, and \ref{fig:opacsize}).
\item The high opacity at mm wavelengths due to the Mie interference is unique to compact mm-cm-sized particles. This means that dust mass estimated using porous particles can be a factor of six or higher than compact particles (Figures \ref{fig:f1}, \ref{fig:f0p01}, and \ref{fig:discussion}).

\item For compact particles, only 100-$\mu$m sized particles can provide enough polarization fractions at ALMA bands 7 and 6. For porous particles, the window for a high polarization fraction becomes much broader, leaving larger porous particles ($f$=0.03-0.3, $a_{\mathrm{max}}f$ $\gtrsim 100 \mu$m) a valid solution for the polarization observations (Figures \ref{fig:opacwav}, \ref{fig:opacsize}, \ref{fig:discussion2}, and \ref{fig:af_supplement}).

\item We adopt a fiducial particle size distribution slope $q$=3.5. We find that the major impact on changing $q$ from 3.5 to 2.5 is that the best solution's particle sizes can go down from 10 cm-1 m to 1 mm-10 cm. The estimated dust mass can also decrease slightly (Figures \ref{fig:f0p01}, \ref{fig:q=2.5}, and \ref{fig:discussion}).

\item Using MCRT, we demonstrate that $f$=0.01 cases cannot provide enough polarization fractions at ALMA B7 and B6, regardless of the particle size (Figure \ref{fig:inferred_supplement}).

\item Combining SED and polarization fittings from analytical approximation and MCRT, we summarize the viable solutions of particle sizes and porosities in Table \ref{tab:summary}. In the table, we study the disk in three regions (1-20 au, 20-60 au, and 60-100 au), separated by small and big particle solutions, and three different filling factors ($f$ = 1, 0.1, and 0.01).

\item In Table \ref{tab:summary}, we also provide predictions for future polarization observations at longer wavelengths (e.g., ALMA B1 and ngVLA) that can distinguish the small and big particle solutions.

\item For radii between 20-60 au, we demonstrate that only $f$ = 0.1, $a_{\mathrm{max}}f$ $\gtrsim\ 100 \mu$m MCRT model can explain both types of observations (Figure \ref{fig:money_plot}). With the help of $P\omega_{\mathrm{eff}}$ (an analytical proxy for polarization fraction), the allowable filling factors can be optimistically relaxed to 0.03-0.3, which translates to porosities between 70\%-97\% (Figure \ref{fig:discussion2}).

\item We predict that we will observe higher linear self-scattering polarization fractions at ALMA B1 and ngVLA bands than ALMA B7 and B6 ($\sim$0.5\%) when they are integrated within 100 au (Figure \ref{fig:money_plot}).

\item We find that the existence of porous particles in HL Tau disk and their levels of porosity are consistent with near-infrared scattered light observations of protoplanetary disks and characterization of the comet 67P from the \textit{Rosetta} mission (Section \ref{sec:connection}).

\end{enumerate}

\acknowledgments
We thank the anonymous reviewer who helped us improve the quality of this manuscript. S.Z. thanks Lee Hartmann for helpful discussion. S.Z. and Z.Z. acknowledge support through the NASA FINESST grant 80NSSC20K1376. S.Z. acknowledges supports from Russell L. and Brenda Frank Scholarship and Barrick Graduate Fellowship. Z. Z. acknowledges support from the National Science Foundation under CAREER grant AST-1753168 and support from NASA award 80NSSC22K1413. T.U. acknowledges the support of the DFG-Grant "Inside: inner regions of protoplanetary disks: simulations and observations" (FL 909/5-1). A.S. acknowledges support from ANID FONDECYT project No. 3220495. C.C.-G. acknowledges support from  UNAM DGAPA-PAPIIT grant IG101321, and CONACyT Ciencia de Frontera grant number 86372.

The National Radio Astronomy Observatory is a facility of the National Science Foundation operated under cooperative agreement by Associated Universities, Inc. This paper makes use of the following ALMA data: ADS/JAO.ALMA\#2011.0.00015.SV,\\ ADS/JAO.ALMA\#2016.1.00115.S,\\  ADS/JAO.ALMA\#2016.1.00162.S.  ALMA is a partnership of ESO (representing its member states), NSF (USA) and NINS (Japan), together with NRC (Canada), MOST and ASIAA (Taiwan), and KASI (Republic of Korea), in cooperation with the Republic of Chile. The Joint ALMA Observatory is operated by ESO, AUI/NRAO and NAOJ.

\software{
{\tt RADMC-3D} \citep{dullemond12},
{\tt dsharp\_opac} \citep{github_dsharp_opac},
{\tt OpTool} \citep{optool},
{\tt Matplotlib} \citep{matplotlib},
{\tt Numpy} \citep{numpy}, 
{\tt Scipy} \citep{scipy}
}

\appendix

\section{Validity of the Effective Medium Approximation \label{sec:applicability} }
In reality, dust particles can have different shapes and porosities, but it is unrealistic to measure the optical properties of each individual particle. Fortunately, the overall optical properties as an ensemble of heterogeneous particles can be well captured by the effective medium approximation (EMA) when 2$\pi a_{\mathrm{inc}}/\lambda \ll$ 1, where $a_{\mathrm{inc}}$ is the characteristic radius of the \textit{inclusion}, rather than the dust particle's radius, $a$ \citep{mishchenko00}. We choose the ``Bruggeman'' mixing rule to calculate all the optical properties throughout the paper, following \citet{birnstiel18}. With this mixing rule, direct comparisons with experiments and direct dipole approximations (DDAs\footnote{a more accurate but more time-consuming method to calculate opacity.}) show that the error of angular dependent extinction property can be as low as 1\% when 2$\pi a_{\mathrm{inc}}/\lambda <$ 0.1, and 10\% even when 2$\pi a_{\mathrm{inc}}/\lambda =$ 2 (\citealt{mishchenko00} and references therein). In our application, for $\lambda \sim$ 1 mm, we need to assume the inclusions to be $\lesssim$ 200 $\mu$m for the EMA to work, which should be a valid assumption as particles in the ISM have much smaller sizes ($< 1 \mu$m). Recently, \citet{tazaki23} constrain the monomer sizes of the small particles on the atmosphere of IM Lup  to be less than 0.4 $\mu$m. If we assume the big particles in the midplane are constituted of similar monomers, the EMA should be very accurate.

\section{Key parameters \label{sec:keyparameters}}
The size parameter, 
\begin{equation}
    x = \frac{2\pi a}{\lambda}, \label{eq:sizeparameter}
\end{equation}
is a dimensionless parameter that characterizes the size of the particle, where $a$ is the physical radius of the particle, and $\lambda$ is the wavelength. We can separate the particle size into three regimes with the size parameter. When $x \ll$1, the particle is small and in the Rayleigh regime. When $x\sim$1, the particle is in the Mie regime. When $x\gg$1, the particle is in the geometric regime. For spherical particles, the complex refractive index $m$ fully determines the dust's optical properties, where
\begin{equation}
    m(\lambda) = n(\lambda) + ik(\lambda), \label{eq:refractiveindex}
\end{equation}
where $n$ is related to the scattering and $k$ is associated with the absorption. However, both the reflective index's real and imaginary parts participate in calculating the dimensionless absorption and scattering properties, namely $Q_{\mathrm{abs}}$ and $Q_{\mathrm{sca}}$, the absorption and scattering efficiencies. They are the ratios between the absorption/scattering cross sections over the geometric cross-sections (e.g., $\pi a^2$, for a spherical particle).

These efficiencies can be converted to the mass opacity commonly used in observations,
\begin{equation}
    \mathrm{\kappa_{abs}} \equiv \frac{\pi a^2}{\mathcal{M}} Q_{\mathrm{abs}} = \frac{3/4}{\rho_{\mathrm{int}}af} Q_{\mathrm{abs}},
\end{equation}
and
\begin{equation}
    \label{eq:scat0}
    \mathrm{\kappa_{sca}} \equiv \frac{\pi a^2}{\mathcal{M}} Q_{\mathrm{sca}} = \frac{3/4}{\rho_{\mathrm{int}}af} Q_{\mathrm{sca}},
\end{equation}
where the $\mathcal{M}$ is the particle's mass, $\rho_{\mathrm{int}}$ is the internal density of the compact monomer that forms dust particles, and $f$ is the filling factor (1-$\mathcal{P}$). It is worth noting that the scattering opacity in Equation \ref{eq:scat0} often needs to be adjusted with the forward-scattering parameter $g$  \citep{ishimaru78, bohren83, birnstiel18}. That is, we should use the effective scattering opacity, $\kappa_{\mathrm{sca, eff}}$ = (1-$g$)$\kappa_{\mathrm{sca}}$. This is because the forward scattering becomes dominant when the size parameter $x > 1$. Since the forward scattering is effectively not scattering, this fraction must be adjusted in all the calculations related to scattering. The (effective) extinction opacity is the sum of the absorption and (effective) scattering opacity,
\begin{equation}
    \kappa_{\mathrm{ext}} = \mathrm{\kappa_{abs}} + \kappa_{\mathrm{sca}},
\end{equation}
and
\begin{equation}
    \kappa_{\mathrm{ext,eff}} = \mathrm{\kappa_{abs}} + \kappa_{\mathrm{sca,eff}},
\end{equation}

and the (effective) albedo is the fraction of (effective) scattered light among the total radiation,
\begin{equation}
    \omega = \frac{\kappa_{\mathrm{sca}}}{\kappa_{\mathrm{sca}} + \mathrm{\kappa_{abs}}},
\end{equation}
and
\begin{equation}
    \omega_{\mathrm{eff}} = \frac{\kappa_{\mathrm{sca,eff}}}{\kappa_{\mathrm{sca,eff}} + \mathrm{\kappa_{abs}}}.
\end{equation}
The albedo is important in both modeling dust continuum when the disk is optically thick (e.g., \citealt{zhu19b}) and the polarization. For the latter, it determines the fraction of the incident light that is scattered. Among the available scattered light, the fraction that produces polarized light is characterized by $P$, where $P$ = -$Z_{12}(a, \theta)/Z_{11}(a, \theta)$ at $\theta$ = 90$^{\circ}$\footnote{We follow the convention by taking the polarization fraction at 90$^{\circ}$. In the Rayleigh regime, this is valid since polarization fraction peaks at 90$^{\circ}$. As $x\gtrsim$ 1, the polarization fraction no longer peaks at 90$^{\circ}$, e.g., see Figure 2 in \citet{kataoka15}. Still, the value at 90$^{\circ}$ can be a good indicator of the amplitude of the polarization fraction at the peak.}. $\theta$ is the scattering angle between the scattered light and the incident light. $Z_{11}$ (relating to the total scattering by $\kappa_\mathrm{sca}(a) = \oint Z_{11}(a,\theta)\,\mathrm{d}\Omega$) and $Z_{12}$ (the linear polarized scattering) are elements in the scattering matrix (or Mueller matrix, e.g., see the appendix of \citealt{kataoka15}).
The product of $P$ and $\mathrm{\omega_{eff}}$ indicates the total polarization fraction. Note that the $P\mathrm{\omega_{eff}}$ is typically much higher than the polarization fraction measured from MCRT synthetic images, so a conversion factor $C$ is often used to convert the analytical prediction to the MCRT value. Thus, the MCRT value used to match observations is $CP\mathrm{\omega_{eff}}$, where $C$ is normalized to be $\sim$ 2\% \citep{kataoka16} for 100-$\mu$m-sized particles in the HL Tau model.

\section{Analytical approximations \label{sec:analyticalapprox}}
Suppose the refractive index of the compact particle mixture is $m_{c}$, the effective medium approximation (Maxwell-Garnett rule\footnote{Throughout the paper, we use the Bruggeman rule to calculate the refractive index as it is more accurate for our problems. Here we use Maxwell-Garnett rule to derive the analytical approximation following \citet{kataoka14}, since it is a much simpler expression. It can also capture the overall behavior of the opacity.}) gives the refractive index of the porous particle as \citep{kataoka14}
\begin{equation}
    m_{p}^{2} = \frac{1+2fF}{1-fF},
\end{equation}
where 
\begin{equation}
    F(\lambda) = \frac{m_{c}^2 -1 }{m_{c}^2 + 2}.
\end{equation}
The squares in these equations are the multiplication of complex numbers, so $m_p$, $m_c$ and $F$ are all complex numbers.
When $f<1$, it can be shown that
\begin{equation}
 n-1 \approx \frac{3}{2}f \mathrm{Re}(F), \label{eq:n-1}
\end{equation}
and
\begin{equation}
 k \approx \frac{3}{2} f \mathrm{Im}(F), \label{eq:k}
\end{equation}
so $n-1$ $\propto$ $f$, and $k$ $\propto$ $f$. For reference, the compact DSHARP mixture \citep{birnstiel18} has the refractive index $n$=2.3 and $k$=0.02 at 1 mm. With increasing porosity, the real part of the refractive index $n$-1 at 1 mm are 1.3, 0.28, 9.0$\times 10^{-2}$, and 8.8 $\times 10^{-3}$ for $f$=1, 0.3, 0.1 and 0.01 . The imaginary part $k$ are 2.0$\times 10^{-2}$, 2.7$\times 10^{-3}$, 8.2$\times 10^{-4}$, 7.9$\times 10^{-5}$. Taking the ratio of the neighbouring refractive indices between two filling factors, we find that the refractive index ratio is closer to the filling factors' ratio (as expected from Equations \ref{eq:n-1} and \ref{eq:k}) when the mixture is more porous. In other words, the approximation is better for smaller $f$. It is also worth noting that for radio wavelengths, $n$-1 $\gg$ $k$.\footnote{For further references, the refractive index of the DSHARP opacity can be found in Figure 2 in \citet{birnstiel18}. Refractive indices across all the wavelengths can be seen in Figure A.1. in \citet{kataoka14}. The dependence of refractive indices on wavelengths and levels of porosity can be reproduced using \citet{dsharp_opac} under \texttt{notebooks/porosity.ipynb}.} A recent analysis on \textit{Rosetta}/MIRO data constrains the refractive index of mm-cm-sized pebbles from sub-surface of comet 67P to be $n$ = 1.074-1.256 and $k$ = (2.580-7.431) $\times 10^{-3}$ \citep{burger23} at 1.594 mm. This is similar to the DSHARP refractive index at 1 mm with $f\sim$ 0.1.

Three regimes for analytical approximations are the \textit{Rayleigh} regime, the \textit{Mie's optically thin} regime, and the \textit{Mie's optically thick} regime. Note that the optical depth here refers to the optical depth inside a particle. The boundaries between these regimes are at $x=1$ and $kx=3/8$ (for absorption) or $(n-1)x = 1$ (for scattering). They are marked by solid vertical lines ($x=1$) and dashed vertical lines ($kx=3/8$ or $(n-1)x=1$) in Figure \ref{fig:demo}, where the x-axes are the wavelengths. For compact particles, the Mie's optically thin regime can be very narrow or even does not exist. It is also worth noting that $kx$ and $(n-1)x$ are all proportional to the product of particle size and filling factor, $af$ (Equations \ref{eq:sizeparameter}, \ref{eq:n-1}, \ref{eq:k}). Thus, the location of the dashed lines is almost constant for different $f$ with the same $af$, as long as $f$ is small.

Combining three regimes, the approximation of the absorption efficiency is \citep{kataoka14, bohren83}
\begin{equation}
\label{eq:abscoeff1}
    \begin{split}
        Q_{\mathrm{abs}} &= \frac{24nkx}{(n^2+2)^2}\\ &\mathrm{for}\ x<1;\\
        &= \frac{8kx}{3n}\big(n^3-(n^2-1)^{3/2}\big) \\ &\mathrm{for}\ x>1\ \mathrm{and}\ kx<3/8;\\
        &= 1 \\ 
        & \mathrm{for}\ x>1\ \mathrm{and}\   kx>3/8.
    \end{split}
\end{equation}
Using Equations \ref{eq:n-1} and \ref{eq:k}, the mass opacity can be written as,
\begin{equation}
\label{eq:abscoeff2}
    \begin{split}
        \mathrm{\kappa_{abs}} &\propto \frac{\mathrm{Im}(F(\lambda))}{\lambda}\ \mathrm{for}\ x<1\ \mathrm{or}\ kx<3/8;\\
        &\propto \frac{1}{af}\ \ \ \ \ \ \mathrm{for}\ x>1\ \mathrm{and}\ kx>3/8.
    \end{split}
\end{equation}
The upper left panel of Figure \ref{fig:demo} shows the single sized ($af$=160 $\mu$m) absorption opacity as a function of wavelength. {\color{blue}Blue} lines represent the compact particles, whereas {\color{orange}orange}, {\color{green}green} and {\color{red}red} lines represent porous particles with filling factors $f$= 0.3, 0.1 and 0.01. Dotted lines show the analytical approximation from Equations \ref{eq:abscoeff1} or \ref{eq:abscoeff2}. The direct calculation and the analytical approximation are in excellent agreement.

In the Rayleigh regime ($x<1$), the mass opacity does not depend on the particle size or porosity. To understand this, we can divide the dust particle into the smallest unit--atom or molecule \citep{moosmuller09}. We call them scatterers. Each scatterer has its own cross section, $\sigma_{\mathrm{abs, i}}$. The total cross section is proportional to the total number of the scatterers, $NV$$\sigma_{\mathrm{abs, i}}$, where $N$ is the number density of the scatterers and $V$ is the volume. $N$=$\rho/m_{\mathrm{mol}}$, where $m_{\mathrm{mol}}$ is the mass of the scatterer. On the other hand, the absorption mass opacity $\mathrm{\kappa_{abs}}$ is the total cross section divided by its mass, which is $NV$$\sigma_{\mathrm{abs, i}}$/$\rho V$ $\propto$ $\sigma_{\mathrm{abs, i}}$/$m_{\mathrm{mol}}$. This means that in the Rayleigh regime, the mass opacity is only dependent on the microscopic property of the particle along the wavelength, independent of the particle size and the filling factor. At long wavelengths, the opacity changes as $\lambda^{-1.7}$, and $\mathrm{Im}(F) \propto \lambda^{-0.7}$ (not shown here, but can be seen using the DSHARP refractive index $m_c$), in agreement with the approximation that $\kappa_{\mathrm{abs}}\propto \mathrm{Im}(F) / \lambda$ (Equation \ref{eq:abscoeff2}).

In the geometric regime (or Mie's optically thick regime), the opacity is inversely proportional to $af$, as the absorption cross section equals the geometric cross section ($Q_{\mathrm{abs}}=1$). The opacity is independent of the wavelength in this regime.

In Mie's optically thin regime, a unique feature for compact particles is the interference feature, in which the opacity is much higher than its porous counterparts. This interference is partially captured by the analytical approximation (dotted lines). From the viewpoint of elementary optics, it is the interference between the incident and forward-scattered light at Mie's optically thin regime. The amplitude of the absorption efficiency is proportional to $m$+1/$m$, and the separation between maxima is 1/2$a$($m$-1) (Equations 4.63 and 4.64 in \citealt{bohren83}). Since $n-1\gg k$, the amplitude ($\approx$$n+n^{-1}$) reaches a minimum when $n$ approaches unity. The $n$ of the more porous particles is closer to unity. The separation between peaks also becomes 1/$f$ wider.

Overall, the analytical approximations closely match the single-sized absorption opacity. The  opacity in all three regimes can be expressed in terms of $af$. For this reason, we use $af$ to represent the particle size throughout the paper, as advocated by \citet{kataoka14}.

\section{Scattering Opacity \label{sec:scat}}

The scattering opacity is shown on the right panels of Figure \ref{fig:demo}. Similar to the absorption opacity, the scattering opacity can also be divided into three regimes, separated by $x=1$ (solid vertical lines) and $(n-1)x=1$ (dashed vertical lines). The analytical approximation fits the single-sized un-adjusted scattering opacity closely, except at $(n-1)x=1$. Here, we emphasis the importance of adjustment for forward scattering by comparing dashed ($\kappa_{\mathrm{sca,eff}}$) and solid lines ($\kappa_{\mathrm{sca}}$). We can clearly see that the forward-scattering-adjusted scattering opacity has significantly lower values in the Mie regime ($x\gtrsim$1). This is why we must use the effective scattering opacity when the particle size is comparable or larger than the observing wavelengths.

The scattering efficiency can also be approximated in the three regimes \citep{kataoka14, bohren83},
\begin{equation}
    \begin{split}
        Q_{\mathrm{sca}} &= \frac{32}{27}x^4\big((n-1)^2+k^2\big)\\
        &\mathrm{for}\ x<1;\\
        &= \frac{32}{27}x^2\big((n-1)^2+k^2\big) \\
        &\mathrm{for}\ x>1\ \mathrm{and}\ x(n-1)<1;\\
        &= 1 \\
        &\mathrm{for}\ x>1\ \mathrm{and}\   x(n-1) > 1.
    \end{split} \label{eq:scat1}
\end{equation}
Using Equations \ref{eq:n-1} and \ref{eq:k}, the mass opacity can be written as,
\begin{equation}
    \begin{split}
        \kappa_{\mathrm{sca}} &\propto \frac{a^3f}{\lambda^4}|F(\lambda)|^2\ \mathrm{for}\ x<1;\\
        &\propto \frac{af}{\lambda^2} |F(\lambda)|^2 \ \mathrm{for}\ x>1\ \mathrm{and}\ x(n-1)<1;\\
        &\propto \frac{1}{af}\ \ \ \ \ \ \mathrm{for}\ x>1\ \mathrm{and}\ x(n-1)>1.
    \end{split}\label{eq:scat2}
\end{equation}
The upper right panel of Figure \ref{fig:demo} shows the analytical approximation against the DSHARP opacity at $af=$160$\mu$m for particles with various levels of porosity. These approximations also match the Mie calculations quite well, except around $(n-1)x$=1 due to the interference pattern.

In the Rayleigh regime, the scattering opacity is no longer proportional to the $af$ (with $\lambda \gtrsim$ 0.1 cm in the right panels of Figure \ref{fig:demo}). Unlike the absorption light, the scattered light is due to individual scatterers radiating \textit{in phase}. The scattered light energy is proportional to the square of the number of scatterers, $(NV)^2$. Thus, the mass opacity is $(NV)^2\sigma_{\mathrm{sca,i}}/\rho V$ $\propto$ $a^3f \sigma_{\mathrm{sca,i}}\rho_{\mathrm{int}}/m_{\mathrm{mol}}^2$. The $\lambda^{-4}$ dependence in Equation \ref{eq:scat2} can also be deduced by dimensional analysis \citep{moosmuller09}.

In the geometric regime (Mie's optically thick regime), the scattering efficiency approaches unity, so the mass opacity is inversely proportional to $af$. The opacity for different levels of porosity overlaps, given the same $af$. The opacity does not depend on the wavelength. However, forward scattering dominates in this regime, so the forward scattering parameter $g$ is almost unity. After removing the forward scattering, the effective scattering opacity is nearly zero. The dashed lines show the effective scattering opacity. They are much lower than the unadjusted scattering opacity, with even lower values for more porous particles. Due to the forward scattering, the scattering opacity becomes wavelength-dependent.

In Mie's optically thin regime, the mass opacity is proportional to $af$ and $\lambda^{-2}$. This regime is narrow for compact particles but wider for more porous particles. The forward scattering starts to dominate in this regime, so forward scattering adjustment is also necessary.

The analytical approximations also closely match the single-sized scattering opacity ($\kappa_{\mathrm{sca}}$, without adjustment). Aside from the Rayleigh scattering, the opacity in other regimes can be expressed in terms of $af$. However, the analytical approximations cannot capture the effective scattering opacity with the adjustment of the forward scattering, due to the $g$ factor.

\section{Polarization Fraction at 90$^\circ$ \label{sec:poldeg}}
\begin{figure*}[t!]
\includegraphics[width=\linewidth]{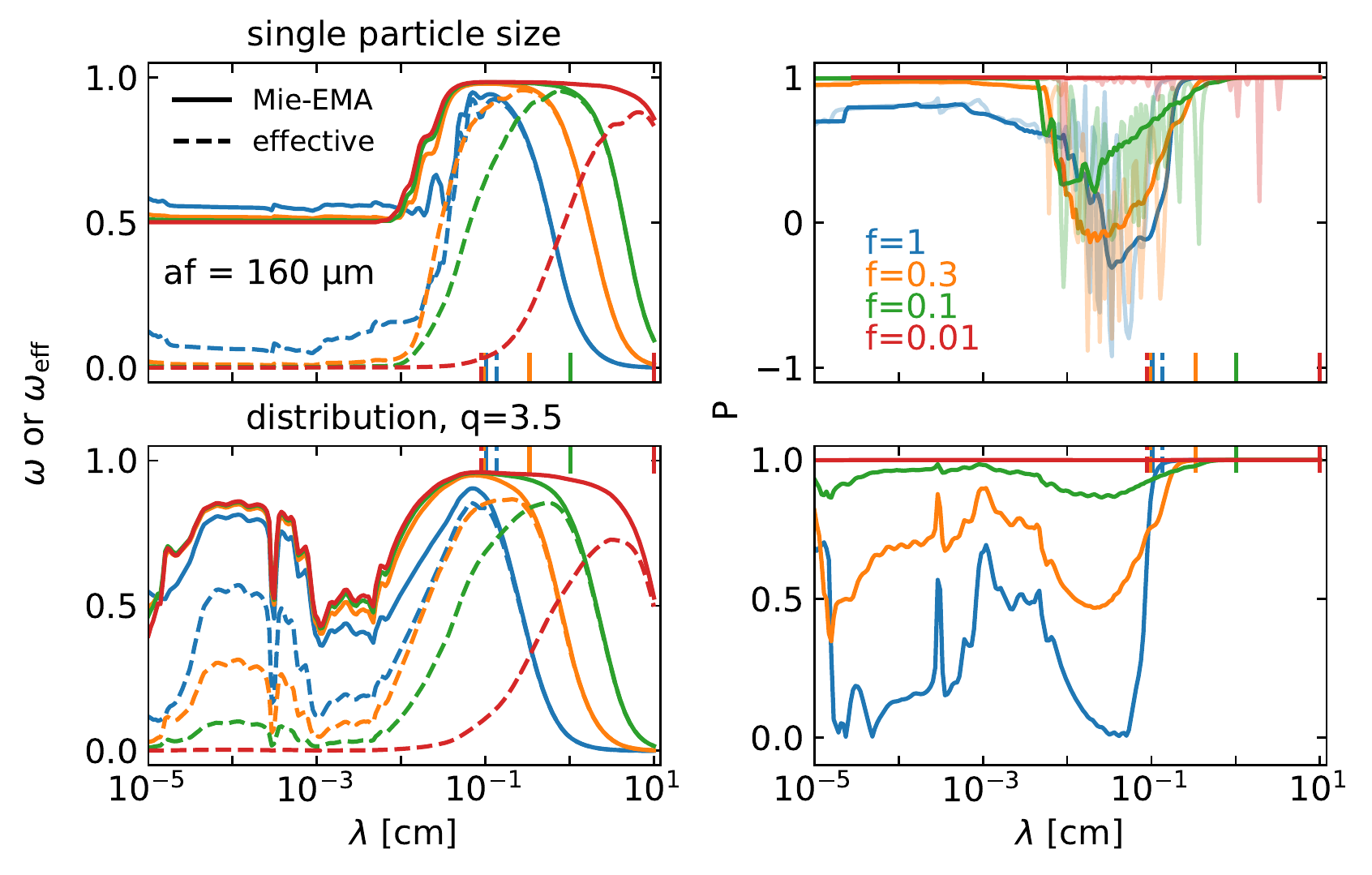}
\figcaption{\label{fig:demo2} The albedo $\omega$ or $\omega_\mathrm{eff}$ (left panels) and polarization fraction $P$ (right panels) for single-sized particles (top panels) and particles with size distribution $q$=3.5 (bottom panels). The dashed lines on the left panels are the effective albedo adjusted for the forward scattering. The more transparent lines for $P$ are the single-sized values. The more opaque lines are the values averaging 20 neighboring particle sizes to smooth out the interference feature.}
\end{figure*}

While the albedo is just the scattering and extinction opacity ratio, the effective albedo, $\mathrm{\omega_{eff}}$, is difficult to characterize using analytical approximation. The same goes for the polarization fraction at 90$^{\circ}$, $P$, which is also a scattering property. Nevertheless, we describe the behavior of albedo (left panels) and polarization fraction (right panels) in Figure \ref{fig:demo2} since they are crucial to predicting the observational polarization fraction. The values approach unity at Mie's optically thin regime (between dashed and solid vertical lines) for un-adjusted scattering albedo. The albedo drops to zero at the Rayleigh regime (longer wavelengths). At the Mie's optically thick regime  (shorter wavelengths), the albedo is 1/2, since the $Q_{\mathrm{abs}}=Q_{\mathrm{sca}}=1$ (Equations \ref{eq:abscoeff1}, \ref{eq:scat1}). After the adjustment of the forward scattering, the albedo in the whole Mie's regime drops to nearly zero, leaving only large values around $x=1$. The situation is similar with a particle size distribution, albeit moderate values at shorter wavelengths contributed from small particles.

The polarization fraction $P$ is shown on the right panels of Figure \ref{fig:demo2}. The value can be either positive or negative. When the value is positive, the polarization direction is perpendicular to the incident direction of the radiation. When the value is negative, the polarization direction is parallel to the incident radiation, also called ``polarization reversal'' (e.g., \citealt{murakawa10, kirchschlager14,kataoka15, yang16, brunngraber19}). The transparent lines on the top panel are the single-sized $P$. They have oscillating behavior between positive and negative values in the Mie regime. After averaging the values of the $Z_{11}$ and $Z_{12}$ in the 20 neighboring sizes, the polarization fraction becomes positive in most of the wavelengths but has zero or negative values around $(n-1)x = 1$. As the particle becomes more porous, this dip becomes narrower with higher values, which means that the available window for high polarization fraction becomes wider for more porous particles. For $f=0.01$, the polarization fraction is unity across all the wavelengths. A particle size distribution makes the polarization fraction in the Mie's optically thick region (at shorter wavelengths) much lower, but it does not affect more porous particles. We restrain the discussion on why more porous particles have larger $P$ in the Appendix \ref{sec:appendixA}. 

For compact particle, the effective albedo peaks at $x(n-1)=1$, and the polarization fraction drops sharply for $x>1$. This leaves only a narrow range of wavelengths and sizes that can explain the high polarization fraction in ALMA observations. The sharp drop of $P$ at $x(n-1)=1$ narrows the range once again. Increasing the porosity does not widen the window of high-value effective albedo (the peak value even decreases for $f$=0.01). Instead, the significant difference for porous particle lies in constantly high values of $P$. Most of the scattered lights are polarized for more porous particles, so only the effective albedo determines the observational polarization fraction. This leaves a much broader window of particle sizes to explain high polarization fractions in observations \citep{tazaki19}.

\section{More on Polarization Fraction \texorpdfstring{$P$}{Lg} \label{sec:appendixA}}
\begin{figure*}[t!]
\includegraphics[width=\linewidth]{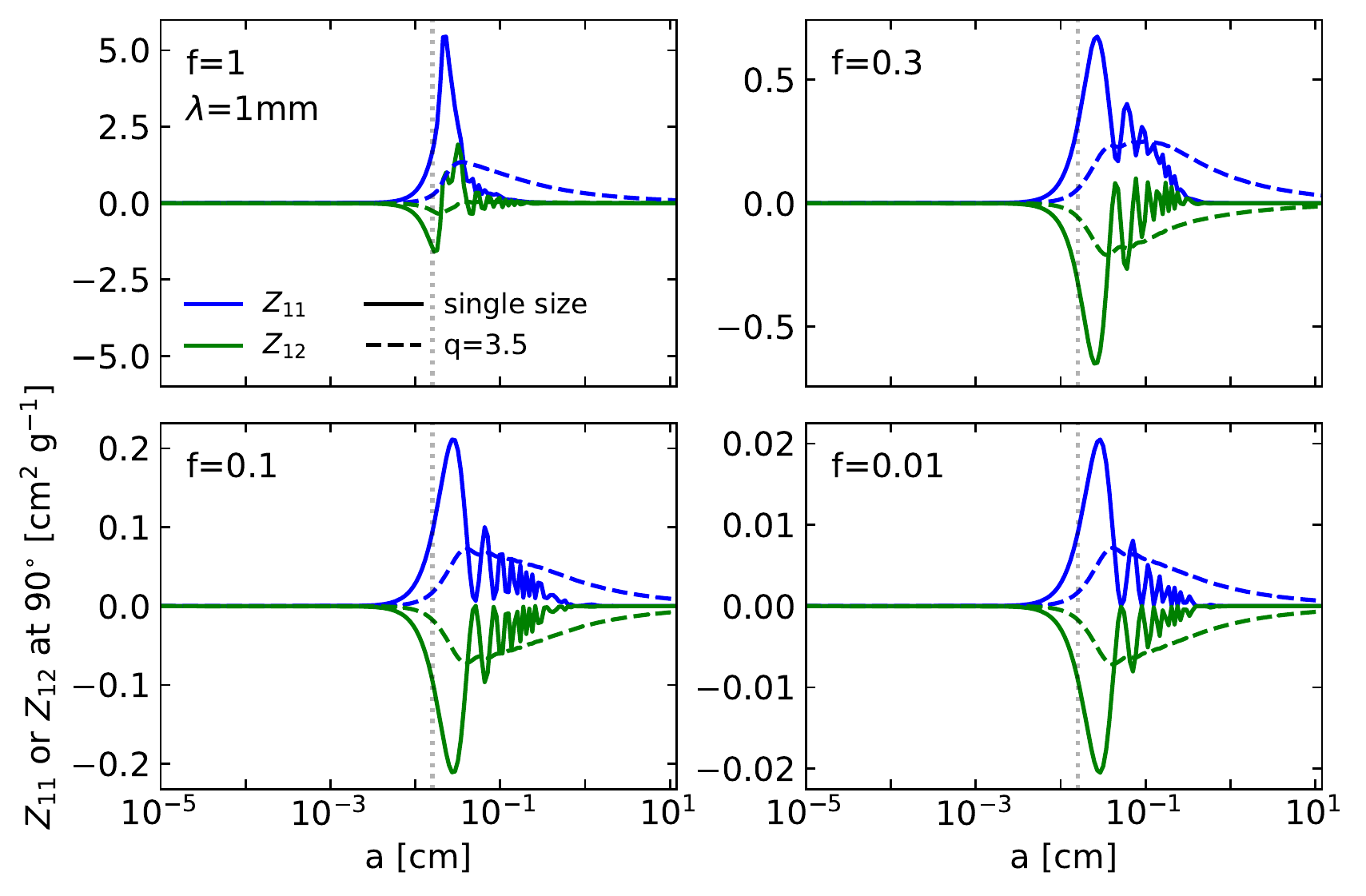}
\figcaption{\label{fig:demo3} $Z_{11}$$|_{90^\circ}$ ({\color{blue}blue curves}) and $Z_{12}$$|_{90^\circ}$ ({\color{green}green curves}) components of the scattering matrix against particle size $a$ at 1 mm wavelength for $f=$ 1, 0.3, 0.1 and 0.01. The solid lines show the single-sized values, whereas the dashed lines show the values with particle size distribution $q$=3.5. The vertical dotted lines are $x=1$.}
\end{figure*}

We use Figure \ref{fig:demo3} to show why more porous particles have constantly higher polarization fraction $P$. The scattering matrix elements $Z_{11}$ ({\color{blue}blue lines}) and $Z_{12}$ ({\color{green}green lines}) at 90$^{\circ}$ are shown for various filling factors at $\lambda$= 1 mm. The single-sized values are represented in solid lines, whereas the values with particle size distributions are represented in dashed lines.
Since $P$=-($Z_{11}/Z_{12}$)$|_{90^\circ}$, the $P$ is unity as long as $Z_{11}$$|_{90^\circ}$ and $Z_{12}$$|_{90^\circ}$ are symmetric against zero. This is the case for $x<1$ (on the left of the vertical dotted line). However, as $x>1$, both values become very large and oscillate with ever smaller amplitudes. The $P$ is very sensitive to the wavelength and particle size for single sized particles. On the other hand, the size distribution can average out this interference pattern, but the asymmetry between the two components is still strong for compact particles. At particular locations, $Z_{12}$$|_{90^\circ}$ can also be positive, so the polarization reversal occurs. As the particles become more porous, $Z_{12}$$|_{90^\circ}$ becomes more symmetric to the $Z_{11}$$|_{90^\circ}$. Thus, the polarization fraction becomes constantly high. 
An intuitive way to understand the polarization fraction is to make an analogy to the scattering between particles, where a single scattering produces polarized light, and multiple scattering depolarizes light. Similarly, within a particle, the Mie solution can be thought of as a summation of the waves scattered by individual discrete dipole sub-elements in which the coherency  comes into play to determine the net behavior of the single particle's polarization. As the size parameter becomes larger, the particle becomes more optically thick within the particle and the summation of individual waves become incoherent.  This is why $P$ is very low when $x>1$ for compact particles. For porous particles, as $f\ll 1$, the refractive index is closer to that of the vacuum and becomes more optically thin, so it preserves the high polarization fraction for a large size parameter \citep{tazaki19}.

\section{Comparison to Carrasco-Gonz\'{a}lez et al.\label{sec:appendixB}}
\begin{figure*}[t!]
\centering
\includegraphics[width=0.66\linewidth]{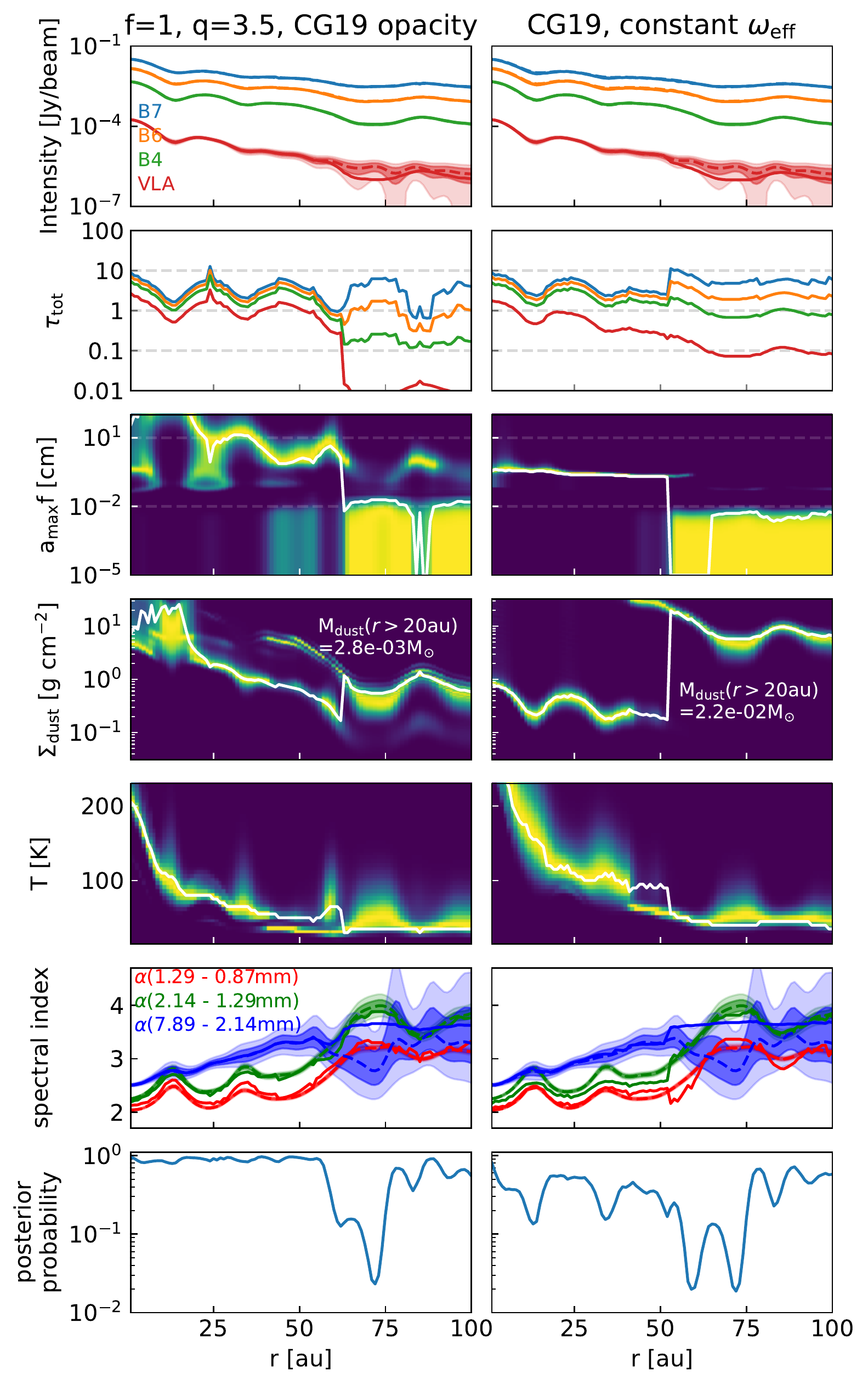}
\figcaption{Analytical fitting of SEDs for the opacity used in \citet{carrasco19} (left), the opacity used in \citet{carrasco19} but with constant $\mathrm{\omega_{eff}}$(=0.9) (right). The layout is similar to Figures \ref{fig:f1} and \ref{fig:f0p01}.
\label{fig:supplement}}
\end{figure*}

In \citet{carrasco19}, the $a_{\mathrm{max}}$ from analytical fitting is around mm, but in this paper the $a_{\mathrm{max}}f$ can be as high as 1 m, under the same assumption of particle size distribution $q$=3.5. We demonstrate that this is due to the different fitting approaches in these papers. We first use the opacity in \citet{carrasco19} to fit the disk, shown in the left panels of Figure \ref{fig:supplement}. The result is very similar to the DSHARP in Figure \ref{fig:f1} since the opacity used in the study originated in \citet{dalessio01} is very similar to the DSHARP one, except for a higher water fraction. Hence, the difference in the results should be in the fitting method. In \citet{carrasco19}, the opacity and albedo are fitted with power laws, and then the radiative equations are solved iteratively using the observations from four bands. In their paper, the albedo is assumed to be constant, $\mathrm{\omega_{eff}}$ = 0.9. However, the result of particle size is very sensitive to the albedo. A constant albedo might not be a good assumption since, for most of the particle sizes, the albedo varies with wavelengths drastically. A constant albedo might be valid only when $a_{\mathrm{max}}$ is $\sim$ 1 mm. That is why their solution is around 1 mm. In the right column of Figure \ref{fig:supplement}, we artificially set the $\mathrm{\omega_{eff}}$ to be 0.9 for all wavelengths. The best-fit $a_{\mathrm{max}}$ becomes $\sim$ 1 mm, very close to the result in \citet{carrasco19}. Additionally, \citet{carrasco19} does not consider the small particle population with $a_{\mathrm{max}}$ $\lesssim$ 100 $\mu$m. This work has no such assumption, and a more expansive parameter space has been explored. That is why the best-fit solution at the outer disk ($\gtrsim$ 60 au) are very small particles $\lesssim$ 100 $\mu$m. Considering these factors, we do not find conflicting results between these papers.

\section{More Constraints on Dust Properties from SED Fitting\label{sec:constraints_SED}}

We use analytical approximations to explore a larger parameter space of filling factors and particle size distributions to complement the observational constraints from MCRT models. The analytical approximations are less accurate than MCRT models, but they are less computational expensive. They can cover a broader and finer parameter space so we can have a better understanding on how different parameters are interconnected. 

\begin{figure*}[t!]
\includegraphics[width=\linewidth]{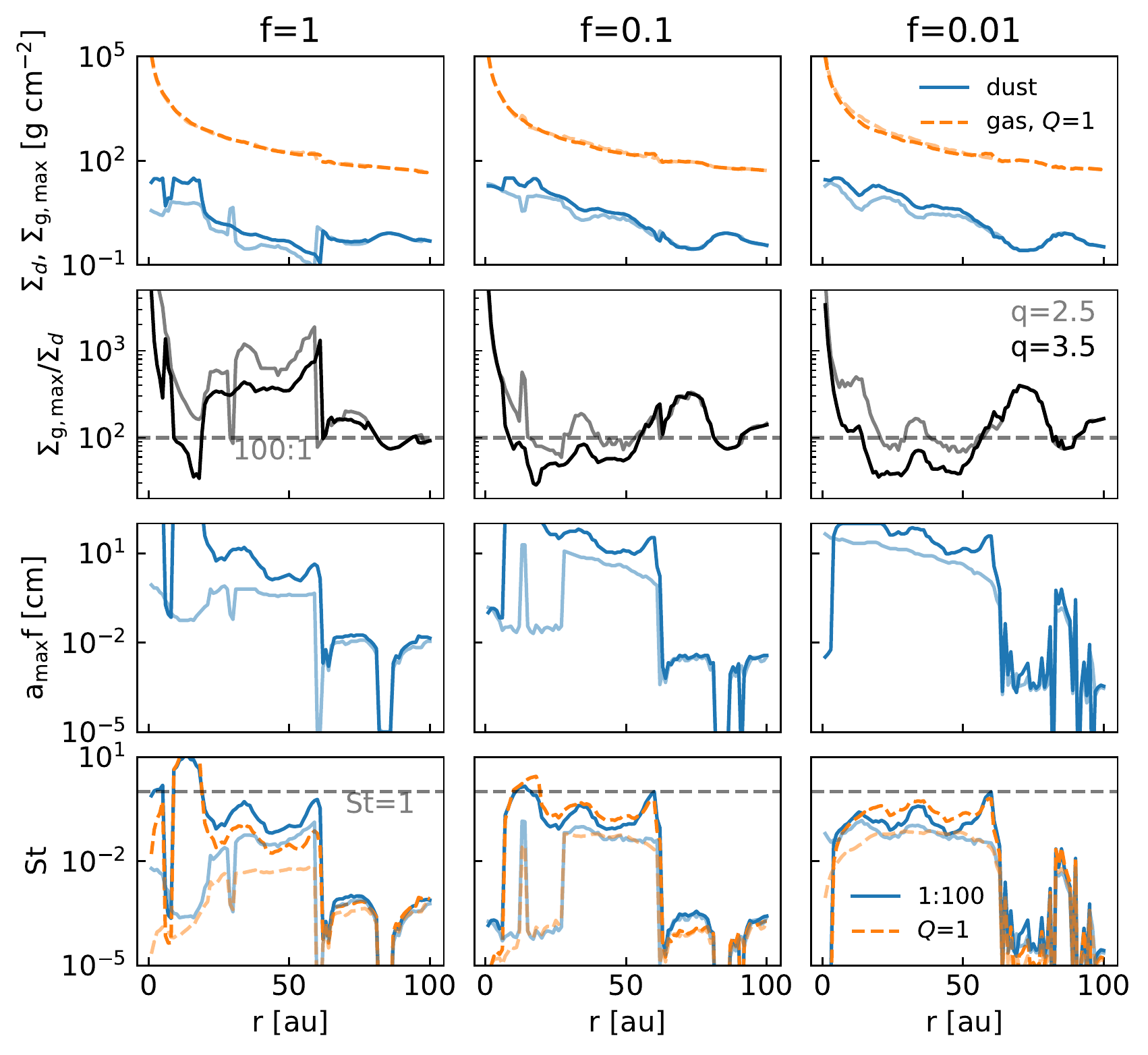}
\figcaption{Gas and dust surface densities (first row), maximally allowed gas-to-dust mass ratios (second row), maximum particle sizes (third row), and Stokes numbers (fourth row) for $f$=1 (left column), 0.1 (middle column), and 0.01 (right column) cases. Cases with a particle size distribution $q=3.5$ are shown in opaque colors, whereas the $q=2.5$ cases are shown in more transparent colors. First row: dust surface densities constrained from analytical fittings ({\color{blue}blue} solid lines, Figures \ref{fig:f1}, \ref{fig:f0p01}, \ref{fig:inferred_supplement}, and \ref{fig:q=2.5}) and the maximally allowed ($Q$=1) gas surface densities ({\color{orange}orange} dashed lines). Bottom row: the Stokes numbers of the dust particles assuming gas-to-dust mass ratio as 100 ({\color{blue}blue} solid lines), and the Stokes numbers assuming the gas has the maximal surface density ({\color{orange}orange} dashed lines). The latter is also the minimal Stokes number a particle can have.
\label{fig:discussion}}
\end{figure*}

We describe the constraints on dust surface density, gas-to-dust mass ratio, maximum particle size, Stokes number in Figure \ref{fig:discussion} using current SED observations. Although the dust surface density and maximum particle size have been reported for individual models (Figures \ref{fig:f1}, \ref{fig:f0p01}, \ref{fig:inferred_supplement}, and \ref{fig:q=2.5}), now we collect them together since they are crucial to deriving other quantities.

\subsection{Dust Surface Density}
The first row of Figure \ref{fig:discussion} shows the dust surface densities in {\color{blue}blue} lines for various filling factors ($f$=1, 0.1, and 0.01, from left to right). They are obtained from the whole-population models (Figures \ref{fig:f1}, and \ref{fig:f0p01} first columns, and Figures \ref{fig:inferred_supplement}, and \ref{fig:q=2.5}). More opaque colors are for $q$=3.5 and more transparent colors are for $q$=2.5. From compact to porous cases, the dust surface densities can increase several times within 60 au. Between $f$=0.1 and 0.01 cases, the dust surface densities are similar. For all three filling factors, the densities are similar beyond 60 au. The densities for $q$=2.5 cases are less than the $q$=3.5 cases by a small amount, but this difference is less significant than the change when we switch from compact particles to porous particles. The inner disk ($r\lesssim$ 20 au) for $f$=1 sees a factor of 5-10 difference between two size distribution slopes, and this is because the $q=2.5$ solution picks up the small-particle population and the $q=2.5$ solution falls into the big-particle population.

\subsection{Gas-to-Dust Mass Ratio}

With the obtained dust surface densities, we can estimate the maximum gas-to-dust mass ratio. This is done by calculating the maximum gas surface density when the disk becomes gravitationally unstable. This happens when $Q \lesssim$ 1, where 
\begin{equation}
    Q = \frac{c_s\Omega_K}{\pi G\Sigma_g}.
\end{equation}
$c_s$ is the sound speed, and $c_s$=$(RT/\mu)^{1/2}$; $R$ is the gas constant, $\mu$ is the mean molecular weight, $G$ is the gravitational constant, and $\Omega_K$ is the Keplarian orbital frequency. 

In Figure \ref{fig:discussion} we show the maximum gas surface densities ({\color{orange}orange} dashed curves) on the first row. The maximum gas surface densities for two different particle size distribution slopes are almost identical. The slight difference is due to their different temperatures. 

The maximum gas-to-dust ratios are shown on the second row. 
For compact particles, the gas-to-dust mass ratio can be above 100, and even 1000 within 60 au where big particles dominate. Beyond 60 au (where small particles dominate), the gas-to-dust mass ratio can only be as high as 100, since the small particles have lower opacities, leading to higher inferred dust mass. 
For porous particles (both $f$=0.1 and 0.01), the gas-to-dust mass ratios can also be very high within 20 au. Between 20 au to 60 au, they can only be as high as 20-80 (high metalicity) for $q$=3.5, since the dust surface densities are higher for porous cases. the gas-to-dust mass ratios can be 100-300 beyond 60 au. Overall, the $q=2.5$ cases can have several times higher gas-to-dust mass ratios between 20-60 au.

\subsection{Maximum Particle Size}
The third row of Figure \ref{fig:discussion} shows the maximum particle size, $a_{\mathrm{max}}f$. Models of all three filling factors show decreasing $a_{\mathrm{max}}f$ with radius. For the $q=3.5$ cases, $a_{\mathrm{max}}f$ can be as high as 1 m within 20 au, regardless of the filling factor. Between 20-60 au, $a_{\mathrm{max}}f$ is between 1-10 cm for $f$=1; and 10 cm-1 m for $f$=0.1 and 0.01. Beyond 60 au, small particles with $a_{\mathrm{max}}f$ $\lesssim$ 100 $\mu$m dominate. Models with $q$=2.5 predict much smaller $a_{\mathrm{max}}f$. They are 1 mm-1 cm for $f$=1; 1 mm-10 cm for $f$=0.1; and 1 cm-10 cm for $f$=0.01.

\subsection{Stokes Number}
The Stokes number (St) is another important parameter that determines the dynamical coupling between dust and gas (introduced in Equation \ref{eq:st}). It is proportional to $a_{\mathrm{max}}f$ over the gas surface density. The dust-gas coupling is strong when St $\ll$ 1, whereas when St $\sim$ 1, the dust particle drifts the fastest in the gas radially. On the vertical direction, a larger St means the dust particles are more settled to the midplane (Equation \ref{eq:hd}, for a fixed turbulent $\alpha_\mathrm{t}$).    

In the fourth row of Figure \ref{fig:discussion}, we plot the St assuming either gas-to-dust mass ratio as 100:1 ({\color{blue}blue} curves) or maximum gas surface density in the first row ({\color{orange}orange} dashed curves, essentially the lower bound of St). 

For all filling factors and size distribution slopes, the Stokes numbers are less than unity except for $r\sim$ 20 au for compact particles $q$=3.5, where St can be as high as 10. This is where $a_{\mathrm{max}}f$ is around 10 m. Overall, the $q$=2.5 cases have smaller St than the $q$=3.5 cases, since $a_{\mathrm{max}}f$ (third row) is smaller. For most of the cases with $q$=2.5 size distribution, the St is less than 0.1 for the whole disk. The St is $\sim 10^{-4}$ beyond 60 au for all cases.

Current dust continuum observations show that the dust particles are highly settled into the disk midplane (e.g., \citealt{pinte16, villenave22}), despite the expected vertical stirring that should be caused by the vertical shear instability \citep{stoll16, lin19, flock20, lehmann22}. \citet{dullemond22} recently show that the only way to reconcile these razor-thin disks with the presence of the VSI is constraining St$>$1. If this condition is met, the dust particles would remain completely decoupled from the gas and hence unaffected by the VSI. Our constraint on HL Tau's Stokes numbers (whether compact or porous, $q$=3.5 or 2.5) are less than unity, which means that the VSI must be somehow quenched in order to reproduce the observed high levels of settling \citep{dullemond22}. However, when the gas-to-dust mass ratio is $\lesssim$ 10, the St can be $\gtrsim$ 1, consistent with the finding in \citet{dullemond22} that a high metalicity is required for St$\gtrsim$ 1 so that the VSI can operate while the dust is settled in the midplane. On the other hand, St can be difficult to reach unity considering particle evolution without a pressure bump. When St approaches unity, the particles will drift very fast onto the star, in contrast with the large disk sizes we observed \citep{brauer08, birnstiel09, estrada22a}. In summary, including porosity does not help explain the non-detection of the dust stir-up predicted by VSI.

\section{SED and polarization fraction fitting with compact 100 micron-sized particles \label{sec:af100um}}
\begin{figure*}[t!]
\includegraphics[width=\linewidth]{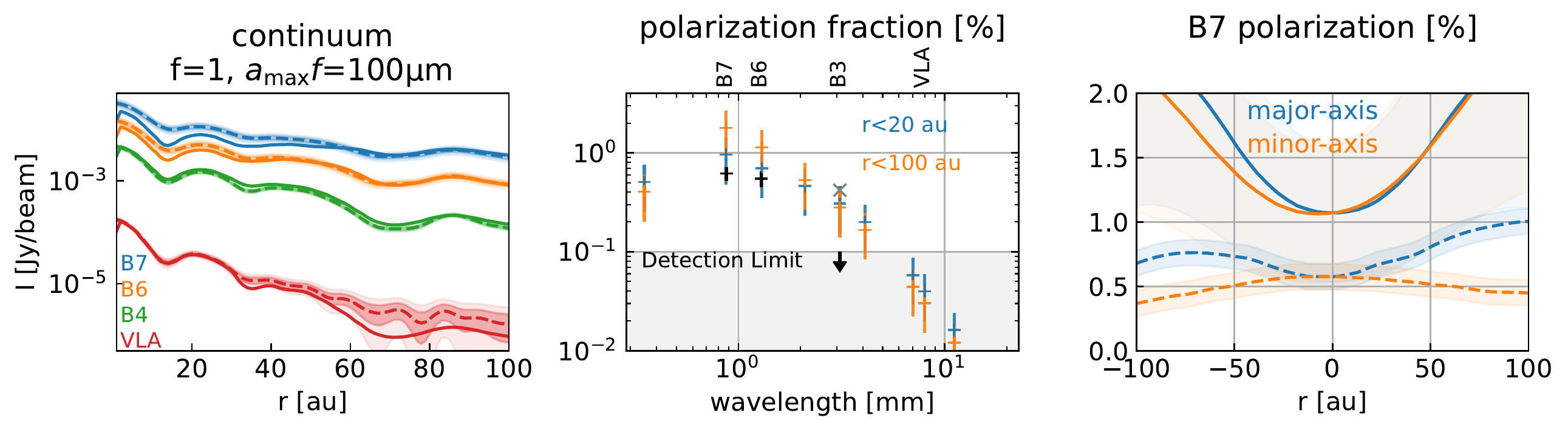}
\figcaption{The RADMC-3D (MCRT) model with constant $a_{\mathrm{max}}f$ = 100 $\mu$m for compact particles. The left panel is the continuum emissions at ALMA bands 7, 6, 4, and VLA K+Qa band. The middle panel shows the linear polarization fractions within 20 au and 100 au at various wavelengths. The right panel shows the linear polarization fractions along major and minor axes ({\color{blue}blue}: major axis; {\color{orange}orange}: minor axis; dashed lines: observation; solid lines: models).
\label{fig:100um}}
\end{figure*}

While the small particles are less favorable solutions from the SED fitting compared to the big-particle model, the solution for 100-$\mu$m particles might have a similar quality to the 0.1-$\mu$m solution by examining Figure \ref{fig:f1}. Since the constraint on $a_{\mathrm{max}}f$ is weak (the marginalized posterior probability is constantly high from 0.1-50 $\mu$m), and the best-fit is found at the parameter boundary, $a_{\mathrm{max}}f$ = 100 $\mu$m might have a similar fitting score as 0.1-$\mu$m particles. To demonstrate this, we run another simulation with constant $a_{\mathrm{max}}f$ = 100 $\mu$m, using the small-particle model's best-fit surface density and temperature. The continuum and polarization fractions are shown in Figure \ref{fig:100um}. It is clear that the model has similar SEDs as the best-fit small-particle model (Figure \ref{fig:radmcintens}) and can provide enough polarization fractions within 20 and 100 au at ALMA bands 7 and 6. If we neglect the poor quality of the SED fitting, the prediction for this solution is that the polarization fraction is negligible at longer wavelengths, e.g., ALMA bands 1 ($\sim$ 7 mm), 2 ($\sim$ 4 mm), or ngVLA bands (3 mm-2 m).

\clearpage


\clearpage

\begin{thebibliography}{}
\expandafter\ifx\csname natexlab\endcsname\relax\def\natexlab#1{#1}\fi
\providecommand{\url}[1]{\href{#1}{#1}}
\providecommand{\dodoi}[1]{doi:~\href{http://doi.org/#1}{\nolinkurl{#1}}}
\providecommand{\doeprint}[1]{\href{http://ascl.net/#1}{\nolinkurl{http://ascl.net/#1}}}
\providecommand{\doarXiv}[1]{\href{https://arxiv.org/abs/#1}{\nolinkurl{https://arxiv.org/abs/#1}}}

\bibitem[{Andrews(2020)}]{andrews20}
Andrews, S.~M. 2020, Annual Review of Astronomy and Astrophysics, 58, 483,
  \dodoi{10.1146/annurev-astro-031220-010302}

\bibitem[{{Andrews} {et~al.}(2013){Andrews}, {Rosenfeld}, {Kraus}, \&
  {Wilner}}]{andrews13}
{Andrews}, S.~M., {Rosenfeld}, K.~A., {Kraus}, A.~L., \& {Wilner}, D.~J. 2013,
  \apj, 771, 129, \dodoi{10.1088/0004-637X/771/2/129}

\bibitem[{{Ansdell} {et~al.}(2016){Ansdell}, {Williams}, {van der Marel},
  {Carpenter}, {Guidi}, {Hogerheijde}, {Mathews}, {Manara}, {Miotello},
  {Natta}, {Oliveira}, {Tazzari}, {Testi}, {van Dishoeck}, \& {van
  Terwisga}}]{ansdell16}
{Ansdell}, M., {Williams}, J.~P., {van der Marel}, N., {et~al.} 2016, \apj,
  828, 46, \dodoi{10.3847/0004-637X/828/1/46}

\bibitem[{{Bacciotti} {et~al.}(2018){Bacciotti}, {Girart}, {Padovani}, {Podio},
  {Paladino}, {Testi}, {Bianchi}, {Galli}, {Codella}, {Coffey}, {Favre}, \&
  {Fedele}}]{bacciotti18}
{Bacciotti}, F., {Girart}, J.~M., {Padovani}, M., {et~al.} 2018, \apjl, 865,
  L12, \dodoi{10.3847/2041-8213/aadf87}

\bibitem[{{Bentley} {et~al.}(2016){Bentley}, {Schmied}, {Mannel}, {Torkar},
  {Jeszenszky}, {Romstedt}, {Levasseur-Regourd}, {Weber}, {Jessberger},
  {Ehrenfreund}, {Koeberl}, \& {Havnes}}]{bentley16}
{Bentley}, M.~S., {Schmied}, R., {Mannel}, T., {et~al.} 2016, \nat, 537, 73,
  \dodoi{10.1038/nature19091}

\bibitem[{{Birnstiel}(2018)}]{dsharp_opac}
{Birnstiel}, T. 2018, {birnstiel/dsharp\_opac: revised release of package},
  v1.1.0, Zenodo,  Zenodo, \dodoi{10.5281/zenodo.1495277}

\bibitem[{Birnstiel(2018)}]{github_dsharp_opac}
Birnstiel, T. 2018, {dsharp\_opac: The DSHARP Mie-Opacity Library}, 1.1.0,
  Zenodo, \dodoi{10.5281/zenodo.1495277}

\bibitem[{{Birnstiel} {et~al.}(2009){Birnstiel}, {Dullemond}, \&
  {Brauer}}]{birnstiel09}
{Birnstiel}, T., {Dullemond}, C.~P., \& {Brauer}, F. 2009, \aap, 503, L5,
  \dodoi{10.1051/0004-6361/200912452}

\bibitem[{{Birnstiel} {et~al.}(2012){Birnstiel}, {Klahr}, \&
  {Ercolano}}]{birnstiel12}
{Birnstiel}, T., {Klahr}, H., \& {Ercolano}, B. 2012, \aap, 539, A148,
  \dodoi{10.1051/0004-6361/201118136}

\bibitem[{{Birnstiel} {et~al.}(2011){Birnstiel}, {Ormel}, \&
  {Dullemond}}]{birnstiel11}
{Birnstiel}, T., {Ormel}, C.~W., \& {Dullemond}, C.~P. 2011, \aap, 525, A11,
  \dodoi{10.1051/0004-6361/201015228}

\bibitem[{{Birnstiel} {et~al.}(2018){Birnstiel}, {Dullemond}, {Zhu}, {Andrews},
  {Bai}, {Wilner}, {Carpenter}, {Huang}, {Isella}, {Benisty}, {P{\'e}rez}, \&
  {Zhang}}]{birnstiel18}
{Birnstiel}, T., {Dullemond}, C.~P., {Zhu}, Z., {et~al.} 2018, \apjl, 869, L45,
  \dodoi{10.3847/2041-8213/aaf743}

\bibitem[{{Blum}(2018)}]{blum18}
{Blum}, J. 2018, \ssr, 214, 52, \dodoi{10.1007/s11214-018-0486-5}

\bibitem[{{Blum} {et~al.}(2022){Blum}, {Bischoff}, \& {Gundlach}}]{blum22}
{Blum}, J., {Bischoff}, D., \& {Gundlach}, B. 2022, Universe, 8, 381,
  \dodoi{10.3390/universe8070381}

\bibitem[{{Blum} {et~al.}(2017){Blum}, {Gundlach}, {Krause}, {Fulle},
  {Johansen}, {Agarwal}, {von Borstel}, {Shi}, {Hu}, {Bentley}, {Capaccioni},
  {Colangeli}, {Della Corte}, {Fougere}, {Green}, {Ivanovski}, {Mannel},
  {Merouane}, {Migliorini}, {Rotundi}, {Schmied}, \& {Snodgrass}}]{blum17}
{Blum}, J., {Gundlach}, B., {Krause}, M., {et~al.} 2017, \mnras, 469, S755,
  \dodoi{10.1093/mnras/stx2741}

\bibitem[{{Bohren} \& {Huffman}(1983)}]{bohren83}
{Bohren}, C.~F., \& {Huffman}, D.~R. 1983, {Absorption and scattering of light
  by small particles}

\bibitem[{{Brauer} {et~al.}(2008){Brauer}, {Dullemond}, \&
  {Henning}}]{brauer08}
{Brauer}, F., {Dullemond}, C.~P., \& {Henning}, T. 2008, \aap, 480, 859,
  \dodoi{10.1051/0004-6361:20077759}

\bibitem[{{Brunngr{\"a}ber} \& {Wolf}(2019)}]{brunngraber19}
{Brunngr{\"a}ber}, R., \& {Wolf}, S. 2019, \aap, 627, L10,
  \dodoi{10.1051/0004-6361/201935169}

\bibitem[{{Brunngr{\"a}ber} \& {Wolf}(2021)}]{brunngraber21}
---. 2021, \aap, 648, A87, \dodoi{10.1051/0004-6361/202040033}

\bibitem[{{Bukhari Syed} {et~al.}(2017){Bukhari Syed}, {Blum}, {Wahlberg
  Jansson}, \& {Johansen}}]{bukhari17}
{Bukhari Syed}, M., {Blum}, J., {Wahlberg Jansson}, K., \& {Johansen}, A. 2017,
  \apj, 834, 145, \dodoi{10.3847/1538-4357/834/2/145}

\bibitem[{{B{\"u}rger} {et~al.}(2023){B{\"u}rger}, {Gli{\ss}mann},
  {Lethuillier}, {Bischoff}, {Gundlach}, {Mutschke}, {H{\"o}fer}, {Wolf}, \&
  {Blum}}]{burger23}
{B{\"u}rger}, J., {Gli{\ss}mann}, T., {Lethuillier}, A., {et~al.} 2023, \mnras,
  519, 641, \dodoi{10.1093/mnras/stac3420}

\bibitem[{{Calvet} {et~al.}(2002){Calvet}, {D'Alessio}, {Hartmann}, {Wilner},
  {Walsh}, \& {Sitko}}]{calvet02}
{Calvet}, N., {D'Alessio}, P., {Hartmann}, L., {et~al.} 2002, \apj, 568, 1008,
  \dodoi{10.1086/339061}

\bibitem[{{Carrasco-Gonz{\'a}lez} {et~al.}(2016){Carrasco-Gonz{\'a}lez},
  {Henning}, {Chandler}, {Linz}, {P{\'e}rez}, {Rodr{\'\i}guez},
  {Galv{\'a}n-Madrid}, {Anglada}, {Birnstiel}, {van Boekel}, {Flock}, {Klahr},
  {Macias}, {Menten}, {Osorio}, {Testi}, {Torrelles}, \& {Zhu}}]{carrasco16}
{Carrasco-Gonz{\'a}lez}, C., {Henning}, T., {Chandler}, C.~J., {et~al.} 2016,
  \apjl, 821, L16, \dodoi{10.3847/2041-8205/821/1/L16}

\bibitem[{{Carrasco-Gonz{\'a}lez} {et~al.}(2019){Carrasco-Gonz{\'a}lez},
  {Sierra}, {Flock}, {Zhu}, {Henning}, {Chandler}, {Galv{\'a}n-Madrid},
  {Mac{\'\i}as}, {Anglada}, {Linz}, {Osorio}, {Rodr{\'\i}guez}, {Testi},
  {Torrelles}, {P{\'e}rez}, \& {Liu}}]{carrasco19}
{Carrasco-Gonz{\'a}lez}, C., {Sierra}, A., {Flock}, M., {et~al.} 2019, \apj,
  883, 71, \dodoi{10.3847/1538-4357/ab3d33}

\bibitem[{{Ciarletti} {et~al.}(2017){Ciarletti}, {Herique}, {Lasue},
  {Levasseur-Regourd}, {Plettemeier}, {Lemmonier}, {Guiffaut}, {Pasquero}, \&
  {Kofman}}]{ciarletti17}
{Ciarletti}, V., {Herique}, A., {Lasue}, J., {et~al.} 2017, \mnras, 469, S805,
  \dodoi{10.1093/mnras/stx3132}

\bibitem[{{Cieza} {et~al.}(2019){Cieza}, {Ru{\'\i}z-Rodr{\'\i}guez}, {Hales},
  {Casassus}, {P{\'e}rez}, {Gonzalez-Ruilova}, {C{\'a}novas}, {Williams},
  {Zurlo}, {Ansdell}, {Avenhaus}, {Bayo}, {Bertrang}, {Christiaens}, {Dent},
  {Ferrero}, {Gamen}, {Olofsson}, {Orcajo}, {Pe{\~n}a Ram{\'\i}rez},
  {Principe}, {Schreiber}, \& {van der Plas}}]{cieza19}
{Cieza}, L.~A., {Ru{\'\i}z-Rodr{\'\i}guez}, D., {Hales}, A., {et~al.} 2019,
  \mnras, 482, 698, \dodoi{10.1093/mnras/sty2653}

\bibitem[{{D'Alessio} {et~al.}(2001){D'Alessio}, {Calvet}, \&
  {Hartmann}}]{dalessio01}
{D'Alessio}, P., {Calvet}, N., \& {Hartmann}, L. 2001, \apj, 553, 321,
  \dodoi{10.1086/320655}

\bibitem[{{Dent} {et~al.}(2019){Dent}, {Pinte}, {Cortes}, {M{\'e}nard},
  {Hales}, {Fomalont}, \& {de Gregorio-Monsalvo}}]{dent19}
{Dent}, W.~R.~F., {Pinte}, C., {Cortes}, P.~C., {et~al.} 2019, \mnras, 482,
  L29, \dodoi{10.1093/mnrasl/sly181}

\bibitem[{{Dohnanyi}(1969)}]{dohnanyi69}
{Dohnanyi}, J.~S. 1969, \jgr, 74, 2531, \dodoi{10.1029/JB074i010p02531}

\bibitem[{{Dominik} {et~al.}(2021){Dominik}, {Min}, \& {Tazaki}}]{optool}
{Dominik}, C., {Min}, M., \& {Tazaki}, R. 2021, {OpTool: Command-line driven
  tool for creating complex dust opacities}, Astrophysics Source Code Library,
  record ascl:2104.010.
\newblock \doeprint{2104.010}

\bibitem[{{Draine}(2003)}]{draine03}
{Draine}, B.~T. 2003, \araa, 41, 241,
  \dodoi{10.1146/annurev.astro.41.011802.094840}

\bibitem[{{Draine}(2006)}]{draine06}
---. 2006, \apj, 636, 1114, \dodoi{10.1086/498130}

\bibitem[{{Dubrulle} {et~al.}(1995){Dubrulle}, {Morfill}, \&
  {Sterzik}}]{dubrulle95}
{Dubrulle}, B., {Morfill}, G., \& {Sterzik}, M. 1995, Icarus, 114, 237,
  \dodoi{10.1006/icar.1995.1058}

\bibitem[{{Dullemond} {et~al.}(2012){Dullemond}, {Juhasz}, {Pohl}, {Sereshti},
  {Shetty}, {Peters}, {Commercon}, \& {Flock}}]{dullemond12}
{Dullemond}, C.~P., {Juhasz}, A., {Pohl}, A., {et~al.} 2012, {RADMC-3D: A
  multi-purpose radiative transfer tool}.
\newblock \doeprint{1202.015}

\bibitem[{{Dullemond} {et~al.}(2022){Dullemond}, {Ziampras}, {Ostertag}, \&
  {Dominik}}]{dullemond22}
{Dullemond}, C.~P., {Ziampras}, A., {Ostertag}, D., \& {Dominik}, C. 2022,
  \aap, 668, A105, \dodoi{10.1051/0004-6361/202244218}

\bibitem[{{Estrada} \& {Cuzzi}(2022)}]{estrada22b}
{Estrada}, P.~R., \& {Cuzzi}, J.~N. 2022, \apj, 936, 40,
  \dodoi{10.3847/1538-4357/ac81c6}

\bibitem[{{Estrada} {et~al.}(2016){Estrada}, {Cuzzi}, \& {Morgan}}]{estrada16}
{Estrada}, P.~R., {Cuzzi}, J.~N., \& {Morgan}, D.~A. 2016, \apj, 818, 200,
  \dodoi{10.3847/0004-637X/818/2/200}

\bibitem[{{Estrada} {et~al.}(2022){Estrada}, {Cuzzi}, \&
  {Umurhan}}]{estrada22a}
{Estrada}, P.~R., {Cuzzi}, J.~N., \& {Umurhan}, O.~M. 2022, \apj, 936, 42,
  \dodoi{10.3847/1538-4357/ac7ffd}

\bibitem[{{Finkbeiner} {et~al.}(1999){Finkbeiner}, {Davis}, \&
  {Schlegel}}]{finkbeiner99}
{Finkbeiner}, D.~P., {Davis}, M., \& {Schlegel}, D.~J. 1999, \apj, 524, 867,
  \dodoi{10.1086/307852}

\bibitem[{{Flock} {et~al.}(2020){Flock}, {Turner}, {Nelson}, {Lyra}, {Manger},
  \& {Klahr}}]{flock20}
{Flock}, M., {Turner}, N.~J., {Nelson}, R.~P., {et~al.} 2020, \apj, 897, 155,
  \dodoi{10.3847/1538-4357/ab9641}

\bibitem[{{Fulle} \& {Blum}(2017)}]{fulle17}
{Fulle}, M., \& {Blum}, J. 2017, \mnras, 469, S39, \dodoi{10.1093/mnras/stx971}

\bibitem[{{Ginski} {et~al.}(2023){Ginski}, {Tazaki}, {Dominik}, \&
  {Stolker}}]{ginski23}
{Ginski}, C., {Tazaki}, R., {Dominik}, C., \& {Stolker}, T. 2023, arXiv
  e-prints, arXiv:2301.04617.
\newblock \doarXiv{2301.04617}

\bibitem[{{Golabek} \& {Jutzi}(2021)}]{golabek21}
{Golabek}, G.~J., \& {Jutzi}, M. 2021, \icarus, 363, 114437,
  \dodoi{10.1016/j.icarus.2021.114437}

\bibitem[{{Grant} {et~al.}(2021){Grant}, {Espaillat}, {Wendeborn}, {Tobin},
  {Mac{\'\i}as}, {Rilinger}, {Ribas}, {Megeath}, {Fischer}, {Calvet}, \& {Hee
  Kim}}]{grant21}
{Grant}, S.~L., {Espaillat}, C.~C., {Wendeborn}, J., {et~al.} 2021, \apj, 913,
  123, \dodoi{10.3847/1538-4357/abf432}

\bibitem[{{Groussin} {et~al.}(2019){Groussin}, {Attree}, {Brouet}, {Ciarletti},
  {Davidsson}, {Filacchione}, {Fischer}, {Gundlach}, {Knapmeyer},
  {Knollenberg}, {Kokotanekova}, {K{\"u}hrt}, {Leyrat}, {Marshall}, {Pelivan},
  {Skorov}, {Snodgrass}, {Spohn}, \& {Tosi}}]{groussin19}
{Groussin}, O., {Attree}, N., {Brouet}, Y., {et~al.} 2019, \ssr, 215, 29,
  \dodoi{10.1007/s11214-019-0594-x}

\bibitem[{{Guidi} {et~al.}(2022){Guidi}, {Isella}, {Testi}, {Chandler}, {Liu},
  {Schmid}, {Rosotti}, {Meng}, {Jennings}, {Williams}, {Carpenter}, {de
  Gregorio-Monsalvo}, {Li}, {Liu}, {Ortolani}, {Quanz}, {Ricci}, \&
  {Tazzari}}]{guidi22}
{Guidi}, G., {Isella}, A., {Testi}, L., {et~al.} 2022, \aap, 664, A137,
  \dodoi{10.1051/0004-6361/202142303}

\bibitem[{{Guillet} {et~al.}(2020){Guillet}, {Girart}, {Maury}, \&
  {Alves}}]{guillet20}
{Guillet}, V., {Girart}, J.~M., {Maury}, A.~J., \& {Alves}, F.~O. 2020, \aap,
  634, L15, \dodoi{10.1051/0004-6361/201937314}

\bibitem[{{G{\"u}ttler} {et~al.}(2010){G{\"u}ttler}, {Blum}, {Zsom}, {Ormel},
  \& {Dullemond}}]{guttler10}
{G{\"u}ttler}, C., {Blum}, J., {Zsom}, A., {Ormel}, C.~W., \& {Dullemond},
  C.~P. 2010, \aap, 513, A56, \dodoi{10.1051/0004-6361/200912852}

\bibitem[{{G{\"u}ttler} {et~al.}(2019){G{\"u}ttler}, {Mannel}, {Rotundi},
  {Merouane}, {Fulle}, {Bockel{\'e}e-Morvan}, {Lasue}, {Levasseur-Regourd},
  {Blum}, {Naletto}, {Sierks}, {Hilchenbach}, {Tubiana}, {Capaccioni},
  {Paquette}, {Flandes}, {Moreno}, {Agarwal}, {Bodewits}, {Bertini}, {Tozzi},
  {Hornung}, {Langevin}, {Kr{\"u}ger}, {Longobardo}, {Della Corte}, {T{\'o}th},
  {Filacchione}, {Ivanovski}, {Mottola}, \& {Rinaldi}}]{guttler19}
{G{\"u}ttler}, C., {Mannel}, T., {Rotundi}, A., {et~al.} 2019, \aap, 630, A24,
  \dodoi{10.1051/0004-6361/201834751}

\bibitem[{Harris {et~al.}(2020)Harris, Millman, van~der Walt, Gommers,
  Virtanen, Cournapeau, Wieser, Taylor, Berg, Smith, Kern, Picus, Hoyer, van
  Kerkwijk, Brett, Haldane, del R{\'{i}}o, Wiebe, Peterson,
  G{\'{e}}rard-Marchant, Sheppard, Reddy, Weckesser, Abbasi, Gohlke, \&
  Oliphant}]{numpy}
Harris, C.~R., Millman, K.~J., van~der Walt, S.~J., {et~al.} 2020, Nature, 585,
  357, \dodoi{10.1038/s41586-020-2649-2}

\bibitem[{{Henning} \& {Stognienko}(1996)}]{henning96}
{Henning}, T., \& {Stognienko}, R. 1996, \aap, 311, 291

\bibitem[{{Herique} {et~al.}(2016){Herique}, {Kofman}, {Beck}, {Bonal},
  {Buttarazzi}, {Heggy}, {Lasue}, {Levasseur-Regourd}, {Quirico}, \&
  {Zine}}]{herique16}
{Herique}, A., {Kofman}, W., {Beck}, P., {et~al.} 2016, \mnras, 462, S516,
  \dodoi{10.1093/mnras/stx040}

\bibitem[{{Huang} {et~al.}(2018){Huang}, {Andrews}, {Cleeves}, {{\"O}berg},
  {Wilner}, {Bai}, {Birnstiel}, {Carpenter}, {Hughes}, {Isella}, {P{\'e}rez},
  {Ricci}, \& {Zhu}}]{huang18}
{Huang}, J., {Andrews}, S.~M., {Cleeves}, L.~I., {et~al.} 2018, \apj, 852, 122,
  \dodoi{10.3847/1538-4357/aaa1e7}

\bibitem[{{Huang} {et~al.}(2020){Huang}, {Andrews}, {Dullemond}, {{\"O}berg},
  {Qi}, {Zhu}, {Birnstiel}, {Carpenter}, {Isella}, {Mac{\'\i}as}, {McClure},
  {P{\'e}rez}, {Teague}, {Wilner}, \& {Zhang}}]{huang20}
{Huang}, J., {Andrews}, S.~M., {Dullemond}, C.~P., {et~al.} 2020, \apj, 891,
  48, \dodoi{10.3847/1538-4357/ab711e}

\bibitem[{{Hull} {et~al.}(2018){Hull}, {Yang}, {Li}, {Kataoka}, {Stephens},
  {Andrews}, {Bai}, {Cleeves}, {Hughes}, {Looney}, {P{\'e}rez}, \&
  {Wilner}}]{hull18}
{Hull}, C. L.~H., {Yang}, H., {Li}, Z.-Y., {et~al.} 2018, \apj, 860, 82,
  \dodoi{10.3847/1538-4357/aabfeb}

\bibitem[{Hunter(2007)}]{matplotlib}
Hunter, J.~D. 2007, Computing In Science \& Engineering, 9, 90

\bibitem[{{Ishimaru}(1978)}]{ishimaru78}
{Ishimaru}, A. 1978, {Wave propagation and scattering in random media. Volume 1
  - Single scattering and transport theory}, Vol.~1,
  \dodoi{10.1016/B978-0-12-374701-3.X5001-7}

\bibitem[{{Johansen} {et~al.}(2014){Johansen}, {Blum}, {Tanaka}, {Ormel},
  {Bizzarro}, \& {Rickman}}]{johansen14}
{Johansen}, A., {Blum}, J., {Tanaka}, H., {et~al.} 2014, in Protostars \&
  Planets VI, eds.~H.~Beuther, R.~Klessen, C.~Dullemond, \& Th.~Henning
  (Univ.~Arizona Press: Tucson), in press.
\newblock \doarXiv{1402.1344}

\bibitem[{{Jorda} {et~al.}(2016){Jorda}, {Gaskell}, {Capanna}, {Hviid}, {Lamy},
  {{\v{D}}urech}, {Faury}, {Groussin}, {Guti{\'e}rrez}, {Jackman}, {Keihm},
  {Keller}, {Knollenberg}, {K{\"u}hrt}, {Marchi}, {Mottola}, {Palmer},
  {Schloerb}, {Sierks}, {Vincent}, {A'Hearn}, {Barbieri}, {Rodrigo}, {Koschny},
  {Rickman}, {Barucci}, {Bertaux}, {Bertini}, {Cremonese}, {Da Deppo},
  {Davidsson}, {Debei}, {De Cecco}, {Fornasier}, {Fulle}, {G{\"u}ttler}, {Ip},
  {Kramm}, {K{\"u}ppers}, {Lara}, {Lazzarin}, {Lopez Moreno}, {Marzari},
  {Naletto}, {Oklay}, {Thomas}, {Tubiana}, \& {Wenzel}}]{jorda16}
{Jorda}, L., {Gaskell}, R., {Capanna}, C., {et~al.} 2016, \icarus, 277, 257,
  \dodoi{10.1016/j.icarus.2016.05.002}

\bibitem[{{Jutzi} \& {Asphaug}(2015)}]{jutzi15}
{Jutzi}, M., \& {Asphaug}, E. 2015, Science, 348, 1355,
  \dodoi{10.1126/science.aaa4747}

\bibitem[{{Jutzi} \& {Benz}(2017)}]{jutzi17}
{Jutzi}, M., \& {Benz}, W. 2017, \aap, 597, A62,
  \dodoi{10.1051/0004-6361/201628964}

\bibitem[{{Kaeufer} {et~al.}(2023){Kaeufer}, {Woitke}, {Min}, {Kamp}, \&
  {Pinte}}]{kaeufer23}
{Kaeufer}, T., {Woitke}, P., {Min}, M., {Kamp}, I., \& {Pinte}, C. 2023, \aap,
  672, A30, \dodoi{10.1051/0004-6361/202245461}

\bibitem[{{Kataoka} {et~al.}(2016{\natexlab{a}}){Kataoka}, {Muto}, {Momose},
  {Tsukagoshi}, \& {Dullemond}}]{kataoka16}
{Kataoka}, A., {Muto}, T., {Momose}, M., {Tsukagoshi}, T., \& {Dullemond},
  C.~P. 2016{\natexlab{a}}, \apj, 820, 54, \dodoi{10.3847/0004-637X/820/1/54}

\bibitem[{{Kataoka} {et~al.}(2014){Kataoka}, {Okuzumi}, {Tanaka}, \&
  {Nomura}}]{kataoka14}
{Kataoka}, A., {Okuzumi}, S., {Tanaka}, H., \& {Nomura}, H. 2014, \aap, 568,
  A42, \dodoi{10.1051/0004-6361/201323199}

\bibitem[{{Kataoka} {et~al.}(2013){Kataoka}, {Tanaka}, {Okuzumi}, \&
  {Wada}}]{kataoka13}
{Kataoka}, A., {Tanaka}, H., {Okuzumi}, S., \& {Wada}, K. 2013, \aap, 557, L4,
  \dodoi{10.1051/0004-6361/201322151}

\bibitem[{{Kataoka} {et~al.}(2017){Kataoka}, {Tsukagoshi}, {Pohl}, {Muto},
  {Nagai}, {Stephens}, {Tomisaka}, \& {Momose}}]{kataoka2017}
{Kataoka}, A., {Tsukagoshi}, T., {Pohl}, A., {et~al.} 2017, \apjl, 844, L5,
  \dodoi{10.3847/2041-8213/aa7e33}

\bibitem[{{Kataoka} {et~al.}(2015){Kataoka}, {Muto}, {Momose}, {Tsukagoshi},
  {Fukagawa}, {Shibai}, {Hanawa}, {Murakawa}, \& {Dullemond}}]{kataoka15}
{Kataoka}, A., {Muto}, T., {Momose}, M., {et~al.} 2015, \apj, 809, 78,
  \dodoi{10.1088/0004-637X/809/1/78}

\bibitem[{{Kataoka} {et~al.}(2016{\natexlab{b}}){Kataoka}, {Tsukagoshi},
  {Momose}, {Nagai}, {Muto}, {Dullemond}, {Pohl}, {Fukagawa}, {Shibai},
  {Hanawa}, \& {Murakawa}}]{kataoka16b}
{Kataoka}, A., {Tsukagoshi}, T., {Momose}, M., {et~al.} 2016{\natexlab{b}},
  \apjl, 831, L12, \dodoi{10.3847/2041-8205/831/2/L12}

\bibitem[{{Kempf} {et~al.}(1999){Kempf}, {Pfalzner}, \& {Henning}}]{kempf99}
{Kempf}, S., {Pfalzner}, S., \& {Henning}, T.~K. 1999, \icarus, 141, 388,
  \dodoi{10.1006/icar.1999.6171}

\bibitem[{{Kirchschlager} \& {Bertrang}(2020)}]{kirchschlager20}
{Kirchschlager}, F., \& {Bertrang}, G. H.~M. 2020, \aap, 638, A116,
  \dodoi{10.1051/0004-6361/202037943}

\bibitem[{{Kirchschlager} {et~al.}(2019){Kirchschlager}, {Bertrang}, \&
  {Flock}}]{kirchschlager19}
{Kirchschlager}, F., {Bertrang}, G. H.~M., \& {Flock}, M. 2019, \mnras, 488,
  1211, \dodoi{10.1093/mnras/stz1763}

\bibitem[{{Kirchschlager} \& {Wolf}(2014)}]{kirchschlager14}
{Kirchschlager}, F., \& {Wolf}, S. 2014, \aap, 568, A103,
  \dodoi{10.1051/0004-6361/201323176}

\bibitem[{{Kofman} {et~al.}(2015){Kofman}, {Herique}, {Barbin}, {Barriot},
  {Ciarletti}, {Clifford}, {Edenhofer}, {Elachi}, {Eyraud}, {Goutail}, {Heggy},
  {Jorda}, {Lasue}, {Levasseur-Regourd}, {Nielsen}, {Pasquero}, {Preusker},
  {Puget}, {Plettemeier}, {Rogez}, {Sierks}, {Statz}, {Svedhem}, {Williams},
  {Zine}, \& {Van Zyl}}]{kofman15}
{Kofman}, W., {Herique}, A., {Barbin}, Y., {et~al.} 2015, Science, 349, 2.639,
  \dodoi{10.1126/science.aab0639}

\bibitem[{{Krause} \& {Blum}(2004)}]{krause04}
{Krause}, M., \& {Blum}, J. 2004, \prl, 93, 021103,
  \dodoi{10.1103/PhysRevLett.93.021103}

\bibitem[{{Krijt} {et~al.}(2015){Krijt}, {Ormel}, {Dominik}, \&
  {Tielens}}]{krijt15}
{Krijt}, S., {Ormel}, C.~W., {Dominik}, C., \& {Tielens}, A.~G.~G.~M. 2015,
  \aap, 574, A83, \dodoi{10.1051/0004-6361/201425222}

\bibitem[{{Lehmann} \& {Lin}(2022)}]{lehmann22}
{Lehmann}, M., \& {Lin}, M.~K. 2022, \aap, 658, A156,
  \dodoi{10.1051/0004-6361/202142378}

\bibitem[{{Lesur} {et~al.}(2022){Lesur}, {Ercolano}, {Flock}, {Lin}, {Yang},
  {Barranco}, {Benitez-Llambay}, {Goodman}, {Johansen}, {Klahr}, {Laibe},
  {Lyra}, {Marcus}, {Nelson}, {Squire}, {Simon}, {Turner}, {Umurhan}, \&
  {Youdin}}]{lesur22}
{Lesur}, G., {Ercolano}, B., {Flock}, M., {et~al.} 2022, arXiv e-prints,
  arXiv:2203.09821.
\newblock \doarXiv{2203.09821}

\bibitem[{{Li} \& {Draine}(2001)}]{li01}
{Li}, A., \& {Draine}, B.~T. 2001, \apj, 554, 778, \dodoi{10.1086/323147}

\bibitem[{{Lichtenberg} \& {Krijt}(2021)}]{lichtenberg21}
{Lichtenberg}, T., \& {Krijt}, S. 2021, \apjl, 913, L20,
  \dodoi{10.3847/2041-8213/abfdce}

\bibitem[{{Lin}(2019)}]{lin19}
{Lin}, M.-K. 2019, \mnras, 485, 5221, \dodoi{10.1093/mnras/stz701}

\bibitem[{{Lin} {et~al.}(2020){Lin}, {Li}, {Yang}, {Looney}, {Stephens}, \&
  {Hull}}]{lin20}
{Lin}, Z.-Y.~D., {Li}, Z.-Y., {Yang}, H., {et~al.} 2020, \mnras, 496, 169,
  \dodoi{10.1093/mnras/staa1499}

\bibitem[{{Lin} {et~al.}(2022){Lin}, {Li}, {Yang}, {Stephens}, {Looney},
  {Harrison}, \& {Fern{\'a}ndez-L{\'o}pez}}]{lin22a}
---. 2022, \mnras, 512, 3922, \dodoi{10.1093/mnras/stac753}

\bibitem[{{Lin} {et~al.}(2023){Lin}, {Li}, {Yang}, {Mu{\~n}oz}, {Looney},
  {Stephens}, {Hull}, {Fern{\'a}ndez-L{\'o}pez}, \& {Harrison}}]{lin23}
---. 2023, \mnras, 520, 1210, \dodoi{10.1093/mnras/stad173}

\bibitem[{{Liu}(2019)}]{liu19}
{Liu}, H.~B. 2019, \apjl, 877, L22, \dodoi{10.3847/2041-8213/ab1f8e}

\bibitem[{{Liu} {et~al.}(2022){Liu}, {Linz}, {Fang}, {Henning}, {Wolf},
  {Flock}, {Rosotti}, {Wang}, \& {Li}}]{liu22}
{Liu}, Y., {Linz}, H., {Fang}, M., {et~al.} 2022, \aap, 668, A175,
  \dodoi{10.1051/0004-6361/202244505}

\bibitem[{{Lorek} {et~al.}(2018){Lorek}, {Lacerda}, \& {Blum}}]{lorek18}
{Lorek}, S., {Lacerda}, P., \& {Blum}, J. 2018, \aap, 611, A18,
  \dodoi{10.1051/0004-6361/201630175}

\bibitem[{{Mac{\'\i}as} {et~al.}(2021){Mac{\'\i}as}, {Guerra-Alvarado},
  {Carrasco-Gonz{\'a}lez}, {Ribas}, {Espaillat}, {Huang}, \&
  {Andrews}}]{macias21}
{Mac{\'\i}as}, E., {Guerra-Alvarado}, O., {Carrasco-Gonz{\'a}lez}, C., {et~al.}
  2021, \aap, 648, A33, \dodoi{10.1051/0004-6361/202039812}

\bibitem[{{Mannel} {et~al.}(2016){Mannel}, {Bentley}, {Schmied}, {Jeszenszky},
  {Levasseur-Regourd}, {Romstedt}, \& {Torkar}}]{mannel16}
{Mannel}, T., {Bentley}, M.~S., {Schmied}, R., {et~al.} 2016, \mnras, 462,
  S304, \dodoi{10.1093/mnras/stw2898}

\bibitem[{{Mannel} {et~al.}(2019){Mannel}, {Bentley}, {Boakes}, {Jeszenszky},
  {Ehrenfreund}, {Engrand}, {Koeberl}, {Levasseur-Regourd}, {Romstedt},
  {Schmied}, {Torkar}, \& {Weber}}]{mannel19}
{Mannel}, T., {Bentley}, M.~S., {Boakes}, P.~D., {et~al.} 2019, \aap, 630, A26,
  \dodoi{10.1051/0004-6361/201834851}

\bibitem[{{Mathis} {et~al.}(1977){Mathis}, {Rumpl}, \& {Nordsieck}}]{mathis77}
{Mathis}, J.~S., {Rumpl}, W., \& {Nordsieck}, K.~H. 1977, \apj, 217, 425,
  \dodoi{10.1086/155591}

\bibitem[{{Merouane} {et~al.}(2016){Merouane}, {Zaprudin}, {Stenzel},
  {Langevin}, {Altobelli}, {Della Corte}, {Fischer}, {Fulle}, {Hornung},
  {Sil{\'e}n}, {Ligier}, {Rotundi}, {Ryno}, {Schulz}, {Hilchenbach}, {Kissel},
  \& {COSIMA Team}}]{merouane16}
{Merouane}, S., {Zaprudin}, B., {Stenzel}, O., {et~al.} 2016, \aap, 596, A87,
  \dodoi{10.1051/0004-6361/201527958}

\bibitem[{{Miotello} {et~al.}(2022){Miotello}, {Kamp}, {Birnstiel}, {Cleeves},
  \& {Kataoka}}]{miotello22}
{Miotello}, A., {Kamp}, I., {Birnstiel}, T., {Cleeves}, L.~I., \& {Kataoka}, A.
  2022, arXiv e-prints, arXiv:2203.09818.
\newblock \doarXiv{2203.09818}

\bibitem[{{Mishchenko} {et~al.}(2000){Mishchenko}, {Hovenier}, \&
  {Travis}}]{mishchenko00}
{Mishchenko}, M.~I., {Hovenier}, J.~W., \& {Travis}, L.~D. 2000, {Light
  scattering by nonspherical particles : theory, measurements, and
  applications}

\bibitem[{{Miyake} \& {Nakagawa}(1993)}]{miyake93}
{Miyake}, K., \& {Nakagawa}, Y. 1993, Icarus, 106, 20,
  \dodoi{10.1006/icar.1993.1156}

\bibitem[{Moosmüller \& Arnott(2009)}]{moosmuller09}
Moosmüller, H., \& Arnott, W.~P. 2009, Journal of the Air \& Waste Management
  Association, 59, 1028, \dodoi{10.3155/1047-3289.59.9.1028}

\bibitem[{{Mori} \& {Kataoka}(2021)}]{mori21}
{Mori}, T., \& {Kataoka}, A. 2021, \apj, 908, 153,
  \dodoi{10.3847/1538-4357/abd08a}

\bibitem[{{Mori} {et~al.}(2019){Mori}, {Kataoka}, {Ohashi}, {Momose}, {Muto},
  {Nagai}, \& {Tsukagoshi}}]{mori19}
{Mori}, T., {Kataoka}, A., {Ohashi}, S., {et~al.} 2019, \apj, 883, 16,
  \dodoi{10.3847/1538-4357/ab3575}

\bibitem[{{Mousis} {et~al.}(2017){Mousis}, {Drouard}, {Vernazza}, {Lunine},
  {Monnereau}, {Maggiolo}, {Altwegg}, {Balsiger}, {Berthelier}, {Cessateur},
  {De Keyser}, {Fuselier}, {Gasc}, {Korth}, {Le Deun}, {Mall}, {Marty},
  {R{\`e}me}, {Rubin}, {Tzou}, {Waite}, \& {Wurz}}]{mousis17}
{Mousis}, O., {Drouard}, A., {Vernazza}, P., {et~al.} 2017, \apjl, 839, L4,
  \dodoi{10.3847/2041-8213/aa6839}

\bibitem[{{Murakawa}(2010)}]{murakawa10}
{Murakawa}, K. 2010, \aap, 518, A63, \dodoi{10.1051/0004-6361/201014159}

\bibitem[{{Ohashi} \& {Kataoka}(2019)}]{ohashi19}
{Ohashi}, S., \& {Kataoka}, A. 2019, \apj, 886, 103,
  \dodoi{10.3847/1538-4357/ab5107}

\bibitem[{{Ohashi} {et~al.}(2018){Ohashi}, {Kataoka}, {Nagai}, {Momose},
  {Muto}, {Hanawa}, {Fukagawa}, {Tsukagoshi}, {Murakawa}, \&
  {Shibai}}]{ohashi18}
{Ohashi}, S., {Kataoka}, A., {Nagai}, H., {et~al.} 2018, \apj, 864, 81,
  \dodoi{10.3847/1538-4357/aad632}

\bibitem[{{Okuzumi} {et~al.}(2012){Okuzumi}, {Tanaka}, {Kobayashi}, \&
  {Wada}}]{okuzumi12}
{Okuzumi}, S., {Tanaka}, H., {Kobayashi}, H., \& {Wada}, K. 2012, \apj, 752,
  106, \dodoi{10.1088/0004-637X/752/2/106}

\bibitem[{{Ossenkopf}(1993)}]{ossenkopf93}
{Ossenkopf}, V. 1993, \aap, 280, 617

\bibitem[{{Pascucci} {et~al.}(2016){Pascucci}, {Testi}, {Herczeg}, {Long},
  {Manara}, {Hendler}, {Mulders}, {Krijt}, {Ciesla}, {Henning}, {Mohanty},
  {Drabek-Maunder}, {Apai}, {Sz{\H u}cs}, {Sacco}, \& {Olofsson}}]{pascucci16}
{Pascucci}, I., {Testi}, L., {Herczeg}, G.~J., {et~al.} 2016, \apj, 831, 125,
  \dodoi{10.3847/0004-637X/831/2/125}

\bibitem[{{P{\"a}tzold} {et~al.}(2016){P{\"a}tzold}, {Andert}, {Hahn}, {Asmar},
  {Barriot}, {Bird}, {H{\"a}usler}, {Peter}, {Tellmann}, {Gr{\"u}n},
  {Weissman}, {Sierks}, {Jorda}, {Gaskell}, {Preusker}, \&
  {Scholten}}]{patzold16}
{P{\"a}tzold}, M., {Andert}, T., {Hahn}, M., {et~al.} 2016, \nat, 530, 63,
  \dodoi{10.1038/nature16535}

\bibitem[{{Perez} {et~al.}(2015){Perez}, {Casassus}, {M{\'e}nard}, {Roman},
  {van der Plas}, {Cieza}, {Pinte}, {Christiaens}, \& {Hales}}]{perez15}
{Perez}, S., {Casassus}, S., {M{\'e}nard}, F., {et~al.} 2015, \apj, 798, 85,
  \dodoi{10.1088/0004-637X/798/2/85}

\bibitem[{{Pinte} {et~al.}(2016){Pinte}, {Dent}, {M{\'e}nard}, {Hales}, {Hill},
  {Cortes}, \& {de Gregorio-Monsalvo}}]{pinte16}
{Pinte}, C., {Dent}, W.~R.~F., {M{\'e}nard}, F., {et~al.} 2016, \apj, 816, 25,
  \dodoi{10.3847/0004-637X/816/1/25}

\bibitem[{{Ribas} {et~al.}(2020){Ribas}, {Espaillat}, {Mac{\'\i}as}, \&
  {Sarro}}]{ribas20}
{Ribas}, {\'A}., {Espaillat}, C.~C., {Mac{\'\i}as}, E., \& {Sarro}, L.~M. 2020,
  \aap, 642, A171, \dodoi{10.1051/0004-6361/202038352}

\bibitem[{{Rilinger} {et~al.}(2023){Rilinger}, {Espaillat}, {Xin}, {Ribas},
  {Mac{\'\i}as}, \& {Luettgen}}]{rilinger22}
{Rilinger}, A.~M., {Espaillat}, C.~C., {Xin}, Z., {et~al.} 2023, \apj, 944, 66,
  \dodoi{10.3847/1538-4357/aca905}

\bibitem[{{Schwartz} {et~al.}(2018){Schwartz}, {Michel}, {Jutzi}, {Marchi},
  {Zhang}, \& {Richardson}}]{schwartz18}
{Schwartz}, S.~R., {Michel}, P., {Jutzi}, M., {et~al.} 2018, Nature Astronomy,
  2, 379, \dodoi{10.1038/s41550-018-0395-2}

\bibitem[{{Sierra} \& {Lizano}(2020)}]{sierra20}
{Sierra}, A., \& {Lizano}, S. 2020, \apj, 892, 136,
  \dodoi{10.3847/1538-4357/ab7d32}

\bibitem[{{Sierra} {et~al.}(2019){Sierra}, {Lizano}, {Mac{\'\i}as},
  {Carrasco-Gonz{\'a}lez}, {Osorio}, \& {Flock}}]{sierra19}
{Sierra}, A., {Lizano}, S., {Mac{\'\i}as}, E., {et~al.} 2019, \apj, 876, 7,
  \dodoi{10.3847/1538-4357/ab1265}

\bibitem[{{Sierra} {et~al.}(2021){Sierra}, {P{\'e}rez}, {Zhang}, {Law},
  {Guzm{\'a}n}, {Qi}, {Bosman}, {{\"O}berg}, {Andrews}, {Long}, {Teague},
  {Booth}, {Walsh}, {Wilner}, {M{\'e}nard}, {Cataldi}, {Czekala}, {Bae},
  {Huang}, {Bergner}, {Ilee}, {Benisty}, {Le Gal}, {Loomis}, {Tsukagoshi},
  {Liu}, {Yamato}, \& {Aikawa}}]{sierra21}
{Sierra}, A., {P{\'e}rez}, L.~M., {Zhang}, K., {et~al.} 2021, \apjs, 257, 14,
  \dodoi{10.3847/1538-4365/ac1431}

\bibitem[{{Skorov} \& {Blum}(2012)}]{skorov12}
{Skorov}, Y., \& {Blum}, J. 2012, \icarus, 221, 1,
  \dodoi{10.1016/j.icarus.2012.01.012}

\bibitem[{{Stephens} {et~al.}(2014){Stephens}, {Looney}, {Kwon},
  {Fern{\'a}ndez-L{\'o}pez}, {Hughes}, {Mundy}, {Crutcher}, {Li}, \&
  {Rao}}]{stephens14}
{Stephens}, I.~W., {Looney}, L.~W., {Kwon}, W., {et~al.} 2014, \nat, 514, 597,
  \dodoi{10.1038/nature13850}

\bibitem[{{Stephens} {et~al.}(2017){Stephens}, {Yang}, {Li}, {Looney},
  {Kataoka}, {Kwon}, {Fern{\'a}ndez-L{\'o}pez}, {Hull}, {Hughes}, {Segura-Cox},
  {Mundy}, {Crutcher}, \& {Rao}}]{stephens17}
{Stephens}, I.~W., {Yang}, H., {Li}, Z.-Y., {et~al.} 2017, \apj, 851, 55,
  \dodoi{10.3847/1538-4357/aa998b}

\bibitem[{{Stoll} \& {Kley}(2016)}]{stoll16}
{Stoll}, M. H.~R., \& {Kley}, W. 2016, \aap, 594, A57,
  \dodoi{10.1051/0004-6361/201527716}

\bibitem[{{Tanaka} {et~al.}(1996){Tanaka}, {Inaba}, \& {Nakazawa}}]{tanaka96}
{Tanaka}, H., {Inaba}, S., \& {Nakazawa}, K. 1996, \icarus, 123, 450,
  \dodoi{10.1006/icar.1996.0170}

\bibitem[{{Tazaki} {et~al.}(2023){Tazaki}, {Ginski}, \& {Dominik}}]{tazaki23}
{Tazaki}, R., {Ginski}, C., \& {Dominik}, C. 2023, \apjl, 944, L43,
  \dodoi{10.3847/2041-8213/acb824}

\bibitem[{{Tazaki} \& {Tanaka}(2018)}]{tazaki18}
{Tazaki}, R., \& {Tanaka}, H. 2018, \apj, 860, 79,
  \dodoi{10.3847/1538-4357/aac32d}

\bibitem[{{Tazaki} {et~al.}(2019{\natexlab{a}}){Tazaki}, {Tanaka}, {Kataoka},
  {Okuzumi}, \& {Muto}}]{tazaki19}
{Tazaki}, R., {Tanaka}, H., {Kataoka}, A., {Okuzumi}, S., \& {Muto}, T.
  2019{\natexlab{a}}, \apj, 885, 52, \dodoi{10.3847/1538-4357/ab45f0}

\bibitem[{{Tazaki} {et~al.}(2019{\natexlab{b}}){Tazaki}, {Tanaka}, {Muto},
  {Kataoka}, \& {Okuzumi}}]{tazaki19a}
{Tazaki}, R., {Tanaka}, H., {Muto}, T., {Kataoka}, A., \& {Okuzumi}, S.
  2019{\natexlab{b}}, \mnras, 485, 4951, \dodoi{10.1093/mnras/stz662}

\bibitem[{{Tazaki} {et~al.}(2016){Tazaki}, {Tanaka}, {Okuzumi}, {Kataoka}, \&
  {Nomura}}]{tazaki16}
{Tazaki}, R., {Tanaka}, H., {Okuzumi}, S., {Kataoka}, A., \& {Nomura}, H. 2016,
  \apj, 823, 70, \dodoi{10.3847/0004-637X/823/2/70}

\bibitem[{{Tazzari} {et~al.}(2016){Tazzari}, {Testi}, {Ercolano}, {Natta},
  {Isella}, {Chandler}, {P{\'e}rez}, {Andrews}, {Wilner}, {Ricci}, {Henning},
  {Linz}, {Kwon}, {Corder}, {Dullemond}, {Carpenter}, {Sargent}, {Mundy},
  {Storm}, {Calvet}, {Greaves}, {Lazio}, \& {Deller}}]{tazzari16}
{Tazzari}, M., {Testi}, L., {Ercolano}, B., {et~al.} 2016, \aap, 588, A53,
  \dodoi{10.1051/0004-6361/201527423}

\bibitem[{{Ueda} {et~al.}(2020){Ueda}, {Kataoka}, \& {Tsukagoshi}}]{ueda20}
{Ueda}, T., {Kataoka}, A., \& {Tsukagoshi}, T. 2020, \apj, 893, 125,
  \dodoi{10.3847/1538-4357/ab8223}

\bibitem[{{Ueda} {et~al.}(2022){Ueda}, {Kataoka}, \& {Tsukagoshi}}]{ueda22}
---. 2022, \apj, 930, 56, \dodoi{10.3847/1538-4357/ac634d}

\bibitem[{{Ueda} {et~al.}(2021){Ueda}, {Kataoka}, {Zhang}, {Zhu},
  {Carrasco-Gonz{\'a}lez}, \& {Sierra}}]{ueda21}
{Ueda}, T., {Kataoka}, A., {Zhang}, S., {et~al.} 2021, \apj, 913, 117,
  \dodoi{10.3847/1538-4357/abf7b8}

\bibitem[{{Villenave} {et~al.}(2022){Villenave}, {Stapelfeldt}, {Duch{\^e}ne},
  {M{\'e}nard}, {Lambrechts}, {Sierra}, {Flores}, {Dent}, {Wolff}, {Ribas},
  {Benisty}, {Cuello}, \& {Pinte}}]{villenave22}
{Villenave}, M., {Stapelfeldt}, K.~R., {Duch{\^e}ne}, G., {et~al.} 2022, \apj,
  930, 11, \dodoi{10.3847/1538-4357/ac5fae}

\bibitem[{Virtanen {et~al.}(2020)Virtanen, Gommers, Oliphant, Haberland, Reddy,
  Cournapeau, Burovski, Peterson, Weckesser, Bright, {van der Walt}, Brett,
  Wilson, Millman, Mayorov, Nelson, Jones, Kern, Larson, Carey, Polat, Feng,
  Moore, {VanderPlas}, Laxalde, Perktold, Cimrman, Henriksen, Quintero, Harris,
  Archibald, Ribeiro, Pedregosa, {van Mulbregt}, \& {SciPy 1.0
  Contributors}}]{scipy}
Virtanen, P., Gommers, R., Oliphant, T.~E., {et~al.} 2020, Nature Methods, 17,
  261, \dodoi{10.1038/s41592-019-0686-2}

\bibitem[{{Wahlberg Jansson} {et~al.}(2017){Wahlberg Jansson}, {Johansen},
  {Bukhari Syed}, \& {Blum}}]{wahlberg17}
{Wahlberg Jansson}, K., {Johansen}, A., {Bukhari Syed}, M., \& {Blum}, J. 2017,
  \apj, 835, 109, \dodoi{10.3847/1538-4357/835/1/109}

\bibitem[{{Warren} \& {Brandt}(2008)}]{warren08}
{Warren}, S.~G., \& {Brandt}, R.~E. 2008, Journal of Geophysical Research
  (Atmospheres), 113, D14220, \dodoi{10.1029/2007JD009744}

\bibitem[{{Weidenschilling} \& {Cuzzi}(1993)}]{weidenschilling93}
{Weidenschilling}, S.~J., \& {Cuzzi}, J.~N. 1993, in Protostars and Planets
  III, ed. E.~H. {Levy} \& J.~I. {Lunine}, 1031

\bibitem[{{Weidling} {et~al.}(2009){Weidling}, {G{\"u}ttler}, {Blum}, \&
  {Brauer}}]{weidling09}
{Weidling}, R., {G{\"u}ttler}, C., {Blum}, J., \& {Brauer}, F. 2009, \apj, 696,
  2036, \dodoi{10.1088/0004-637X/696/2/2036}

\bibitem[{{Weissman} {et~al.}(2020){Weissman}, {Morbidelli}, {Davidsson}, \&
  {Blum}}]{weissman20}
{Weissman}, P., {Morbidelli}, A., {Davidsson}, B., \& {Blum}, J. 2020, \ssr,
  216, 6, \dodoi{10.1007/s11214-019-0625-7}

\bibitem[{{Wurm} \& {Blum}(1998)}]{wurm98}
{Wurm}, G., \& {Blum}, J. 1998, \icarus, 132, 125,
  \dodoi{10.1006/icar.1998.5891}

\bibitem[{{Xin} {et~al.}(2023){Xin}, {Espaillat}, {Rilinger}, {Ribas}, \&
  {Mac{\'\i}as}}]{xin23}
{Xin}, Z., {Espaillat}, C.~C., {Rilinger}, A.~M., {Ribas}, {\'A}., \&
  {Mac{\'\i}as}, E. 2023, \apj, 942, 4, \dodoi{10.3847/1538-4357/aca52b}

\bibitem[{{Yang} \& {Li}(2020)}]{yang20}
{Yang}, H., \& {Li}, Z.-Y. 2020, \apj, 889, 15,
  \dodoi{10.3847/1538-4357/ab5f08}

\bibitem[{{Yang} {et~al.}(2016){Yang}, {Li}, {Looney}, \& {Stephens}}]{yang16}
{Yang}, H., {Li}, Z.-Y., {Looney}, L., \& {Stephens}, I. 2016, \mnras, 456,
  2794, \dodoi{10.1093/mnras/stv2633}

\bibitem[{{Yang} {et~al.}(2017){Yang}, {Li}, {Looney}, {Girart}, \&
  {Stephens}}]{yang17}
{Yang}, H., {Li}, Z.-Y., {Looney}, L.~W., {Girart}, J.~M., \& {Stephens}, I.~W.
  2017, \mnras, 472, 373, \dodoi{10.1093/mnras/stx1951}

\bibitem[{{Yang} {et~al.}(2019){Yang}, {Li}, {Stephens}, {Kataoka}, \&
  {Looney}}]{yang19}
{Yang}, H., {Li}, Z.-Y., {Stephens}, I.~W., {Kataoka}, A., \& {Looney}, L.
  2019, \mnras, 483, 2371, \dodoi{10.1093/mnras/sty3263}

\bibitem[{{Youdin} \& {Goodman}(2005)}]{youdin2005}
{Youdin}, A.~N., \& {Goodman}, J. 2005, \apj, 620, 459, \dodoi{10.1086/426895}

\bibitem[{{Youdin} \& {Lithwick}(2007)}]{youdin2007}
{Youdin}, A.~N., \& {Lithwick}, Y. 2007, \icarus, 192, 588,
  \dodoi{10.1016/j.icarus.2007.07.012}

\bibitem[{{Zhang} {et~al.}(2021){Zhang}, {Hu}, {Zhu}, \& {Bae}}]{zhang21}
{Zhang}, S., {Hu}, X., {Zhu}, Z., \& {Bae}, J. 2021, \apj, 923, 70,
  \dodoi{10.3847/1538-4357/ac2c82}

\bibitem[{{Zhu} {et~al.}(2019){Zhu}, {Zhang}, {Jiang}, {Kataoka}, {Birnstiel},
  {Dullemond}, {Andrews}, {Huang}, {P{\'e}rez}, {Carpenter}, {Bai}, {Wilner},
  \& {Ricci}}]{zhu19b}
{Zhu}, Z., {Zhang}, S., {Jiang}, Y.-F., {et~al.} 2019, \apjl, 877, L18,
  \dodoi{10.3847/2041-8213/ab1f8c}

\bibitem[{{Zsom} {et~al.}(2010){Zsom}, {Ormel}, {G{\"u}ttler}, {Blum}, \&
  {Dullemond}}]{zsom10}
{Zsom}, A., {Ormel}, C.~W., {G{\"u}ttler}, C., {Blum}, J., \& {Dullemond},
  C.~P. 2010, \aap, 513, A57, \dodoi{10.1051/0004-6361/200912976}

\end{thebibliography}
\end{document}